\pdfoutput=1

\documentclass[11pt,twoside,a4paper,cmspaper,final,collab]{cms-tdr}
\def\svnVersion{264846}\def\cmsCernNo{2013-037}\def\cmsCernDate{\today}\def\cmsMessage{Published in the European Physical Journal C as \href{http://dx.doi.org/10.1140/epjc/s10052-014-3076-z}{\doi{10.1140/epjc/s10052-014-3076-z.}}}

\begin{document}\cmsNoteHeader{HIG-13-001}

\hyphenation{had-ron-i-za-tion}
\hyphenation{cal-or-i-me-ter}
\hyphenation{de-vices}
\RCS$Revision: 258840 $
\RCS$HeadURL: svn+ssh://svn.cern.ch/reps/tdr2/papers/HIG-13-001/trunk/HIG-13-001.tex $
\RCS$Id: HIG-13-001.tex 258840 2014-09-03 18:29:03Z alverson $
\newlength\cmsFigWidth
\ifthenelse{\boolean{cms@external}}{\setlength\cmsFigWidth{0.95\columnwidth}}{\setlength\cmsFigWidth{0.6\textwidth}}
\ifthenelse{\boolean{cms@external}}{\providecommand{\cmsLeft}{top}}{\providecommand{\cmsLeft}{left}}
\ifthenelse{\boolean{cms@external}}{\providecommand{\cmsRight}{bottom}}{\providecommand{\cmsRight}{right}}
\newlength\cmsFigWidthOne
\ifthenelse{\boolean{cms@external}}{\setlength\cmsFigWidthOne{0.45\textwidth}}{\setlength\cmsFigWidthOne{0.45\textwidth}}

\newcommand{\mgg}{\ensuremath{m_{\gamma\gamma}}\xspace}
\newcommand{\mjj}{\ensuremath{m_\mathrm{jj}}\xspace}
\newcommand{\mH}{\ensuremath{m_\PH\xspace}}
\newcommand{\ptga}{\ensuremath{p_\mathrm{T}^{\gamma1}}\xspace}
\newcommand{\ptgb}{\ensuremath{p_\mathrm{T}^{\gamma2}}\xspace}
\newcommand{\ptja}{\ensuremath{p_\mathrm{T}^\mathrm{j1}}\xspace}
\newcommand{\ptjb}{\ensuremath{p_\mathrm{T}^\mathrm{j2}}\xspace}
\newcommand{\ptgg}{\ensuremath{p_{\mathrm{T}}^{\gamma\gamma}}\xspace}
\newcommand{\ptZ}{\ensuremath{p_{\mathrm{T}}^\mathrm{Z}}\xspace}
\newcommand{\Dggjj}{\ensuremath{\Delta\phi_{\Pgg\Pgg\mathrm{jj}}}\xspace}
\newcommand{\Hgg}{\ensuremath{\PH\to\Pgg\Pgg}\xspace}
\newcommand{\Zee}{\ensuremath{\cPZ\to\Pep\Pem}\xspace}
\newcommand{\Zmm}{\ensuremath{\cPZ\to\Pgmp\Pgmm}\xspace}
\newcommand{\Zmmg}{\ensuremath{\cPZ\to\Pgmp\Pgmm\Pgg}\xspace}
\newcommand{\Wenu}{\ensuremath{\PW\to\Pe\Pgn}\xspace}
\newcommand{\ttH}{\ensuremath{\cPqt\cPaqt\PH}\xspace}
\newcommand{\CLs}{\ensuremath{\mathrm{CL}_\mathrm{s}}\xspace}
\newcommand{\HqT}{{\textsc{h}q\textsc{t}}\xspace}
\newcommand{\musm}{\ensuremath{\sigma/\sigma_\mathrm{SM}}\xspace}
\newcommand{\muhat}{\ensuremath{\hat{\mu}}\xspace}
\newcommand{\mHhat}{\ensuremath{\widehat{m}_\PH}\xspace}
\newcommand{\costhetastar}{\ensuremath{\cos\theta^{\ast}_{\mbox{\tiny{CS}}}}\xspace}
\newcommand{\abscostheta}{\ensuremath{|{\cos\theta^{\ast}_{\mbox{\tiny{CS}}}}|}\xspace}
\newcommand{\absetamax}{\ensuremath{\abs{\eta}_{\mbox{\tiny{max}}}}\xspace}
\newcommand{\fqq}{\ensuremath{f_{\cPq\cPaq}}\xspace}
\newcommand{\zerop}{\ensuremath{0^{+}}\xspace}
\newcommand{\twomp}{\ensuremath{2^{+}_{m}}\xspace}

\newcommand{\kf}{\ensuremath{\kappa_\mathrm{f}}\xspace}
\newcommand{\kV}{\ensuremath{\kappa_\mathrm{V}}\xspace}
\newcommand{\kg}{\ensuremath{\kappa_\mathrm{g}}\xspace}
\newcommand{\kgm}{\ensuremath{\kappa_{\gamma}}\xspace}

\newcommand{\Lk}{\ensuremath{\mathcal{L}}\xspace}
\newcommand{\Lt}{\ensuremath{\widetilde{\mathcal{L}}}\xspace}

\newcommand{\ze}{\ensuremath{z_\Pe}\xspace}
\newcommand{\zvtx}{\ensuremath{z_\mathrm{vtx}}\xspace}
\newcommand{\gconv}{\ensuremath{g_\text{conv}}\xspace}
\newcommand{\ptvecgg}{\ensuremath{\vec{p}^{\Pgg\Pgg}_\mathrm{T}}\xspace}

\newcommand{\musmx}{\ensuremath{\sigma\mathrm{'}/\sigma_\mathrm{SM}}\xspace}
\newcommand{\mHx}{\ensuremath{m_{\PH'}\xspace}}

\cmsNoteHeader{HIG-13-001}
\title{Observation of the diphoton decay of the Higgs boson and measurement of its properties}

\date{\today}

\abstract{
Observation of the diphoton decay mode of the recently discovered Higgs boson
and measurement of some of its properties are reported.
The analysis uses the entire dataset collected by the CMS experiment in
proton-proton collisions during the 2011 and 2012 LHC running periods.
The data samples correspond to integrated
luminosities of 5.1\fbinv at $\sqrt{s}=7\TeV$ and 19.7\fbinv at 8\TeV.
A clear signal is observed in the diphoton channel at a mass close to 125\GeV
with a local significance of $5.7\,\sigma$, where a significance of $5.2\,\sigma$ is expected for the standard model Higgs boson.
The mass is measured to be $124.70\pm0.34\GeV=124.70\pm0.31\stat\pm0.15\syst\GeV$, and the best-fit signal strength
relative to the standard model prediction is $1.14^{+0.26}_{-0.23}=1.14\pm0.21\stat$ $^{+0.09}_{-0.05}\syst$ $^{+0.13}_{-0.09}\thy$.
Additional measurements include the signal strength modifiers associated with
different production mechanisms, and hypothesis tests between spin-0 and spin-2
models.
}
\hypersetup{%
pdfauthor={CMS Collaboration},%
pdftitle={Observation of the diphoton decay of the Higgs boson and measurement of its properties},%
pdfsubject={CMS Higgs search; diphoton channel},%
pdfkeywords={CMS, physics, Higgs, photon, diphoton, ECAL}}

\maketitle 
\section{Introduction}
\label{sec:intro}

In 2012 the ATLAS and CMS Collaborations announced the observation~\cite{Aad:2012tfa, Chatrchyan:2012ufa} of a
new boson with a mass, $\mH$, of about 125\GeV and properties consistent, within uncertainties, with expectations for a
standard model (SM) Higgs boson.
The Higgs boson is the particle predicted to exist as a
consequence of the spontaneous symmetry breaking mechanism acting in the electroweak sector of the
SM~\cite{Glashow:1961tr,Weinberg:1967tq,sm_salam}.
This mechanism was first suggested nearly fifty years
ago~\cite{Englert:1964et,Higgs:1964ia,Higgs:1964pj,Guralnik:1964eu,Higgs:1966ev,Kibble:1967sv},
and introduces a complex scalar field, which also gives masses to the
fundamental fermions through a Yukawa interaction.
Results using the full available dataset have recently been published by
CMS~\cite{Chatrchyan:2013zna,Chatrchyan:2013iaa,Chatrchyan:2013mxa,Chatrchyan:2014nva,CMSttHbb7TeV,CMSMassParity2012,CMSHzg2013,CMSInvisible2014},
and by ATLAS~\cite{Aad:2013wqa,Aad:2013xqa,ATLASMassLegacyRun1,ATLASHzg2014,ATLASInvisible2014,ATLASHgg}.

The diphoton decay channel provides a clean final-state topology that allows
the mass of the decaying object to be reconstructed with high precision.
Having in mind the discovery of a low mass Higgs boson in the diphoton channel,
the electromagnetic calorimeter performance was a design priority for CMS.
The diphoton decay is mediated by loop diagrams containing charged particles.
The top quark loop and the W boson loop diagrams dominate the decay amplitude, though they contribute with opposite sign.
The branching fraction is small, reaches a maximum value of 0.23\% at $\mH=125\GeV$ and falls steeply to values less than 0.1\% above 150\GeV~\cite{Actis:2008ts}.
As a consequence the search reported in this paper is limited to the
mass range, $110<\mH<150\GeV$.
Despite the small branching fraction and the presence of a large diphoton continuum
background, the diphoton decay mode provides an expected signal significance
for the 125\GeV SM Higgs boson that is one of the highest among all the decay modes.

This paper presents the analysis performed on the full dataset
collected in 2011 and 2012, reconstructed with the final detector calibration values,
in $\Pp\Pp$ collisions at the Large Hadron Collider (LHC),
with an integrated luminosity of 5.1\fbinv at a centre-of-mass energy of 7\TeV
(herein referred to as the ``7\TeV dataset'') and
19.7\fbinv at 8\TeV (``8\TeV dataset'').
The results supersede those previously reported by CMS for this decay mode~\cite{Chatrchyan:2012twa, Chatrchyan:2013lba}.

The primary production mechanism of the Higgs boson at the LHC is
gluon-gluon fusion (ggH)~\cite{Georgi:1977gs} with additional smaller contributions from vector boson
fusion (VBF)~\cite{Cahn:1986zv} and production in association with a $\PW$
or $\cPZ$ boson (VH)~\cite{Glashow:1978ab} or a
$\cPqt\cPaqt$ pair (\ttH)~\cite{Raitio:1978pt,Kunszt:1984ri}.
Events from specific production mechanisms are identified and classified by the presence of additional objects in the final state.
Requiring the presence of two forward jets, in addition to the photon pair, favours events produced by the VBF mechanism, while event classes designed to preferentially select
VH or \ttH production require the presence of muons, electrons, missing transverse energy from neutrinos,
or jets arising from the hadronization of b quarks.
To achieve the best sensitivity, the remaining events, and also the dijet events selected as having a VBF signature, are further separated
using multivariate classifiers that provide measures of their probability to be signal rather than background.
The signal is measured performing a simultaneous fit to the diphoton invariant mass distributions in the various event classes.
The signal model is derived from simulation, while the background is obtained from the fit to data.
A very large sample of events is available in which a \cPZ\ boson decays to a pair of electrons;
treating the electron showers in these events as if they were from photons allows precise and detailed knowledge to be obtained concerning
the accuracy of the simulation of the signal, specifically the simulation of the energy reconstruction and selection of photons, and the simulation of the selection and classification of diphoton events.

With respect to analyses of this decay mode previously reported by CMS there are refinements in methodology, which are described in the main body of the paper. In addition, the analysis uses an improved intercalibration of the electromagnetic calorimeter channels and an improved energy regression algorithm to correct the clustered energy, resulting in better energy resolution.
The simulation of the signal and Z boson samples is also improved.
The changes in the energy-equivalent noise in the electromagnetic calorimeter during the data-taking period are simulated, and
a significantly increased time window is used to simulate the effect of deposited energy coming from interactions in earlier bunch crossings.

The paper is organized as follows.
After a brief description of the CMS detector and event reconstruction in
Section~\ref{sec:detector} and of the data and simulated samples in Section~\ref{sec:dataSim}, the reconstruction and identification
of photons is detailed in Section~\ref{sec:photon}.
The issue of identifying the diphoton vertex is covered in Section~\ref{sec:vertex}.
In Section~\ref{sec:classes} the event classification is described.
The section first describes the construction of a multivariate event classifier which
takes as input quantities associated with the two photons,
and then goes on to describe the tagging of events by the presence of objects in the final state,
in addition to the photon pair, that give the event a signature characteristic of one of the production processes.
It concludes by detailing the use of two multivariate event classifiers to additionally subdivide into classes
both the untagged events, and the events tagged as coming from the VBF process.
Sections~\ref{sec:smodelling} and~\ref{sec:bmodelling} describe, respectively, the signal and background models used
in the statistical procedures which provide the results of the analysis, and Section~\ref{sec:systematics}
discusses the systematic uncertainties taken into account in those procedures.
Section~\ref{sec:crosscheck} outlines three alternative analyses that use specific variations of methodology that
provide corroboration of particular aspects of the main analysis.
Finally, in Section~\ref{sec:fitresults} the results of the measurements of the Higgs boson production and its properties are presented and discussed.

\section{CMS detector} 
\label{sec:detector}
The central feature of the CMS apparatus is a superconducting solenoid,
13\unit{m} in length and with an inner diameter of 6\unit{m},
which provides an axial magnetic field of 3.8\unit{T}.
The bore of the solenoid is instrumented with both the central tracker and the calorimeters.
The steel flux-return yoke outside the solenoid hosts gas ionization detectors
used to identify and reconstruct muons.

{\tolerance=800
The CMS experiment uses a right-handed coordinate system, with the origin at the nominal interaction point,
the $x$ axis pointing to the centre of the LHC, the $y$ axis pointing up (perpendicular to the LHC plane),
and the $z$ axis along the anticlockwise-beam direction.
The polar angle $\theta$ is measured from the positive $z$ axis and the azimuthal angle $\phi$ is measured in the $x$-$y$ plane.
Transverse energy, denoted by $\ET$, is defined as the product of energy and $\sin\theta$, with $\theta$ being measured
with respect to the nominal interaction point.
Charged-particle trajectories are measured by the silicon pixel and strip tracker,
with full azimuthal coverage within $\abs{\eta} < 2.5$, where the pseudorapidity
$\eta$ is defined as $\eta = -\ln[\tan(\theta/2)]$.
A lead tungstate crystal electromagnetic calorimeter (ECAL) and a brass/scintillator
hadron calorimeter (HCAL) surround the tracking volume and cover the region $\abs{\eta} < 3$.
The ECAL barrel extends to $\abs{\eta} < 1.48$ while the ECAL endcaps cover the region $1.48 < \abs{\eta} < 3.0$.
A lead/silicon-strip preshower detector is located in front of the
ECAL endcap in the region $1.65 < \abs{\eta} < 2.6$.
The preshower detector includes two planes of silicon sensors measuring the
$x$ and $y$ coordinates of the impinging particles.
A steel/quartz-fibre Cherenkov forward calorimeter extends the
calorimetric coverage to $\abs{\eta} < 5.0$.
In the region $\abs{\eta} < 1.74$, the HCAL cells have widths of 0.087
in both $\eta$ and $\phi$.
In the $\eta$-$\phi$ plane, and for $\abs{\eta} < 1.48$, the HCAL cells map on to 5$\times$5
ECAL crystal arrays to form calorimeter towers projecting radially
outwards from points slightly offset from the nominal interaction point.
In the endcap, the ECAL arrays matching the HCAL cells contain fewer crystals.
\par
}

Calibration of the ECAL is achieved exploiting the $\phi$-symmetry of the energy flow, and using photons from $\Pgpz\to\Pgg\Pgg$ and $\eta\to\Pgg\Pgg$ decays, and electrons from $\Wenu$ and $\Zee$ decays~\cite{Chatrchyan:2013dga}.
Changes in the transparency of the ECAL crystals due to irradiation during the LHC
running periods and their subsequent recovery are monitored continuously,
and corrected for, using light injected from a laser system~\cite{Chatrchyan:2013dga}.

The first level of the CMS trigger system, composed of custom hardware processors, uses information from the calorimeters and muon detectors to select the most interesting events in a fixed time interval of less than 4\mus. The high-level trigger processor farm further decreases the event rate from around 100\unit{kHz} to around 400\unit{Hz}, before data storage.

A more detailed description of the CMS detector can be found in Ref.~\cite{Chatrchyan:2008zzk}.

Reconstruction of the photons used in this analysis is described in Section~\ref{sec:photon}, and uses a clustering of the energy recorded in the ECAL,
known as a ``supercluster'', which may be extended in the $\phi$ direction to form an extended cluster or group of clusters.

The global event reconstruction (also called particle-flow event reconstruction) consists of reconstructing and identifying each particle with an optimized combination of all subdetector information~\cite{CMS-PAS-PFT-09-001,CMS-PAS-PFT-10-001}. In this process, the identification of the particle type (photon, electron, muon, charged hadron, neutral hadron) plays an important role in the determination of the particle direction and energy.
Photons
are identified as ECAL energy clusters not linked to the extrapolation of any charged-particle trajectory to the ECAL.
Electrons
are identified as a primary charged-particle track associated with ECAL energy clusters corresponding to this track's
 extrapolation to the ECAL and to possible bremsstrahlung photons emitted along the way through the tracker material. Muons are identified as a track in the central tracker consistent with either a track or several hits in the muon system, associated with less energy in the calorimeters than would be deposited by a charged hadron or electron. Charged hadrons are identified as charged-particle tracks neither identified as electrons, nor as muons. Finally, neutral hadrons are identified as HCAL energy clusters not linked to any charged hadron trajectory, or as ECAL and HCAL energy excesses with respect to the expected energy deposited
by a matching charged hadron.

The energy of photons used in the global event reconstruction is directly obtained from the ECAL measurement. The energy of electrons is determined from a combination of the track momentum at the main interaction vertex, the corresponding ECAL cluster energy, and the energy sum of all bremsstrahlung photons attached to the track. The energy of muons is obtained from the corresponding track momentum. The energy of charged hadrons is determined from a combination of the track momentum and the corresponding ECAL and HCAL energy, calibrated for the nonlinear response of the calorimeters. Finally, the energy of neutral hadrons is obtained from the corresponding calibrated ECAL and HCAL energies.

For each event, hadronic jets are clustered from these reconstructed particles using the infrared- and collinear-safe anti-\kt algorithm~\cite{Cacciari:2008gp} with a size parameter of 0.5. The jet momentum is determined as the vectorial sum of all particle momenta in the jet, and the scale is found in the simulation to be within 5\% to 10\% of the true momentum over the whole transverse momentum spectrum and detector acceptance. Jet energy corrections are derived from simulation, and are confirmed with in situ measurements using
the energy balance of dijet and $\Pgg/\cPZ+\text{jet}$ events~\cite{Chatrchyan:2011ds}.
The jet energy resolution typically amounts to 15\% (8\%) at 10 (100)\GeV, to be compared to about 40\% (12\%) obtained when the calorimeters alone are used for jet clustering.

To identify jets originating from the hadronization of bottom quarks, the combined secondary vertex b-tagging algorithm~\cite{Chatrchyan:2012jua} is employed.
The algorithm tags jets from b-hadron decays by identifying their displaced decay vertex.
The working point of the tagging algorithm used provides an efficiency for identifying b-quark jets of about 70\% and a misidentification
probability for jets from light quarks and gluons of about 1\%.

The missing transverse energy vector is taken as the negative vector sum of all reconstructed particle candidate
transverse momenta in the global event reconstruction, and its magnitude is referred to as $\MET$.

\section{Data sample and simulated events}
\label{sec:dataSim}

The events used in the analysis were selected by diphoton triggers
with asymmetric transverse energy thresholds and complementary photon selections.
One selection requires a loose calorimetric identification based on the shape
of the electromagnetic shower and loose isolation requirements on the photon candidates,
while the other requires only that the photon candidate has a high value of the
\RNINE~shower shape variable.
High trigger efficiency is maintained by allowing both photons to satisfy either selection.
The \RNINE~variable is defined as the energy sum of
3$\times$3 crystals centred on the most energetic crystal in the
supercluster divided by the energy of the supercluster.
Photons that convert before reaching the calorimeter tend to have wider showers and lower values of \RNINE\
than unconverted photons.
To cover the entire data taking period two trigger threshold configurations are used:  $\ET>26\ (18)\GeV$ on the leading (trailing) photon, and $\ET>36\ (22)\GeV$.
The measured trigger efficiency is $99.4\%$ for events
satisfying the diphoton preselection required for events entering the analysis, as described in Section~\ref{sec:photon}.

The Monte Carlo (MC) simulation of detector response employs a detailed description of the CMS detector,
and uses $\GEANT4$ version~9.4 (patch 03)~\cite{Agostinelli:2002hh}.
Simulated events include simulation of the multiple $\Pp\Pp$ interactions taking place in each bunch crossing and
are weighted to reproduce the distribution of the number of interactions in data.
They thus simulate the effects of pileup --- the presence of signals from multiple
$\Pp\Pp$ interactions, in multiple bunch crossings, in each recorded event.
The interactions used to simulate pileup are generated with the same versions of
\PYTHIA~\cite{Sjostrand:2006za}, 6.424 or 6.426, that are used for other purposes as described below.
The \PYTHIA tunes used for the underlying event activity are Z2 and Z2* for the 7 and 8\TeV samples, respectively~\cite{Chatrchyan:2011id}.
Simulated Higgs boson signal events are used both for training of multivariate discriminants and to construct
the signal model used in the statistical procedures employed to extract the results.
Sufficient samples have been produced to
ensure that the samples of simulated signal events used for construction of the signal model (Section~\ref{sec:smodelling})
are not used for training the multivariate discriminants.
The MC signal event samples for the ggH and VBF processes are obtained using the next-to-leading order (NLO)
matrix-element generator \POWHEG~(version 1.0)~\cite{powheg1,powheg2,powheg3,powheg-ggH,powheg-VBF}
interfaced with \PYTHIA.
For the 7\TeV samples, events are weighted so that the transverse momentum spectrum of Higgs bosons
produced by the ggH process agrees with the next-to-next-to-leading logarithm
+ NLO distribution computed by \HqT~(version 1.0)~\cite{HqT1,HqT2,deFlorian:2011xf}.
At 8\TeV, \POWHEG has been tuned following the recommendations of the
LHC Higgs Cross Section Working Group~\cite{LHCHiggsCrossSectionWorkingGroup2} and reproduces the \HqT spectrum.
The ggH process cross section is reduced by 2.5\% for all values of \mH\ to
account for the interference with nonresonant diphoton production~\cite{Dixon:2003yb}.
For the VH and \ttH processes \PYTHIA is used alone;
processes are generated at leading-order by \PYTHIA, and higher order diagrams are accounted for only by
\PYTHIA's ``parton showering'' model.
The SM Higgs boson cross sections and branching fractions used are taken
from Ref.~\cite{LHCHiggsCrossSectionWorkingGroup3}.
Samples used for the testing of spin hypotheses were generated with leading-order accuracy by
\textsc{jhugen}~\cite{Gao2010,Bolognesi:2012mm}, interfaced to \PYTHIA.

Simulated samples of \Zee, \Zmm, and \Zmmg events used for comparison with data, and for the derivation of energy scale and resolution smearing
corrections are generated with \MADGRAPH, \SHERPA, and \POWHEG~\cite{powheg-Zjj}, allowing comparisons to be made between
the different generators.

{\tolerance=600
Simulated background samples are used only for training multivariate discriminants and defining selection and classification criteria.
The background is simulated using a combination of samples.
At $\sqrt{s}=7\TeV$ the diphoton processes are simulated using a combination of \MADGRAPH\,5 \cite{Alwall:2011uj}
interfaced to \PYTHIA for processes apart from the gluon-fusion box diagram, and \PYTHIA alone for the box diagram.
At $\sqrt{s}=8\TeV$ the diphoton continuum processes involving two prompt photons are simulated
using \SHERPA\,1.4.2 \cite{Gleisberg:2008ta}.
The \SHERPA samples give a noticeably improved description of diphoton continuum events accompanied by one or two jets,
and enable training of a more effective multivariate discriminant in the case of diphoton-plus-dijet events.
The remaining processes where one of the photon candidates arises from misidentified jet fragments are simulated using \PYTHIA alone,
the cross sections of the processes are scaled by $K$-factors derived from CMS measurements~\cite{Chatrchyan:2011qt,CMS-jj}.
\par
}

\section{Photon reconstruction and identification}
\label{sec:photon}

Photon candidates for the analysis are reconstructed from energy deposits in the ECAL using
algorithms that constrain the superclusters in $\eta$ and $\phi$ to the shapes expected from electrons and
photons with high \pt.
The algorithms do not make any hypothesis as to whether the particle originating from the interaction
point is a photon or an electron; when reconstructed in this way, electrons from $\Zee$ events provide
measurements of the photon trigger, reconstruction, and identification efficiencies, and of
the photon energy scale and resolution.
The clustering algorithms achieve a rather complete ($\approx$95\%) collection of the energy
of photons and electrons,
even those that undergo conversion and bremsstrahlung in the material in front of the ECAL.
In the barrel region, superclusters are formed from five-crystal-wide strips
in $\eta$, centred on the locally most energetic crystal (seed), and
have a variable extension in $\phi$.
In the endcaps, where the crystals are arranged according to an $x$-$y$ rather than an $\eta$-$\phi$ geometry,
matrices of 5$\times$5 crystals, which may partially overlap and are
centred on a locally most energetic crystal, are summed
if they lie within a narrow $\phi$ road.
The photon candidates are required to be within the fiducial region $\abs{\eta}<2.5$, excluding the
barrel-endcap transition region $1.44 < \abs{\eta} < 1.57$, where the photon reconstruction is
suboptimal.
The fiducial region requirement is applied to the supercluster position in the ECAL, \ie the value of $\eta$ is
calculated with respect to the origin of the coordinate system.
The exclusion of the barrel-endcap transition region ensures complete
clustering of the accepted showers in either the ECAL barrel or endcaps.

About half of the photons convert in the material upstream of the ECAL.
If the resulting charged particle tracks originate sufficiently close to the interaction point so as to
pass through three or more tracking layers,
conversion track pairs may be reconstructed and matched to the photon candidate.

\subsection{Photon energy}
\label{sec:photonE}
The photon energy is computed from the signals recorded by the ECAL.
In the region covered by the preshower detector ($\abs{\eta} > 1.65$) the signals recorded in it are also considered.
In order to obtain the best energy resolution, the calorimeter signals are
calibrated and corrected for several detector effects~\cite{Chatrchyan:2013dga}.
The variation of crystal transparency during the run is continuously monitored and corrected for
using a factor based on the measured change in response to the light from the laser system,
with the response for each crystal being computed approximately every 40 minutes.
The single-channel response of the ECAL is equalized exploiting the $\phi$-symmetry of the energy flow,
the mass constraint on the energy of the two photons in $\Pgpz$ and $\Pgh$ decays,
and the momentum constraint on the energy of isolated electrons from $\PW$-
and $\cPZ$-boson decays.
Finally, the containment of the shower in the clustered crystals, the shower losses for photons that convert
in the material upstream of the calorimeter, and the effects of pileup, are corrected using a multivariate regression technique.
The photon energy response distribution is parameterized by a function with a Gaussian core and two power law tails,
an extended form of the Crystal Ball function~\cite{CrystalBall}.
The regression provides a per-photon estimate of the parameters of the function, and therefore a
prediction of the distribution of the ratio of true energy to uncorrected supercluster energy.
The most probable value of this distribution is taken as the corrected photon energy.
The width of the Gaussian core is further used as a per-photon estimator of the energy uncertainty.
The regression input variables are a collection of shower shape variables including \RNINE~of the supercluster,
the ratio of the 5$\times$5 crystal energy centred around the seed crystal to the uncorrected supercluster energy sum,
the energy-weighted $\eta$-width and $\phi$-width of the supercluster,
and the ratio between the hadronic energy behind the supercluster and the electromagnetic energy of the cluster.
The global $\eta$ coordinate of the supercluster is included,
and for the barrel the global $\phi$ coordinate and the coordinates of the seed cluster with respect to the crystal centre
are also included.
In the endcap, the ratio of preshower energy to raw supercluster energy is included.
Finally, the number of primary vertices and the median energy density $\rho$~\cite{Cacciari:2007fd} in the event are included
in order to allow for the correction of residual energy scale effects due to pileup.

A multistep procedure has been implemented to correct the energy scale in data, and to determine the parameters of
Gaussian smearing to be applied to showers in simulated events so as to reproduce the energy resolution seen in data.
First, the energy scale in data is equalized with that in simulated events, and
residual long-term drifts in the response are corrected,
using $\Zee$ decays in which the electron showers are reconstructed as photons.
The data are corrected as a function of the time at which they were taken,
using 8 epochs in the 7\TeV dataset and 51 epochs in the 8\TeV dataset.
Following this, the photon energy resolution predicted by the simulation is made more realistic
by adding a Gaussian smearing determined from the comparison between the
$\Zee$ line-shape in data and in simulated events.
The amount of smearing required is extracted differentially in $\abs{\eta}$ (two bins in the barrel and two in the endcap) and \RNINE\ (two bins).
In the fits from which the required amount of smearing is extracted, the data energy scale is allowed to float, and a residual scale correction for the data is extracted in the same eight bins.
A sufficient number of $\Zee$ events is available in the 8\TeV data to allow a third step,
in which the energy scale for the ECAL barrel is further corrected in 20 bins defined by ranges in $\abs{\eta}$, $\RNINE$, and $\ET$,
and the smearing magnitude is allowed to have an energy dependence; the additional
energy resolution ($\sigma/E$) is parameterized
as the quadratic sum of a constant term and a term proportional to $1/\sqrt{\ET}$, and the relative
magnitude of the two components extracted from the fits.

\begin{figure*}[hbtp]
  \centering
    \includegraphics[width=0.8\textwidth]{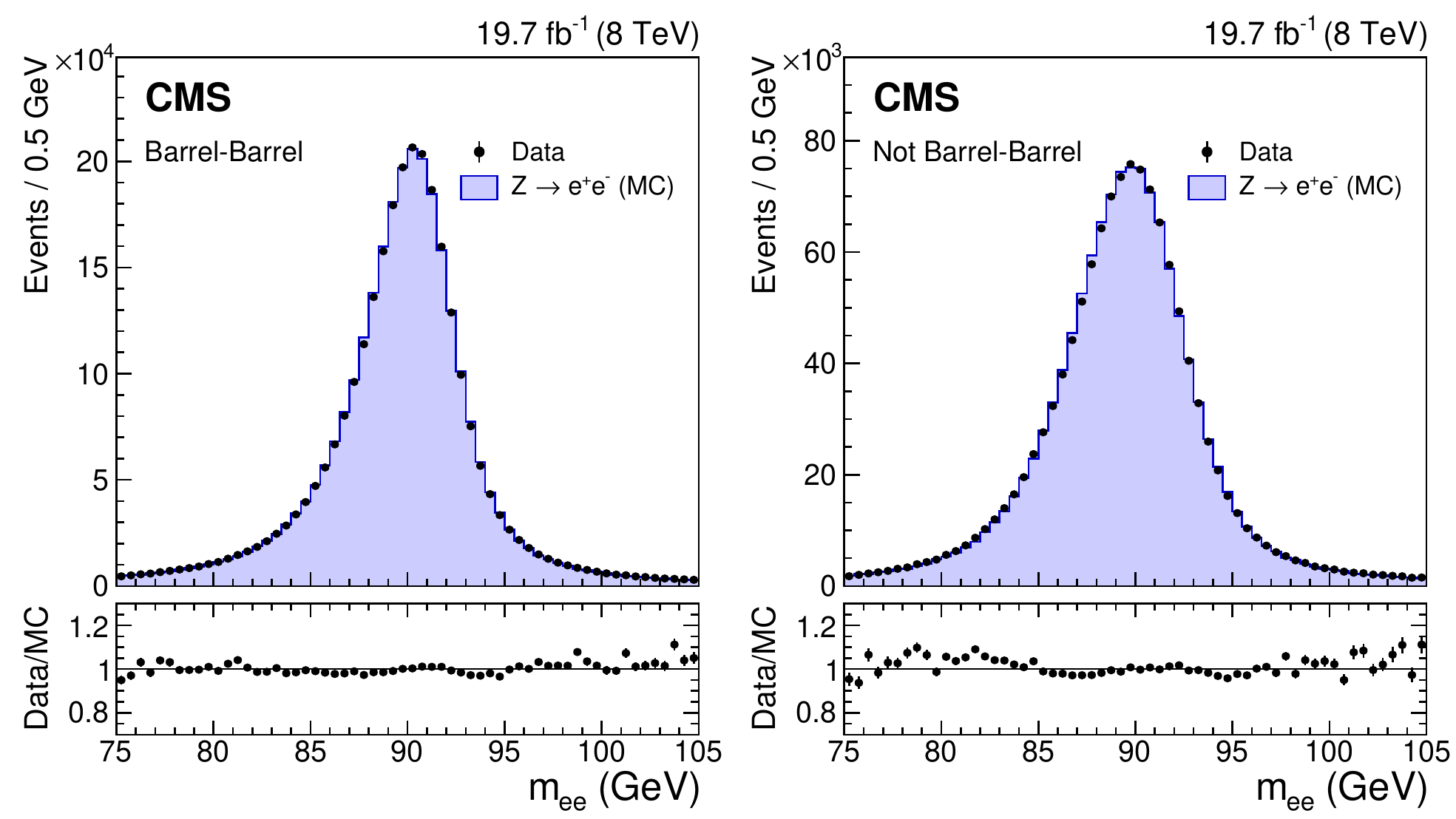} 
    \caption{Invariant mass of $\Pep\Pem$ pairs in $\Zee$ events in the 8\TeV data (points),
and in simulated events (histogram), in which the electron showers are reconstructed as photons, and the full set of
photon corrections and smearings are applied. The comparison is shown for (left) events with both showers
in the barrel, and (right) the remaining events.
For each bin, the ratio of the number of events in data to the number of simulated events
is shown in the lower main plot.
  }
    \label{fig:z-mass}
\end{figure*}

Figure~\ref{fig:z-mass} shows the invariant mass of  electron pairs reconstructed in $\Zee$ events in the 8\TeV data
and in simulated events in which the electron showers are reconstructed as photons, and the full set of
corrections to the data, and smearings of the simulated energies, are applied.
The selection applied to the diphoton candidates is the same, apart from the inversion of the electron veto,
as is applied to diphoton candidates entering the analysis (as described in Section~\ref{sec:classes}).
There is excellent agreement between the data and the simulation in the core of the distributions.
A slight discrepancy is present in the low-mass tail in the endcaps, where the Gaussian smearing
is not enough to account for some noticeable non-Gaussian energy loss.
The mass peaks are shifted from the true  $\cPZ$-boson mass, both in data and
simulation, because the electron showers are reconstructed as photons.

\subsection{Photon preselection}

The continuum background to the \HGG\ process is mainly due to prompt diphoton
production, with a reducible contribution from $\Pp\Pp\to\GAMJET$ and
dijet processes where at least one of the objects reconstructed as a photon comes from a jet.
Typically these photon candidates come
from one or more neutral mesons that take a substantial fraction
of the total jet \pt and are thus relatively isolated from hadronic activity in the detector.
In the transverse momentum range of interest, the photons from neutral pion decays
are rather collimated and are reconstructed as a single photon.
In the events used for the analysis, \ie after all selection and classification criteria are applied,
MC simulation predicts that about 70\% of the total background is due to the irreducible prompt diphoton production.

The photons entering the analysis are required to satisfy preselection
criteria similar to, but slightly more stringent than, the trigger requirements.
These consist of
\begin{itemize}
\item $\ptga>33\GeV$ and $\ptgb>25\GeV$, where \ptga and \ptgb are the
transverse momenta of the leading (in $\pt$) and subleading photons, respectively.
\item a selection on the hadronic leakage of the shower, measured as the ratio of energy in HCAL cells
behind the supercluster to the energy in the supercluster,
\item a loose selection based on isolation and the shape of the shower,
\item an electron veto, which removes the photon candidate if its supercluster is matched to an electron track
with no missing hits in the innermost tracker layers, thus excluding almost all $\Zee$ events.
\end{itemize}
The selection requirements are applied with different
stringency in four categories defined to match the different selections
used in the trigger.
The four categories are shown in Table~\ref{tab:PhotEff}.

The efficiency of the photon preselection is measured in data using a
``tag-and-probe'' technique~\cite{springerlink:10.1007/JHEP10(2011)132}.
The efficiency of all preselection criteria, except the electron veto
requirement, is measured using $\Zee$ events.
The efficiency for photons to satisfy the electron veto requirement is measured
using $\Zmmg$ events, in which the photon is produced by final-state
radiation, which provide a more than $99\%$ pure source of prompt photons.
The ratio of the photon efficiency measured in data to that found in simulated $\Zee$ events,
$\epsilon_\text{data}/\epsilon_\mathrm{MC}$, is consistent with unity in all categories.
The complete set of efficiencies, in data and in simulated $\Zee$ events, and the ratios
$\epsilon_\text{data}/\epsilon_\mathrm{MC}$, are shown in Table~\ref{tab:PhotEff}.
The systematic uncertainty in the measurement is included in both the efficiencies and the ratio.
The statistical uncertainties in the efficiencies measured in simulated events are negligible.
The measured $\epsilon_\text{data}/\epsilon_\mathrm{MC}$ ratios are used to correct
the simulated signal sample, and the associated uncertainties are taken into account as systematic uncertainties in the
signal extraction procedure.
For photons in simulated Higgs boson events the efficiency of the preselection criteria
in the four categories ranges from 92\% to 99\%.

\begin{table*}[htbp]
\caption{
Photon preselection efficiencies for both the 7 and 8\TeV datasets measured for $\Zee$ events,
where the electrons are reconstructed as photons, in four photon
categories. The statistical uncertainties in the efficiencies found in simulated events are negligible,
and the uncertainties measured in data are discussed in the text.}
\centering
\begin{tabular}{lccc}
\hline
Preselection category & $\epsilon_\text{data}$ (\%) & $\epsilon_{\mathrm{MC}}$ (\%) & $\epsilon_{\text{data}}$/$\epsilon_{\mathrm{MC}}$ \\
\hline
\multicolumn{4}{ c }{7\TeV dataset}\\
\hline
Barrel; $\RNINE>$0.90 & 98.7 $\pm$ 0.3 & 99.1 & 0.996 $\pm$ 0.003 \\
Barrel; $\RNINE<$0.90 & 96.2 $\pm$ 0.5 & 96.7 & 0.995 $\pm$ 0.006 \\
Endcap; $\RNINE>$0.90 & 99.1 $\pm$ 0.9 & 98.2 & 1.008 $\pm$ 0.009 \\
Endcap; $\RNINE<$0.90 & 96.1 $\pm$ 1.5 & 95.6 & 1.005 $\pm$ 0.018 \\
\hline
\multicolumn{4}{ c }{8\TeV dataset}\\
\hline
Barrel; $\RNINE>$0.90 & 98.8 $\pm$ 0.3 & 98.6 & 0.999 $\pm$ 0.003 \\
Barrel; $\RNINE<$0.90 & 95.7 $\pm$ 0.6 & 96.1 & 0.995 $\pm$ 0.006 \\
Endcap; $\RNINE>$0.90 & 98.4 $\pm$ 0.9 & 97.9 & 1.005 $\pm$ 0.009 \\
Endcap; $\RNINE<$0.90 & 95.5 $\pm$ 1.7 & 94.5 & 1.011 $\pm$ 0.018 \\
\hline
\end{tabular}
\label{tab:PhotEff}
\end{table*}

\subsection{Photon identification}
\label{sec:photonID}

{\tolerance=800
A boosted decision tree (BDT), implemented using the \textsc{tmva}~\cite{Hocker:2007ht} framework,
is trained to separate prompt photons from photon candidates resulting from misidentification of jet fragments passing
the preselection requirements.
The following variables are used as inputs to the photon identification BDT:
\begin{enumerate}
\item Lateral shower shape variables, six of which use data from
the ECAL crystals, and one of which measures the shower spread in the preshower detector (where it is present).
The shape variables obtained in the MC simulation are compared to those observed in $\Zee$ and $\Zmmg$ data samples.
No significant differences are observed.
\item Isolation variables, based on the particle-flow
  algorithm~\cite{CMS-PAS-PFT-10-001}, and using sums of the $\pt$ of photons, and of charged
  hadrons, within regions of $\DR<0.3$ around the candidate, where $\DR=\sqrt{\smash[b]{(\Delta\phi)^2+(\Delta\eta)^2}}$.
  Two charged-hadron isolation variables are used: one that considers charged hadrons coming from the vertex chosen for the
  event (described in Section~\ref{sec:vertex}), and one that is the largest of all such $\pt$ sums among those made for each reconstructed vertex.
  The second variable is effective when a photon candidate originating from misidentification of jet fragments comes from
  a vertex other than the chosen one (Section~\ref{sec:vertex} describes the vertex choice).
\item The energy median density per unit area in the event, $\rho$.
This variable is introduced to allow the BDT classifier to take into account the pileup dependence of the isolation variables.
\item The pseudorapidity and energy of the supercluster corresponding
  to the reconstructed photon. These variables are introduced to allow
  the dependence of the shower topology and isolation variables on $\eta$ and \pt to be taken into account.
\end{enumerate}
\par}

\begin{figure}[hbtp]
  \centering
    \includegraphics[width=\cmsFigWidth]{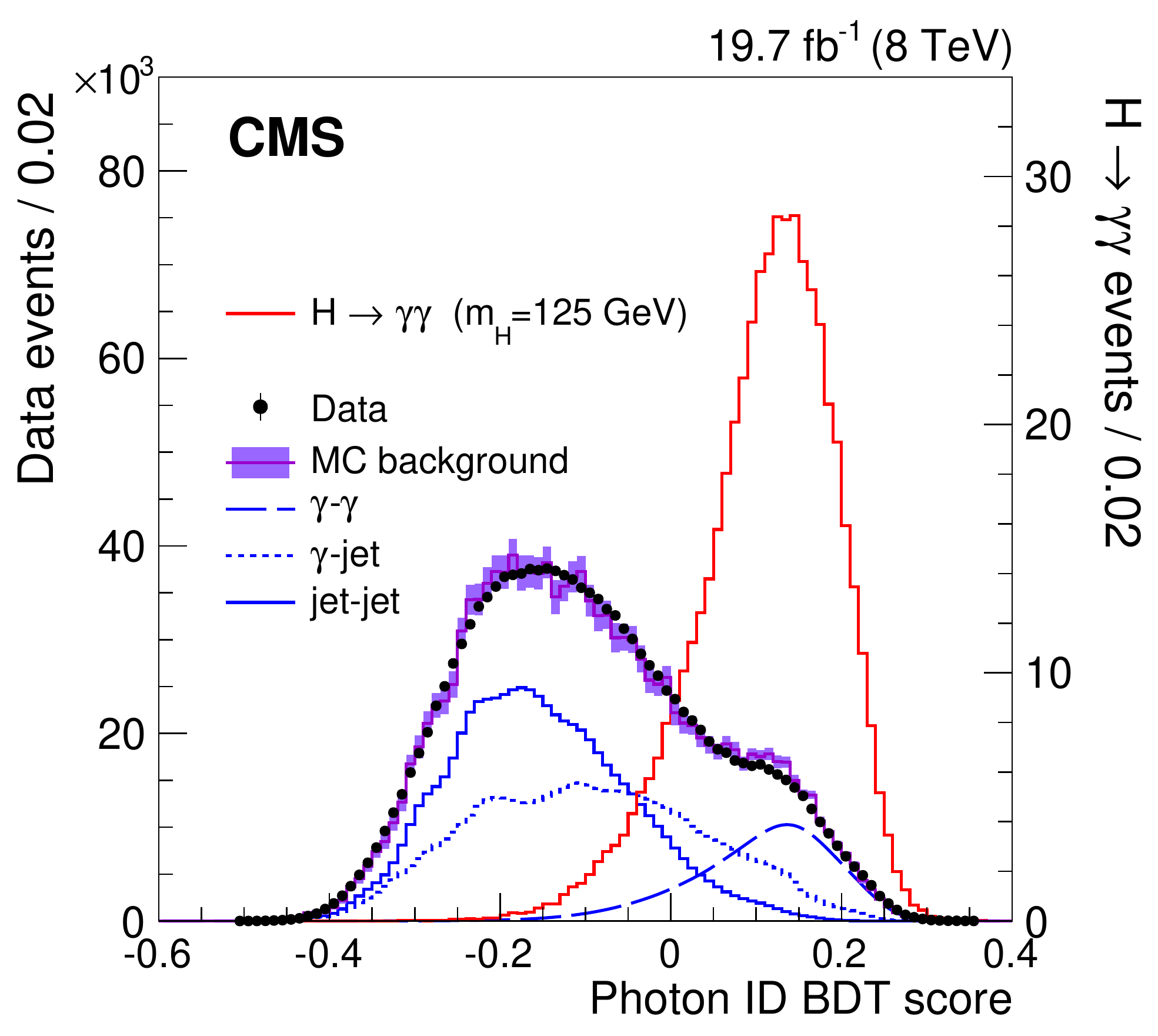} 
    \caption{Photon identification BDT score of the lower-scoring photon of diphoton pairs with an invariant mass
  in the  range $100<\mgg<180\GeV$, for events passing the preselection in the 8\TeV dataset (points),
  and for simulated background events (histogram with shaded error bands showing the statistical uncertainty).
  Histograms are also shown for different components of the simulated background, in which there are either two, one, or
  zero prompt signal-like photons.
  The tall histogram on the right (righthand vertical axis) corresponds to simulated Higgs boson signal events.
  }
    \label{fig:idmva-worst}

\end{figure}

Figure~\ref{fig:idmva-worst} shows the photon identification BDT score of the lower-scoring photon in diphoton pairs with an invariant mass, $\mgg$, in
the range $100<\mgg<180\GeV$, for events passing the preselection in the 8\TeV dataset and for simulated background events
(histogram with shaded error bands showing the statistical uncertainty).
The tall histogram on the right corresponds to  simulated Higgs boson signal events.
Although the simulated background events are only used for training the BDT, it is worth noting that the agreement of their
BDT score distribution with that in data is good.
The bump that can be seen in both distributions at a BDT score of slightly above 0.1 corresponds
to events where both photons are prompt and, therefore, signal-like.

\newcommand{\Nvtx}{\ensuremath{N_\mathrm{vtx}}\xspace}
The agreement between data and simulation for photon identification is
assessed using electrons from $\Zee$ decays, photons from $\Zmmg$ decays, and
the highest-\pt photon in diphoton events with $\mgg > 160\GeV$ in which the relative magnitude
of the contribution from misidentified jet fragments is small.
Figure~\ref{fig:idmva-validation} shows a comparison of the photon identification BDT score for
$\Zee$ electron showers reconstructed as photons
in the barrel, for data and MC simulated events.
The events must pass all the preselection requirements, but the electron veto condition is inverted.
The systematic uncertainty assigned to the photon identification BDT
score is shown as a band, and corresponds to a shift of $\pm$0.01 in the score.
The comparison is made for the 8\TeV dataset, and is shown for two sets of events with different numbers
of primary vertices, $\Nvtx$, to demonstrate the independence of the result
from effects coming from pileup.
The differences between the distributions for the data and the simulation fall within the assigned systematic uncertainties
for both the lower-pileup ($\Nvtx\le15$) and higher-pileup ($\Nvtx>15$) sets of events, and the difference between the distributions
in the two sets is negligible.

\begin{figure*}[hbtp]
  \centering
    \includegraphics[width=0.8\textwidth]{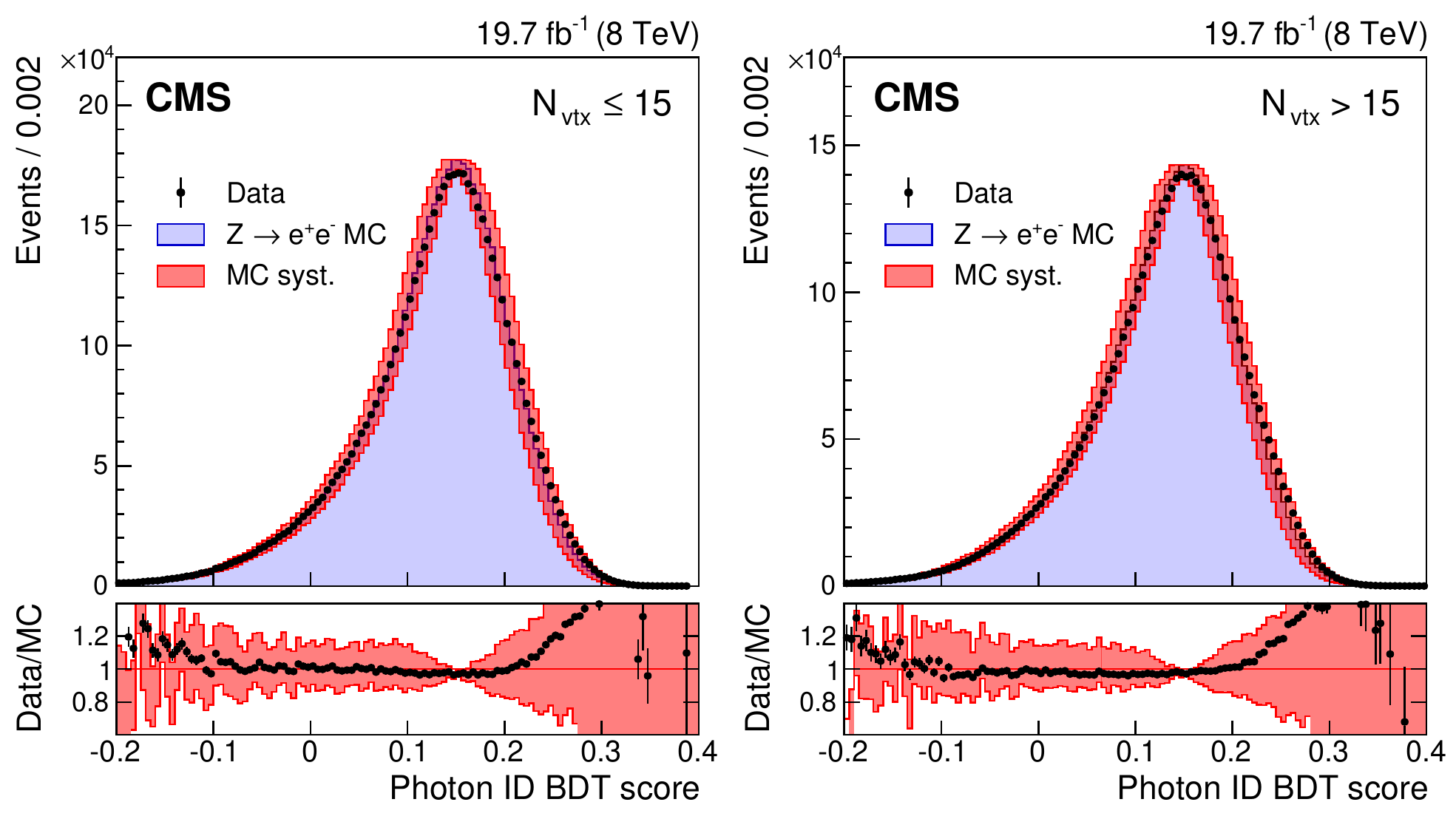} 
    \caption{Comparison of the photon identification BDT score for electron showers in the barrel in $\Zee$
events in the 8\TeV dataset and MC simulated events, for events passing the preselection, but with the electron veto condition inverted.
The systematic uncertainty assigned to the photon identification BDT score is shown as a band.
The comparison is shown for two sets of events with different numbers of primary vertices, $\Nvtx$.
For each bin, the ratio of the number of events in data to the number of simulated events
is shown in the lower plot.
  }
    \label{fig:idmva-validation}
\end{figure*}

\section{Diphoton vertex}
\label{sec:vertex}

The mean number of $\Pp\Pp$ interactions per bunch crossing
is 9 in the 7\TeV dataset and 21 in the 8\TeV dataset.
In the longitudinal direction, $z$, the interaction vertices, built from the reconstructed tracks,
have a distribution with an \textsc{rms} spread of about 6~(5)\cm in the 7~(8)\TeV dataset.

The diphoton mass resolution has contributions from the resolution of
the measurement of the photon energies and the measurement of the angle between the two photons.
If the vertex from which the photons originate is known to within about 10\mm, then the experimental
resolution on the angle between them makes a negligible contribution to the mass resolution.
Thus, if the diphoton is associated with the charged particle vertex corresponding to the interaction in which it originated,
then the mass resolution will be entirely dominated by the photon energy resolution,
since the longitudinal coordinate of the charged particle vertices is known to greater precision than 10\mm.

\subsection{Diphoton vertex identification}
No charged particle tracks result from photons that do not convert, so the diphoton vertex is identified indirectly,
using the kinematic properties of the diphoton system and its
correlations with the kinematic properties of the recoiling tracks.
If either of the photons converts, the direction of the resulting tracks can provide additional information.

Three discriminating variables are calculated for each reconstructed primary vertex:
the sum of the squared transverse momenta of the charged particle tracks associated with the vertex,
and two variables that quantify the vector and scalar balance of \pt between the diphoton system and the charged particle tracks associated with the vertex.
The three variables are:

\begin{enumerate}
\item $\sum\ptvec^{2}$
\item $-\sum (\ptvec\cdot\frac{\ptvecgg}{|\ptvecgg|})$, and
\item $(|\sum\ptvec| - |\ptvecgg|)/(|\sum\ptvec| + |\ptvecgg|)$,
\end{enumerate}

where the sums are over the transverse momentum vectors of the charged tracks, $\ptvec$, and $\ptvecgg$
is the transverse momentum vector of the diphoton system.
In addition, if either photon is associated with any charged particle tracks that have been
identified as resulting from conversion, then a further variable, $\gconv$, is used, as defined below.
An estimate of the primary vertex longitudinal position, $\ze$, is obtained
from the conversion track(s), and the additional variable $\gconv$ is
defined as the pull between $\ze$ and the longitudinal position of the
reconstructed vertex, $\zvtx$: $\gconv=\abs{\ze-\zvtx}/\sigma$,
where $\sigma$ is the uncertainty in $\ze$.
The variables are used as the inputs to a multivariate system based on a BDT
to choose the reconstructed vertex to be associated with the diphoton system.

\begin{figure}
  \centering
    \includegraphics[width=\cmsFigWidth]{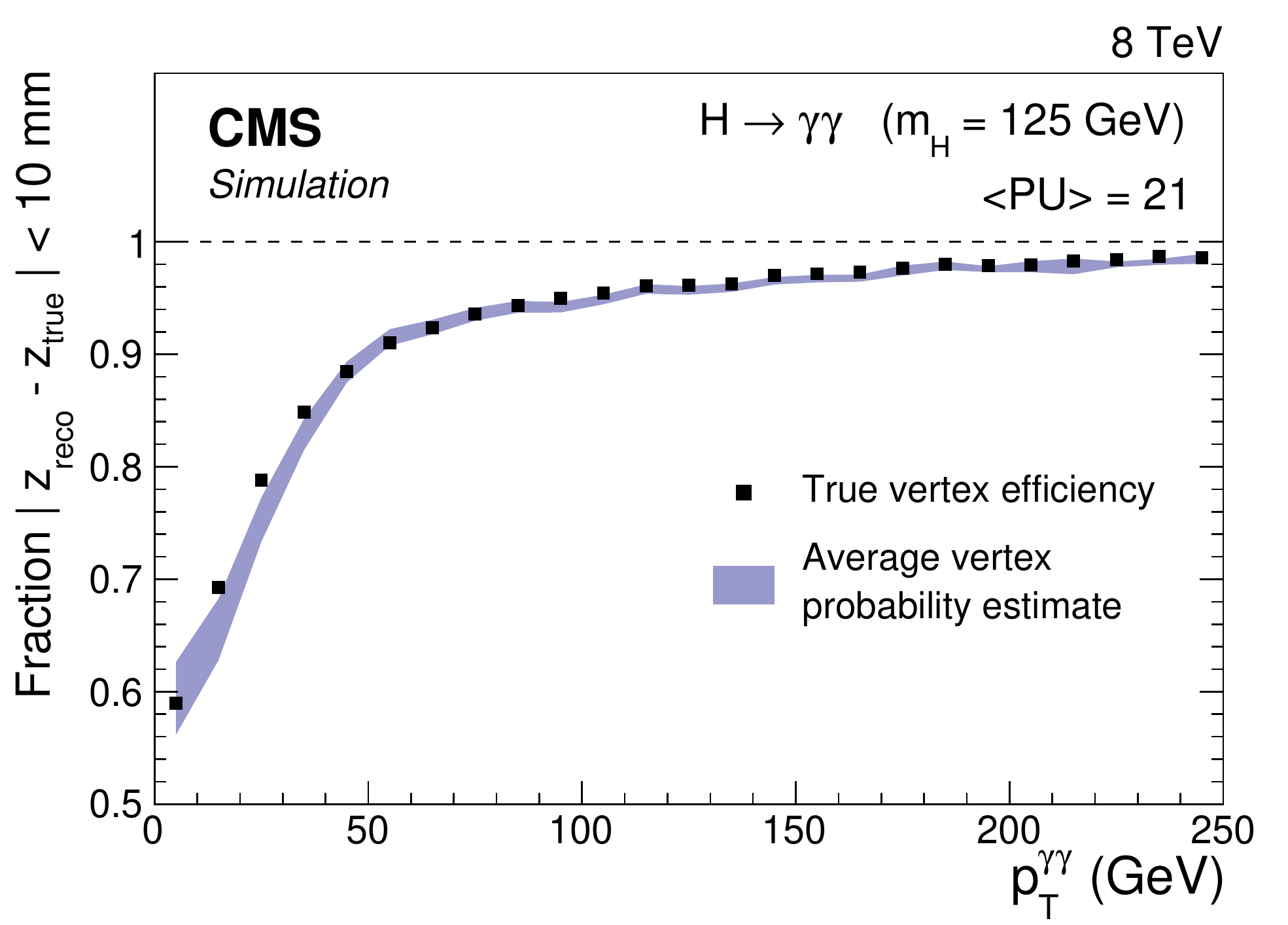} 
    \caption{Fraction of diphoton vertices (solid points) assigned, by the vertex assignment BDT,
to a reconstructed vertex within 10\mm of their true location in simulated
Higgs boson events, $\mH = 125\GeV$, $\sqrt{s}=8\TeV$,
as a function of $\ptgg$.
Also shown is a band, the centre of which is the mean prediction, from the vertex probability BDT (described in Section~\ref{sec:vtx-prob}),
of the probability
of correctly locating the vertex. The mean is calculated in \ptgg\ bins, and the width of the band represents
the event-to-event uncertainty in the estimates.}
    \label{fig:vtx-prob-closure}
\end{figure}

The vertex finding efficiency, defined as the efficiency that the chosen vertex is
within 10\mm of the true vertex location, has been measured using $\Zmm$ events.
The performance of the algorithm is evaluated after re-reconstruction of the vertices following removal of the muon tracks,
so that the event mimics a diphoton event.
The use of tracks from a converted photon to locate the vertex is validated in \GAMJET\ events.
In both cases the ratio of the efficiency measured in data to that measured in MC simulation is within 1\%
of unity when viewed as a function of the number of vertices in the event.
When viewed as a function of the $\cPZ$-boson $\pt$, the deviation of the ratio from unity increases to a
few percent in the region where $\ptZ<15\GeV$.
The measured ratio as a function of the $\cPZ$-boson $\pt$ is used as a correction to the vertex finding efficiency in
simulated Higgs boson signal events.
The vertex finding efficiency for a Higgs boson of mass 125\GeV, integrated over its \pt spectrum, is
computed to be  85.4 (79.6)\% in the 7 (8)\TeV dataset.
Figure~\ref{fig:vtx-prob-closure} shows the efficiency with which a diphoton system is assigned to a
vertex reconstructed within 10\mm of the true diphoton vertex in simulated Higgs boson events
($\mH = 125\GeV$) in the 8\TeV dataset, as a function of the transverse momentum of the
diphoton system.

\subsection{Per-event vertex probability}
\label{sec:vtx-prob}
\begin{figure}
   \centering
       \includegraphics[width=\cmsFigWidth]{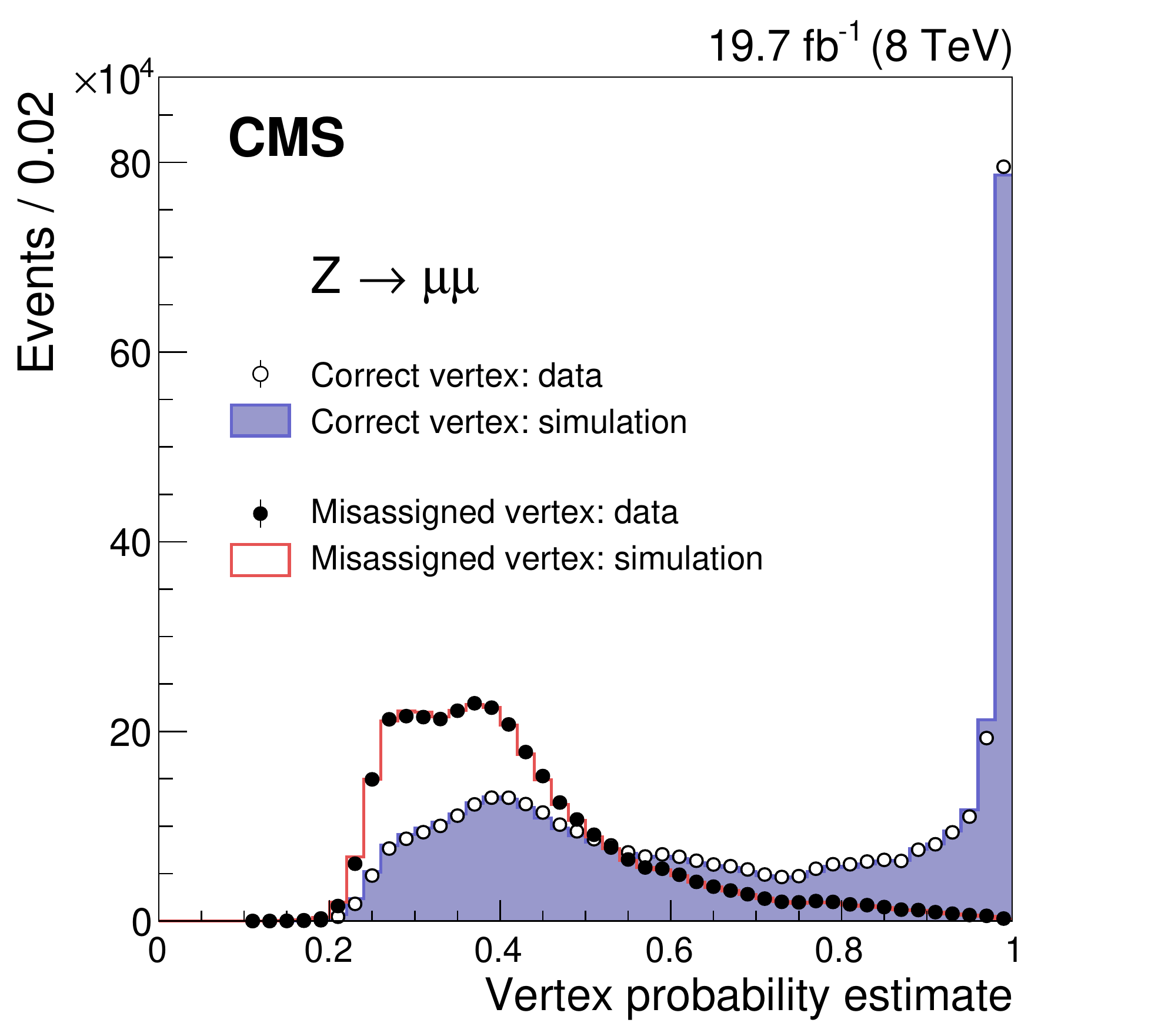} 
       \caption{Distribution of the vertex probability estimate in $\Zmm$ events.
The vertex probability estimates in 8\TeV data (points), are compared to the estimates in MC simulation (histograms).
The comparison is made separately for events in which the vertex is assigned to the same (open circles and filled histogram),
or to a different vertex (filled circles and outlined histogram), as that identified by the muons.
}
       \label{fig:vtx-validation}
\end{figure}

A second vertex-related multivariate discriminant has been designed to
estimate, event-by-event, the probability for the vertex
assignment to be within 10\mm of the diphoton interaction point.
This, in conjunction with the event-by-event estimate of the energy
resolution of each photon, is used to estimate the diphoton mass resolution
for each individual event, and this estimate is used in the event classification, as described in Section~\ref{sec:classes}.
The inputs of the vertex probability BDT are
\begin{itemize}
\item the values of the vertex identification BDT output for the three most likely vertices in the event,
\item the total number of reconstructed vertices in the event,
\item the transverse momentum of the diphoton system, $\ptgg$,
\item the distances between the chosen vertex and the second- and third-best vertices,
\item the number of photons with an associated conversion track or tracks.
\end{itemize}
The vertex probability BDT is tested with simulated signal events as shown in Fig.~\ref{fig:vtx-prob-closure},
and the performance in data is tested using $\Zmm$ events.
Validation of the vertex probability BDT for events in which conversion tracks are present is achieved
using $\GAMJET$ events in which one or more conversion tracks are reconstructed.
The probability to identify a close-enough vertex (vertex probability) has a linear relationship with the vertex probability BDT score, the parameters of which are obtained from a fit using a sample of simulated signal events.
Figure~\ref{fig:vtx-validation} shows the distribution of the vertex probability estimate, obtained from the BDT score,
in $\Zmm$ events.
The charged particle tracks belonging to the muon pair are used to identify the vertex,
and are then removed from the event before re-reconstructing the vertices and passing
them to the vertex identification and the vertex probability BDTs.
The \pt of the dimuon pair is used in the BDT calculation in place of $\ptvecgg$.
The vertex identified by the muons is assumed to be the correct or true vertex,
so that if the vertex assignment BDT chooses that vertex, it chooses the right vertex, otherwise it chooses the wrong vertex.
The vertex probability estimates in data (points), are compared to MC simulation (histograms).
The comparison is made separately for events in which the vertex assignment BDT assigns the right vertex,
and for those in which it assigns a wrong vertex.

\section{Event classification}
\label{sec:classes}

The analysis uses events with two photon candidates satisfying the preselection
requirements (described in Section~\ref{sec:photonID}) with an invariant mass, $\mgg$, in the range
$100<\mgg<180\GeV$, and with $\ptga>\mgg/3$ and $\ptgb>\mgg/4$.
In the rare case of multiple diphoton candidates, the one with the
highest $\ptga+\ptgb$ is selected.
The use of \pt thresholds scaled by \mgg
prevents the distortion of the low end of the \mgg spectrum that results if a fixed threshold is used.
An additional requirement is applied on the photon identification BDT scores for both photons, which
are required to be greater than $-0.2$ (see Fig.~\ref{fig:idmva-worst}).
This requirement retains more than 99\% of simulated signal
events fulfilling the other analysis selection requirements, while removing about 24\% of events in data.
The requirements listed above are referred to as the ``full diphoton preselection''.

To achieve the best analysis performance, the events are separated into classes based on both their mass
resolution and their relative probability to be due to signal rather than background.
The first step in the classification of the events involves the extraction of those tagged
by the presence of objects in the final state, in addition to the photon pair, that give the event a signature characteristic
of one of the production processes.
The remaining untagged events, which constitute the majority ($\approx$99\%) of the
events used in the analysis, are classified according to a variable constructed using multivariate techniques.

The classification procedure, which is described in detail below, results in 11 event classes for the 7\TeV dataset and 14 for the 8\TeV dataset.
The event classes, and the expected number of SM Higgs boson events and estimated background in those classes, are set out
later, in Table~\ref{tab:ClassFracs}, together with the composition of the expected SM Higgs boson signal in terms of the production
processes, and the diphoton mass resolution expected for the signal in each of the classes.
To ensure that the classes are mutually exclusive, events
are tested against the class selection requirements in a fixed order as described in Section~\ref{sec:procedure}.

\subsection{Multivariate event classifier}
\label{sec:diphotonBDT}

A multivariate event classifier, the diphoton BDT, is constructed to satisfy the following criteria:
\begin{enumerate}
\item The diphoton BDT should assign a high score to events that have
\begin{enumerate}
\item good diphoton mass resolution,
\item high probability of being signal rather than background.
\end{enumerate}
\item The classifier should not select events according to the mass of the diphoton system relative to the particular mass of the Higgs boson signal used for training.
\end{enumerate}
The classifier incorporates a per-event estimate of the diphoton
mass resolution, the identification BDT scores of both photons, and
the kinematic properties of the diphoton system, except for $\mgg$.
To avoid any dependence on $\mH$, the transverse momenta and resolutions are divided by $\mgg$.

The complete list of variables used in the BDT is the same as used in previous versions of the analysis~\cite{Chatrchyan:2013lba}:
the scaled photon transverse momenta ($\ptga/\mgg$ and $\ptgb/\mgg$),
the pseudorapidities of both photons,
the photon identification BDT classifier values for both photons,
the cosine of the angle between the two photons in the transverse plane,
the expected relative diphoton mass resolutions under the hypotheses of selecting the correct/a wrong interaction vertex,
and also the probability of selecting the correct vertex.

The diphoton mass resolution depends on several factors: the location
of the associated energy deposits in the calorimeter; whether or not one or both photons
converted in the detector volume in front of the calorimeter; and the
probability that the true diphoton vertex has been identified.
Events in which one of the photons has a low identification BDT score are more likely to be due to background processes.
The Higgs signal-to-background ratio, $S/B$, varies with the kinematic
properties of the diphoton system mainly through the $\eta$ of the photons
(highest $S/B$ when both are in the barrel), and $\ptgg$ (highest $S/B$ for large $\ptgg$).
The BDT is trained using a simulated signal sample having a mass, $\mH=123\GeV$, near the centre of the mass range of the analysis.
The relative abundance of events from different production processes in the sample is set according to the expectations for a SM Higgs boson
with that mass.

\begin{figure}
  \centering
    \includegraphics[width=\cmsFigWidth]{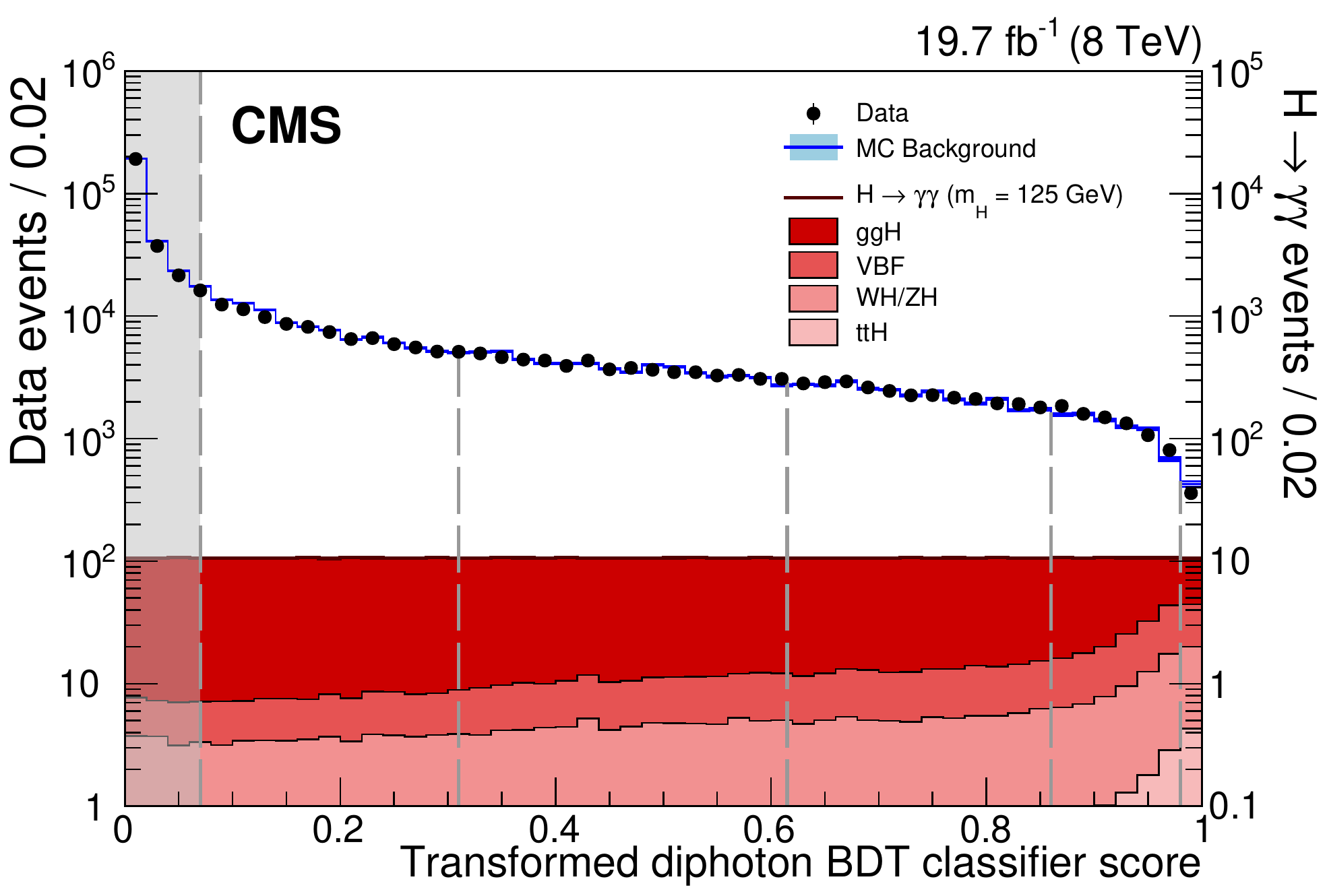} 
    \caption{Transformed diphoton BDT classifier score for events satisfying the full diphoton preselection in the 8\TeV data
(points with error bars, left axis),
and for simulated signal events from the four production processes (solid filled histograms, right axis).
The outlined histogram, following the data points, is for simulated background events.
The vertical dashed lines show the boundaries of the untagged event classes,
with the leftmost dashed line representing the score below which events are
discarded and not used in the final analysis (described in Section~\ref{sec:VBFandUntagged}).
  }
    \label{fig:diphoton-bdt-score}
\end{figure}

The multivariate classifier assigns a score to each event.
It has been verified that selecting simulated background events with high diphoton BDT score does not result in any peak in the
diphoton invariant mass distribution of the selected events.
Figure~\ref{fig:diphoton-bdt-score} shows, for the 8\TeV dataset, how the BDT performs on simulated SM $\Hgg$
signal events with $\mH=125\GeV$, and on data satisfying the full diphoton preselection.
The classifier score has been transformed such that the sum of signal events from all processes has a uniform, flat, distribution.
This transformation assists visualization of the performance of the BDT.
The outlined histogram, following the data points, is for simulated background events.
The vertical dashed lines indicate the boundaries of the untagged event classes, the determination of which is described
in Section~\ref{sec:VBFandUntagged}.
Given that the data are completely dominated by background events, it can be seen that the signal-to-background ratio increases substantially
with the classifier score, and that the VBF, VH, and \ttH processes tend to achieve high scores,
due to their significantly harder \ptgg spectrum~\cite{Cahn:1983ip,Altarelli:1987ue}.

\begin{figure}
  \centering
    \includegraphics[width=\cmsFigWidth]{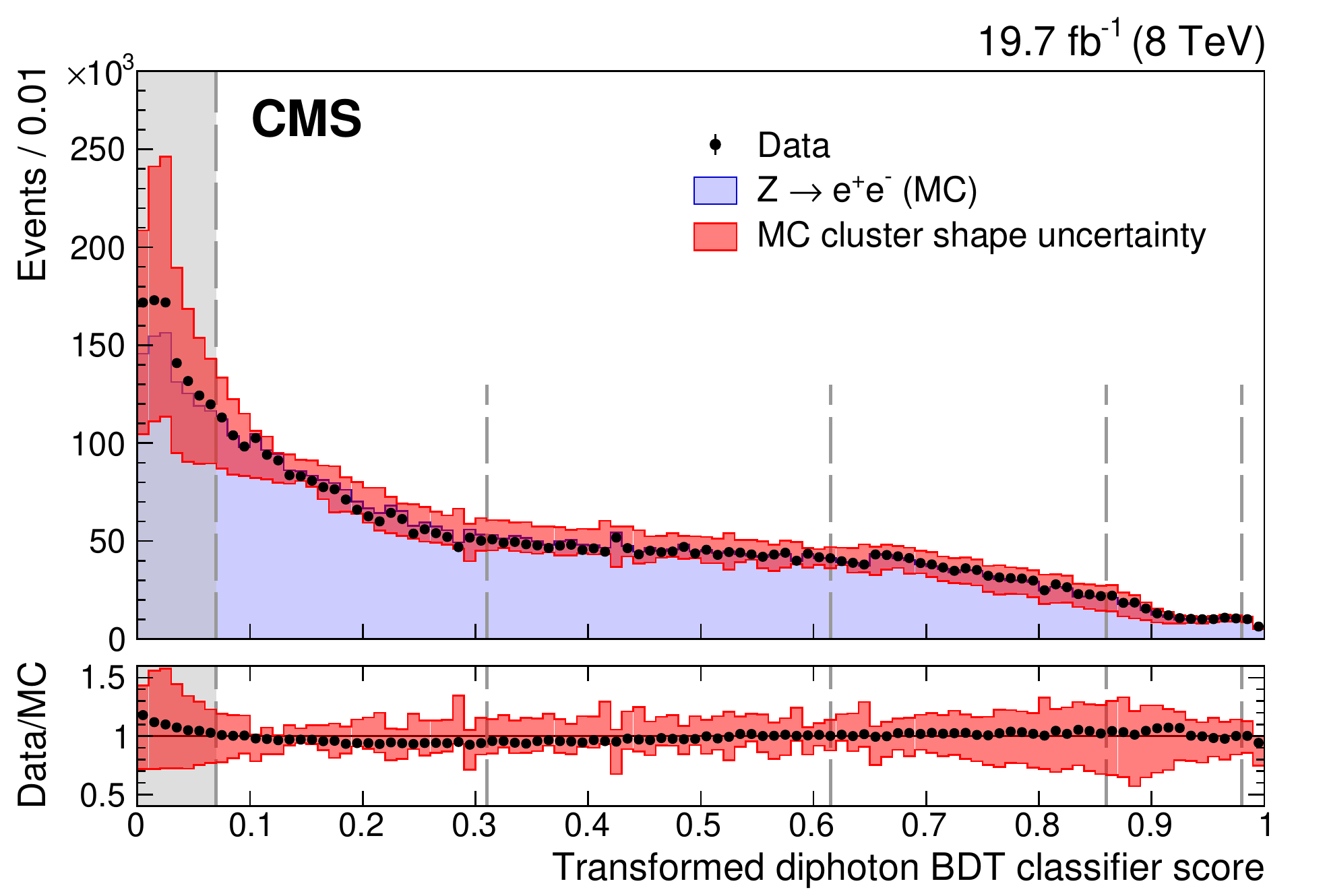} 
    \caption{Transformed diphoton BDT classifier score for $\Zee$ events in 8\TeV data, and
in MC simulation, in which the electrons are reconstructed as photons. The distribution of
simulated events is represented by a histogram, and the data by points with error bars.
For each bin, the ratio of the number of events in data to the number of simulated events
is shown in the lower plot.
The bands in the two plots indicate the systematic uncertainty related to the MC cluster shape uncertainty (see text).
The vertical dashed lines show the boundaries of the untagged event classes,
with the leftmost dashed line representing the score below which events are
discarded and not used in the final analysis (described in Section~\ref{sec:VBFandUntagged}).
  }
    \label{fig:diphoton-bdt-valid}
\end{figure}

Figure~\ref{fig:diphoton-bdt-valid} shows a comparison of the transformed classifier score for $\Zee$ data and
for MC simulated events, in which for both cases the electrons are reconstructed as photons.
The electron showers in the events satisfy the full diphoton preselection requirements with the electron veto condition inverted.
The classifier score has been subjected to the same transformation as
was used for Fig.~\ref{fig:diphoton-bdt-score}.
The score for $\Zee$ events peaks at low values whilst Higgs boson
signal events have a flat distribution, reflecting the differences between the two types of event, but it can be seen that sufficient
numbers of $\Zee$ events are present even at high values of the
classifier score to enable the agreement between data and MC simulation
to be adequately tested there.
The good agreement between MC simulation and data for $\Zee$ events
constitutes an important check that the modeling of the BDT input
variables and their correlations in the simulation of the Higgs boson signal is accurate.
The simulated events have been weighted so that the $\cPZ$-boson \pt distribution matches that observed in $\Zee$ data.
The band indicates the systematic uncertainty resulting from propagating to the diphoton BDT event classifier both the uncertainty associated with the photon identification BDT score (which corresponds to a shift of $\pm$0.01 of the score) and the uncertainty in the per-photon estimate of the energy resolution
(which amounts to a scaling of its value by $\pm$10\%).
Since the magnitudes of these two uncertainties were chosen to cover the discrepancies between data and simulation
in the tails of the distributions of the two variables, the resulting uncertainty in the diphoton BDT event classifier
appears to be slightly overestimated.

\subsection{Events tagged by exclusive signatures}
\label{sec:exclusive-tags}
Selections enriched in Higgs boson production mechanisms other than
ggH can be made by requiring, in addition to the diphoton pair, the presence of other objects
which provide signatures of the production mechanism.
Higgs bosons produced by VBF are accompanied by a pair of jets separated by a large rapidity gap.
Those resulting from the VH production mechanism may be accompanied
by one or more charged leptons, large $\MET$, or jets from the decay of the \PW\ or \cPZ\ boson.
Those resulting from \ttH production are, as a result of the decay of the top quarks,
accompanied by b quarks, and may be accompanied by charged leptons or additional jets.

The tagging of dijet events, targeting VBF production, significantly increases the overall sensitivity of the analysis
and precision on the measured signal strength, and increases the sensitivity to deviations of the Higgs boson couplings from their expected values.
The tagging aimed at the VH process
increases the sensitivity to deviations of the couplings, and the
\ttH tagging further probes the compatibility of the observed signal with a SM Higgs boson.

The \pt spectrum of Higgs bosons produced by the VBF, VH, and \ttH processes
is significantly harder than that of Higgs bosons produced by ggH, or of background diphotons.
This results in a harder leading-photon \pt spectrum.
In the tagged-class selections advantage is taken of this difference by raising the \pt requirement on the leading photon.

\subsubsection{Dijet-tagged event selection and BDT classifiers for VBF production}
\label{sec:VBF}

Vector boson fusion production results in two forward jets, originating from the two scattered quarks.
Separating events tagged by the presence of dijets compatible with the VBF process into specific event classes
not only increases the separation between signal and background,
it also increases the separation between signal production processes.
In the purest VBF dijet-tagged class the signal is expected to have a contribution of only 18\% from ggH production.
A loose preselection of dijet events is defined and a dijet BDT is trained to separate VBF signal from diphoton background
using samples of MC events satisfying this dijet preselection.
Signal events from ggH satisfying the dijet preselection are included as background in the training.
Details of the dijet preselection and the BDT input variables are given below.
A further, ``combined'', BDT is then trained.
This BDT has only three input variables: the score of the dijet BDT, the score of the diphoton BDT,
and the transverse momentum of the diphoton system divided by its mass, $\ptgg/\mgg$.
Events for the VBF dijet-tagged classes are selected, from those satisfying the loose dijet preselection,
by placing a minimum requirement on their combined BDT score,
and the selected events are then classified using that score.

The dijet preselection is applied to diphoton events satisfying the full diphoton preselection
and requires the leading (in $\pt$) and subleading jets in the event, within $\abs{\eta}<4.7$,
to have $\pt>30$ and 20\GeV respectively, and for the pair to have an invariant mass $\mjj>250\GeV$.
The pseudorapidity requirement ($\abs{\eta}<4.7$) is more restrictive than the full detector acceptance ($\abs{\eta}\lesssim5$),
to avoid the use of jets for which the energy corrections are large and less reliable,
and is found to decrease the signal acceptance by $<$2\%.
Additionally, the \pt threshold of the leading photon is raised, requiring $\ptga>\mgg/2$ for VBF dijet-tagged events.

The jet energy measurement is calibrated to correct for detector effects
using samples of dijet, \GAMJET, and $\cPZ + \text{jet}$ events~\cite{Chatrchyan:2011ds}.
The energy from pileup interactions and from the underlying event is also included in the reconstructed jets.
This energy is subtracted using an $\eta$-dependent transverse momentum density calculated with the jet areas technique~\cite{Cacciari:2007fd,Cacciari:2008gn,Cacciari:2011ma}, evaluated on an event-by-event basis.
Particles produced in pileup interactions may be clustered into jets of relatively large \pt, referred to as pileup jets.
These pileup jets are largely removed using selection criteria based on the width of the jet
or the compatibility of the tracks in a jet with the primary vertex~\cite{CMS-PAS-JME-13-005}.
Finally, jets within $\DR<0.5$ of either of the photons are rejected to exclude the possibility of photons having been
included in the reconstruction of the jet.

The variables used in the dijet BDT are
the scaled transverse momenta of the photons, $\ptga/\mgg$ and $\ptgb/\mgg$,
the transverse momenta of the leading and subleading jets, $\ptja$ and $\ptjb$,
the dijet invariant mass, $\mjj$,
the difference between the pseudorapidities of the jets, $|\Delta\eta_\text{jj}|$,
the difference between the average pseudorapidity of the two jets and
the pseudorapidity of the diphoton system, $|\eta_{\gamma\gamma}-(\eta_\text{j1}+\eta_\text{j2})/2|$~\cite{Rainwater:1996ud},
and the absolute difference in the azimuthal angle between the diphoton system and the dijet system, $\Dggjj$.
Because of the large theoretical uncertainty in the cross section
due to higher-order contributions to the ggH process accompanied by two jets
in the region very close to $\Dggjj=\pi$~\cite{LHCHiggsCrossSectionWorkingGroup3,Stewart:2011cf},
the maximum value of the variable is restricted to $\pi-0.2$;
events with $\Dggjj>\pi-0.2$ are treated as if the value was $\pi-0.2$.

\subsubsection{Lepton-, dijet-, and \texorpdfstring{\MET}{MET}-tagged event classes for VH production}

The selection requirements for the classes aimed at selecting events produced by the VH process have
been obtained by minimizing the expected uncertainty in the measurement of signal strength of the process,
using data in control regions to estimate the background and MC signal samples to estimate the signal efficiency.
Four classes are defined: events with a muon or an electron are separated into
two classes, according to whether there is significant $\MET$ or another lepton in the event, or there is not;
a third class selects events with two or more jets; and the fourth class consists of events with large $\MET$.
The leading photon in the events selected for the lepton classes and for the $\MET$-tagged class is required
to satisfy $\ptga>3\mgg/8$; for the dijet-tagged VH class the requirement is tighter, $\ptga>\mgg/2$.

Muons are reconstructed with the particle-flow algorithm and are required to be within $\abs{\eta}<2.4$.
A tight selection is applied, based on the
quality of the track and the number of hits in the tracker and muon spectrometer.
A strict match between
the tracker and the muon spectrometer segments is also applied to reduce the contamination from muons
produced in decays of hadrons and from beam halo interactions.
Finally, a loose particle-flow isolation requirement is applied.

Electrons are identified as clusters of energy deposited in the ECAL matched to tracks.
Electron candidates are required to have an ECAL supercluster within
the same fiducial region as for photons.
Electron identification is based on a multivariate technique~\cite{Chatrchyan:2013mxa}.
The electron track has to fulfil requirements on the transverse and longitudinal
impact parameter with respect to the electron vertex and cannot have more than one missing hit in the innermost layers of the tracker.
Electrons from conversions are excluded as described in Ref.~\cite{CMS-PAS-TRK-10-003}
and a loose particle-flow isolation requirement is applied.

The tightly selected lepton class (``VH tight $\ell$'') is characterised by the full signature of a leptonically decaying \PW\ or \cPZ\ boson,
and requires, in addition to the electron or muon, the
presence of $\MET>45\GeV$ or another lepton of the same flavour as the first and with opposite sign.
For the lepton plus $\MET$ signature the \pt of the lepton is required to be greater than 20\GeV.
For the dilepton signature the lepton \pt requirement is relaxed to $\pt>10\GeV$, but the invariant mass of the pair
is required to be between 70 and 110\GeV.
For the loose lepton class (``VH loose $\ell$'') only a single electron or muon with $\pt>20\GeV$ is required but
additional requirements are made to reduce background from leptonic decays of \cPZ\ bosons with initial- or final-state radiation:
muons and electrons are required to be separated from the closest photon by $\DR>1.0$, and the invariant mass of electron-photon pairs
is required to be more than 10\GeV away from the $\cPZ$-boson mass.
In addition, a conversion veto is applied to the electrons to reduce the number of electrons originating from photon conversions.

Events selected for the dijet-tagged VH class are required to have a pair of jets with $\pt>40\GeV$, within the region $\abs{\eta}<2.4$,
and with an invariant mass within the range $60<\mjj<120\GeV$; additional jets may also be present.
The \pt of the diphoton system is required to satisfy $\ptgg>13\mgg/12$.
The selection also exploits the expected angular distribution of the diphoton pair with respect to the dijet pair from the vector boson decay.
The angle, $\theta^\star$, that the diphoton system makes, in the diphoton-dijet centre-of-mass frame, with respect to the direction of motion of
the diphoton-dijet system in the lab frame is computed.
The distribution of $\cos\theta^\star$ for signal events coming from VH production is rather flat, whereas background and signal events from ggH production result in $\cos\theta^\star$ distributions strongly peaked at $|{\cos\theta^\star}|=1$.
Consequently $|{\cos\theta^\star}|<0.5$ is required.

For the $\MET$ tag, additional selection criteria are applied on the azimuthal angular separation
between the diphoton system and the $\MET$ direction,
$|\Delta\phi_{\Pgg\Pgg\MET}|>2.1$, and between the diphoton system and the leading jet in the event,
$|\Delta\phi_{\Pgg\Pgg\mathrm{j}^1}|<2.7$.
Discrepancies between data and simulated events in the direction and magnitude of the $\MET$ vector have been studied in detail and
a set of corrections derived, some of which need to be applied to simulated events, and others to data.
The corrected $\MET$ is required to satisfy $\MET>70\GeV$.

In addition to the requirements described above, a minimum requirement is also made
on the diphoton BDT classifier score for entry into the event classes tagging VH production.
The severity of the requirement is optimized for each class: 0.17 for the two lepton-tagged classes, 0.62 for the $\MET$-tagged class,
and 0.76 for the VH dijet-tagged class, where the numerical scale is the classifier score shown
in Figs.~\ref{fig:diphoton-bdt-score} and~\ref{fig:diphoton-bdt-valid}.

\subsubsection{Event classes tagged for \texorpdfstring{\ttH}{t t-bar H} production}
The production of Higgs bosons in association with top quarks has a small cross section,
and so the overall cross section times branching fraction of the decay to photons is only 0.3\unit{fb} at NLO.
Therefore, in the full dataset only a handful of events are expected.
To maximize signal efficiency we devise event selections that collect both leptonic and hadronic
decays of the top quarks, defining both a lepton-tagged and a multijet-tagged event class.

As for the VH event classes, the selection requirements for the classes aimed at selecting events produced by the \ttH process have
been obtained by minimizing the expected uncertainty in the measurement of signal strength of the process,
using data in control regions to estimate the background, and MC signal samples to estimate the signal efficiency.
The leading photon is required to have $\ptga>\mgg/2$.
Jets are required to have $\pt>25\GeV$ and both classes require the presence of at least one b-tagged jet.
The lepton tag is then defined by requiring at least one more jet in the event and at least one electron or muon with $\pt>20\GeV$,
and the multijet tag is defined by the requirement of at least four more jets in the event and no lepton.
Requirements are also made on the minimum diphoton BDT classifier score for entry into the two classes tagging $\ttH$:
0.17 for the lepton class, and 0.48 for the multijet class, where the numerical scale is the classifier score shown
in Figs.~\ref{fig:diphoton-bdt-score} and~\ref{fig:diphoton-bdt-valid}.
For the 7\TeV dataset the events in the two classes are combined after selection to form a single $\ttH$ event class.

\subsection{Classification of VBF dijet-tagged and untagged events}
\label{sec:VBFandUntagged}

Classes for the VBF dijet-tagged events and the untagged events are defined using the scores of the classification BDTs:
the combined dijet-diphoton BDT score is used to select and define the dijet-tagged classes, and the diphoton BDT score
defines the untagged class into which the untagged events are placed.
The BDT score requirements that constitute the event class boundaries
are set by an optimization procedure,
using simulated event samples, aimed at minimizing the expected uncertainty in the signal strength.
To avoid biases, the simulated events are divided into three non-overlapping sets, which are then used only for
the training of the BDTs, or the optimization of event class boundaries, or to model the signal in the extraction of the final results.
The number of available simulated events limits the statistical precision in the optimization procedure.
The small number of simulated events for some background processes where one or more of the photon candidates result from misidentified jet fragments,
results in a very uneven and spikey distribution of the event classifier scores for the simulated background
in the range of BDT scores in which there is some contribution from these processes, but it is rare.
So, for the event class boundary optimization procedure, the event classifier BDT scores are smoothed,
using an adaptive-width Gaussian smoothing in the \textsc{RooFit} package~\cite{RooFit}.
Differences in performance of less than about 2\% are indistinguishable from
statistical fluctuations and are regarded as insignificant.

As a result of the optimization procedure, four untagged event classes and two VBF dijet-tagged classes are defined for the 7\TeV dataset.
For the 8\TeV dataset five untagged and three dijet-tagged classes are defined.
Events that fail the requirement on the combined dijet-diphoton BDT score to enter the VBF dijet-tagged classes may enter other event classes.
Untagged events that have a diphoton BDT score less than the lower boundaries of the untagged classes in the two
datasets are not used in the final statistical analysis.
The goal of the optimization setting the diphoton BDT score requirements, which define the untagged classes,
is to minimize the expected uncertainty in the overall signal strength measurement.
The goal of the optimization for the setting of the combined dijet-diphoton BDT score boundaries, which define the VBF dijet-tagged classes,
is to minimize the expected uncertainty in the signal strength associated with the VBF production mechanism.
When optimizing the boundaries for the 7\TeV dataset, for which the number of MC background events available is particularly
limited, the number of dijet-tagged classes is limited to two and the lower boundary of the lowest dijet-tagged class
is fixed so that the same efficiency times acceptance is obtained for VBF signal events as in the 8\TeV dataset.

\begin{figure}
  \centering
    \includegraphics[width=\cmsFigWidth]{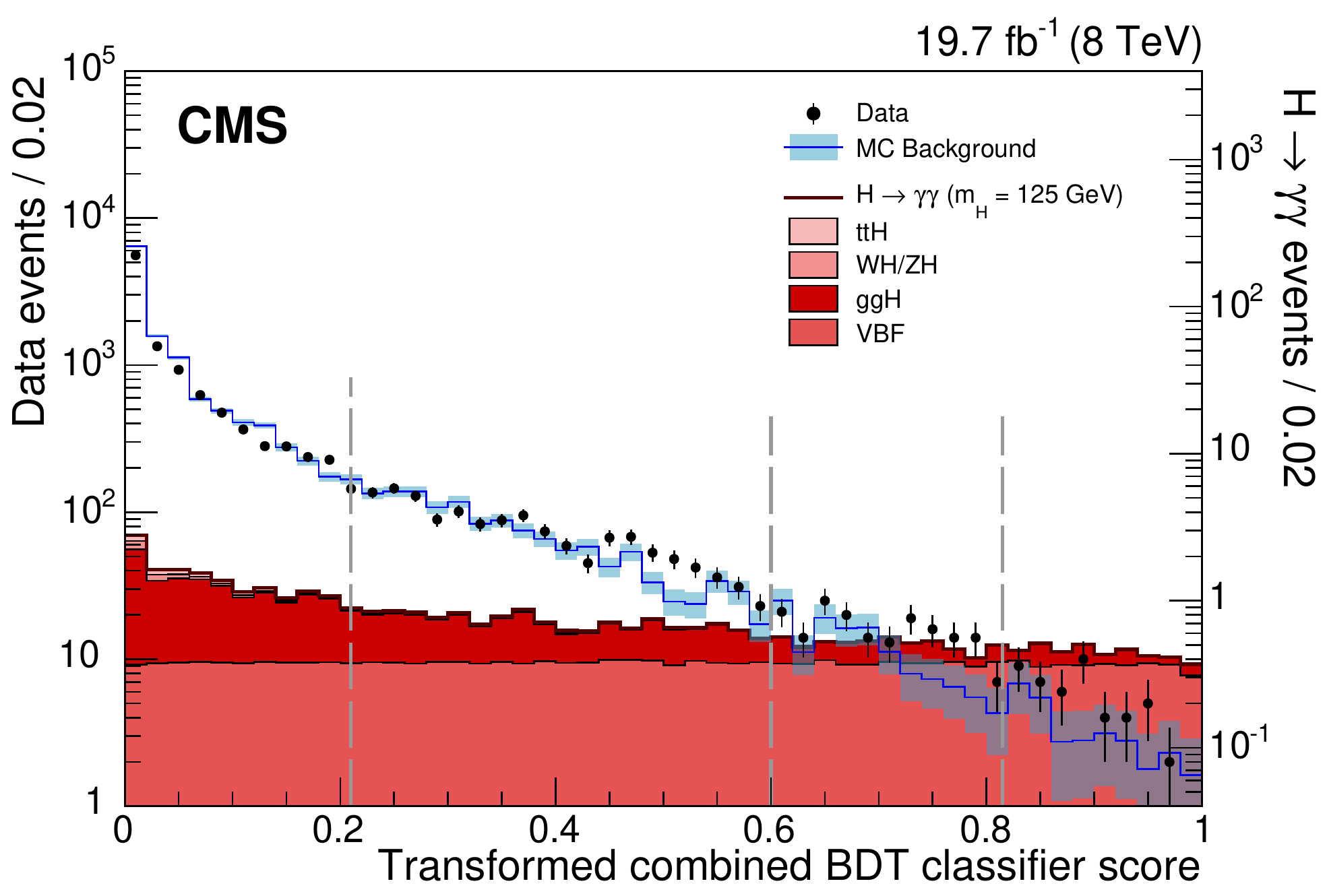} 
    \caption{Score of the combined dijet-diphoton BDT for events satisfying the dijet preselection in 8\TeV data (points with error bars, left axis)
and for simulated signal events from the four production processes (histograms, right axis).
The outlined histogram is for simulated background events; the shaded error bands on the histogram show the statistical uncertainty in the simulation.
The vertical dashed lines show the boundaries of the event classes,
with the leftmost dashed line representing the score below which events are
not included in the VBF dijet-tagged classes, but remain candidates for inclusion in
other classes.
The classifier score is transformed such that signal events produced by the VBF process have a uniform, flat, distribution.
  }
    \label{fig:comb-bdt-score}
\end{figure}

Figure~\ref{fig:comb-bdt-score} shows the combined dijet-diphoton BDT score for events satisfying the dijet preselection
in 8\TeV data, and for simulated signal events from the four production processes.
The outlined histogram is for simulated background events; the shaded error bands on the histogram show the statistical uncertainty in the simulation.
The VBF dijet-tagged class boundaries used for the 8\TeV dataset are shown by vertical dashed lines.
The classifier score is transformed such that signal events produced by the VBF process have a uniform, flat, distribution
across the full range of the score.
This allows the visualization of the extent to which signal events produced by the VBF process are favoured
over background (which predominates in the data), and signal events produced by other processes.
Events with scores below the lower boundary fail the VBF dijet-tagged selection, but remain candidates for inclusion in
other classes.

The lower boundary on the untagged event class with the lowest signal-to-background ratio controls the total number
of events used in the analysis and the overall signal efficiency times acceptance of the analysis (see Fig.~\ref{fig:diphoton-bdt-score}).
The boundary excludes events with very low score in the diphoton BDT for which the background is poorly modelled by MC simulation.
Exclusion of these events has the advantage of allowing a better assessment of the expected sensitivity of the analysis, but the exact
placement of the boundary is of little consequence.

It is found that, within the statistical uncertainty described above, it makes no difference if the optimization goal is
the expected overall uncertainty in signal strength, the expected significance of the signal, or the expected uncertainty in
the measured signal strength associated with the VBF production mechanism.
It is also found that the performance maxima that fix the event class boundaries are rather shallow,
so that the boundaries can be moved without significantly changing the expected performance.
Adding further event classes for either the untagged or the VBF dijet-tagged events does not significantly improve the expected performance.

The overall efficiency times acceptance for SM Higgs boson events with $\mH=125\GeV$ is 49.3\% (48.6\%) in the
8 (7)\TeV analysis.
Investigating the properties of the simulated signal events in the untagged
classes reveals, as expected, that the best untagged class (``untagged 0'') contains events in which the diphoton system has
high \pt\ (almost all events have $\ptgg>80\GeV$), while the second best class (``untagged 1'') is dominated by events in which both photons are
unconverted and situated in the central barrel region of the ECAL.

\subsection{Procedure of classification}
\label{sec:procedure}

In total there are 14 event classes for the analysis of the 8\TeV dataset and 11 for the analysis of the 7\TeV dataset.
To ensure that the classes are mutually exclusive, events
are tested against the class selection requirements in a fixed order: first the production-signature tagged classes
ranked by expected signal-to-background ratio, then the untagged classes.
Once selected, events are no longer candidates for inclusion in
other classes.
The ordering is that shown in Table~\ref{tab:ClassSummary},
which lists the classes together with their key selection requirements.

\begin{table*}[htb]
\caption{Event classes for the 7 and 8\TeV datasets and some of their main selection requirements. Events are tested against the selection requirements of the classes in the order they are listed here.}
\centering
\begin{tabular}{l c c l}
\hline
\multirow{2}{*}{Label} & \multicolumn{2}{l}{No. of classes} & \multirow{2}{*}{Main requirements} \\
 & 7\TeV & 8\TeV & \\
\hline
\hline
\multirow{2}{*}{\ttH\ lepton tag} & \multirow{2}{*}{$\star$} & \multirow{2}{*}{1} & $\ptga>\mgg/2$ \\ 
                                                                               & & & 1 b-tagged jet + 1 electron or muon \\
\hline
\multirow{3}{*}{VH tight $\ell$ tag} & \multirow{3}{*}{1} & \multirow{3}{*}{1} & $\ptga>3\mgg/8$ \\ 
                                                                  & & & [$\Pe$ or $\Pgm$, $\pt>20\GeV$, and $\MET>45\GeV$] or\\
                                                                  & & & [$2\Pe$ or $2\Pgm$, $\pt^\ell>10\GeV$; $70<m_{\ell\ell}<110\GeV$] \\
\hline
\multirow{2}{*}{VH loose $\ell$ tag}& \multirow{2}{*}{1} & \multirow{2}{*}{1}  & $\ptga>3\mgg/8$ \\ 
                                                                   & & & $\Pe$ or $\Pgm$, $\pt>20\GeV$ \\
\hline
\multirow{2}{*}{VBF dijet tag 0-2} & \multirow{2}{*}{2} & \multirow{2}{*}{3} & $\ptga>\mgg/2$ \\
                                                                             & & & 2 jets; classified using combined diphoton-dijet BDT\\
\hline
\multirow{2}{*}{VH \MET\ tag} & \multirow{2}{*}{1} & \multirow{2}{*}{1} & $\ptga>3\mgg/8$\\ 
                                                                    & & & $\MET>70\GeV$ \\
\hline
\multirow{2}{*}{\ttH multijet tag} & \multirow{2}{*}{$\star$} & \multirow{2}{*}{1} & $\ptga>\mgg/2$ \\ 
                                                                       & & & 1 b-tagged jet + 4 more jets \\
\hline
\multirow{2}{*}{VH dijet tag} & \multirow{2}{*}{1} & \multirow{2}{*}{1} & $\ptga>\mgg/2$\\ 
                                                                                         & & & jet pair, $\pt^\mathrm{j}>40\GeV$ and $60<\mjj<120\GeV$ \\
\hline
\multirow{2}{*}{Untagged 0-4} & \multirow{2}{*}{4} & \multirow{2}{*}{5} & The remaining events,\\
                                                                          & &  & classified using diphoton BDT \\
\hline
\multicolumn{4}{ p{15cm} }{$\star$ For the 7\TeV dataset, events in the \ttH lepton tag and multijet tag classes are selected first, and combined to form a single event class.}
\end{tabular}
\label{tab:ClassSummary}
\end{table*}

\section{Signal model}
\label{sec:smodelling}

A parametric signal model is constructed separately for each event class
and for each production mechanism from a fit of the simulated invariant mass shape,
after applying the corrections determined from comparisons of data and simulation for $\Zee$ and $\Zmmg$ events,
for nine values of \mH\ in the range $110\leq\mH\leq150\GeV$, at 5\GeV intervals.
The two possible cases regarding diphoton vertex identification, correct vertex and wrong (misidentified) vertex, are fitted separately.
Good descriptions of the distributions, including the tails, can be achieved using a sum of Gaussian functions,
where the means are not required to be identical.
The fits are first performed for the $\mH=125\GeV$ MC sample to determine the number of Gaussian functions to be used
and the starting values of their parameters for the further fits to the other eight samples.
As many as five Gaussian functions are used, although in most cases the use of two or three results in a good fit.
Signal models for intermediate values of \mH\ are obtained by linear interpolation of the fitted parameters.

Table~\ref{tab:ClassFracs} shows the number of expected signal events from a SM
Higgs boson with $\mH=125\GeV$ as well as the background density at that mass for each of the event classes in the 7 and 8\TeV datasets.
The background estimate is obtained from a fit to the data, as described in Section~\ref{sec:bmodelling},
and is given as the differential rate, $\text{d}N/\text{d}\mgg$ (events/GeV), at $\mgg=125\GeV$.
The table also shows the fraction of each Higgs boson production process (as
predicted by MC simulation) as well as the mass resolution, measured both by
half the width of the narrowest interval containing 68.3\% of the invariant mass distribution,
$\sigma_\text{eff}$, and by the full width at half maximum of the distribution divided by 2.35, $\sigma_\mathrm{HM}$.

It can be seen that in all classes $\sigma_\text{eff}>\sigma_\mathrm{HM}$ since the tails of the signal mass distribution
are always somewhat larger relative to the width of the core of the distribution than would be the case for a Gaussian distribution.
Untagged events with the best mass resolution are selected to the best event classes, and even ignoring the improving mass resolution,
and considering a wide window to include all the signal events,
the signal-to-background ratio improves by an order of magnitude going from the worst to the best
untagged class --- a significantly larger variation than the change in resolution.
The highest signal-to-background ratio is achieved in the tagged classes, many of which manage to also achieve high levels of
purity with respect to contamination from the ggH process.

The mass resolution achieved has improved significantly with respect to analyses of this decay mode previously reported by CMS~\cite{Chatrchyan:2013lba},
due to improved intercalibration of the ECAL, complemented by the improved supercluster energy correction regression
described in Section~\ref{sec:photonE}.
For events in which both photons are in the barrel the $\sigma_\text{eff}$ has been reduced by around 5\%
in 7\TeV data, and by more than 20\% in 8\TeV data.
When at least one photon is in the endcap region the $\sigma_\text{eff}$ has been reduced by around 20\%
in 7\TeV data, and by more than 30\% in 8\TeV data.
The reduction in $\sigma_\mathrm{HM}$, representing the core of the distribution, is slightly larger, generally an additional 5\% better,
when compared to $\sigma_\text{eff}$.
\begin{table*}[htbp]
\centering
\caption{Expected number of SM Higgs boson events ($\mH=125\GeV$) and
estimated background (``Bkg.'') at $\mgg=125\GeV$ for all event classes of the 7
and 8\TeV datasets.
The composition of the SM Higgs boson signal in terms of the production
processes and its mass resolution is also given.
The number corresponding to the production process making the largest contribution to each event class is highlighted in boldface.
Numbers are omitted for production processes representing less than 0.05\% of the total signal.
The variables used to characterize the resolution, $\sigma_\text{eff}$ and $\sigma_\mathrm{HM}$, are defined in the text.
}
\begin{tabular}{|c|r|r|rrrrr|c|c|r|}
\hline
\multicolumn{2}{|c|}{\multirow{3}{*}{Event classes}} & \multicolumn{8}{c|}{Expected SM Higgs boson signal yield (\mH=125\GeV)} & \multicolumn{1}{c|}{Bkg.} \\
\cline{3-10}
\multicolumn{2}{|c|}{} & \multirow{2}{*}{Total} & \multirow{2}{*}{ggH} & \multirow{2}{*}{VBF} & \multirow{2}{*}{WH} & \multirow{2}{*}{ZH} & \multirow{2}{*}{\ttH} & $\sigma_\text{eff}$ & $\sigma_\mathrm{HM}$ & \multicolumn{1}{c|}{\footnotesize{($\GeV^{-1}$)}} \\
\multicolumn{2}{|c|}{} & & & & & & & \footnotesize{(\GeVns)} & \footnotesize{(\GeVns)} &  \\
\hline
\multirow{11}{*}{\begin{sideways}{7\TeV 5.1\fbinv}\end{sideways}}
& Untagged 0 & 5.8 & \textbf{ 79.8\%} & 9.9\% & 6.0\% & 3.5\% & 0.8\% & 1.11 & 0.98 & 11.0 \\
& Untagged 1 & 22.7 &  \textbf{ 91.9\%} & 4.2\% & 2.4\% & 1.3\% & 0.2\% & 1.27 & 1.09 & 69.5 \\
& Untagged 2 & 27.1 &  \textbf{ 91.9\%} & 4.1\% & 2.4\% & 1.4\% & 0.2\% & 1.78 & 1.40 & 135.\phantom{0} \\
& Untagged 3 & 34.1 &  \textbf{ 92.1\%} & 4.0\% & 2.4\% & 1.3\% & 0.2\% & 2.36 & 2.01 & 312.\phantom{0}  \\
\cline{2-11}
& VBF dijet 0 & 1.6 & 19.3\% &  \textbf{ 80.1\%} & 0.3\% & 0.2\% & 0.1\% & 1.41 & 1.17 & 0.5 \\
& VBF dijet 1 & 3.0 & 38.1\% &  \textbf{ 59.5\%} & 1.2\% & 0.7\% & 0.4\% & 1.65 & 1.32 & 3.5 \\
\cline{2-11}
& VH tight $\ell$ & 0.3 & ---~~ & ---~~ &  \textbf{ 77.2\%} & 20.6\% & 2.2\% & 1.61 & 1.31 & 0.1 \\
& VH loose $\ell$ & 0.2 & 3.6\% & 1.1\% &  \textbf{ 79.1\%} & 15.2\% & 1.0\% & 1.63 & 1.32 & 0.2 \\
& VH \MET & 0.3 & 4.5\% & 1.1\% & 41.5\% &  \textbf{ 44.6\%} & 8.2\% & 1.60 & 1.14 & 0.2 \\
& VH dijet & 0.4 & 27.1\% & 2.8\% &  \textbf{ 43.7\%} & 24.3\% & 2.1\% & 1.54 & 1.24 & 0.5 \\
\cline{2-11}
& \ttH tags & 0.2 & 3.1\% & 1.1\% & 2.2\% & 1.3\% &  \textbf{ 92.3\%} & 1.40 & 1.13 & 0.2 \\
\hline
\noalign{\vskip 1mm}
\hline
\multirow{14}{*}{\begin{sideways}{8\TeV 19.7\fbinv}\end{sideways}}
& Untagged 0 & 6.0 &  \textbf{ 75.7\%} & 11.9\% & 6.9\% & 3.6\% & 1.9\% & 1.05 & 0.79 & 4.7 \\
& Untagged 1 & 50.8 &  \textbf{ 85.2\%} & 7.9\% & 4.0\% & 2.4\% & 0.6\% & 1.19 & 1.00 & 120.\phantom{0}  \\
& Untagged 2 & 117.\phantom{0}  &  \textbf{ 91.1\%} & 4.7\% & 2.5\% & 1.4\% & 0.3\% & 1.46 & 1.15 & 418.\phantom{0}  \\
& Untagged 3 & 153.\phantom{0}  &  \textbf{ 91.6\%} & 4.4\% & 2.4\% & 1.4\% & 0.3\% & 2.04 & 1.56 & 870.\phantom{0}  \\
& Untagged 4 & 121.\phantom{0}  &  \textbf{ 93.1\%} & 3.6\% & 2.0\% & 1.1\% & 0.2\% & 2.62 & 2.14 & 1400.\phantom{0}  \\
\cline{2-11}
& VBF dijet 0 & 4.5 & 17.8\% &  \textbf{ 81.8\%} & 0.2\% & 0.1\% & 0.1\% & 1.30 & 0.94 & 0.8 \\
& VBF dijet 1 & 5.6 & 28.5\% &  \textbf{ 70.5\%} & 0.6\% & 0.2\% & 0.2\% & 1.43 & 1.07 & 2.7 \\
& VBF dijet 2 & 13.7 & 43.8\% &  \textbf{ 53.2\%} & 1.4\% & 0.8\% & 0.8\% & 1.59 & 1.24 & 22.1 \\
\cline{2-11}
& VH tight $\ell$ & 1.4 & 0.2\% & 0.2\% &  \textbf{ 76.9\%} & 19.0\% & 3.7\% & 1.63 & 1.24 & 0.4 \\
& VH loose $\ell$ & 0.9 & 2.6\% & 1.1\% &  \textbf{ 77.9\%} & 16.8\% & 1.5\% & 1.60 & 1.16 & 1.2 \\
& VH \MET & 1.8 & 16.3\% & 2.7\% & 34.4\% &  \textbf{ 35.4\%} & 11.1\% & 1.68 & 1.17 & 1.3 \\
& VH dijet & 1.6 & 30.3\% & 3.1\% &  \textbf{ 40.6\%} & 23.4\% & 2.6\% & 1.31 & 1.06 & 1.0 \\
\cline{2-11}
& \ttH lepton & 0.5 & ---~~ & ---~~ & 1.6\% & 1.6\% &  \textbf{ 96.8\%} & 1.34 & 1.03 & 0.2 \\
& \ttH multijet & 0.6 & 4.1\% & 0.9\% & 0.8\% & 0.9\% &  \textbf{ 93.3\%} & 1.34 & 1.03 & 0.6 \\
\hline
\end{tabular}

\label{tab:ClassFracs}
\end{table*}

\section{Statistical methodology}
\label{sec:bmodelling}

To extract a result or measurement a simultaneous binned maximum-likelihood fit
to the diphoton invariant mass distributions in all the event classes is performed
over the range $100<\mgg<180\GeV$.
Binned fits are used for speed of computation, and the bin size chosen, 250\MeV, is sufficiently small
compared to the mass resolution that no information is lost.
It has been verified that a binned fit with this bin size gives the same result as an unbinned fit.
The signal model is derived from MC simulation after applying the corrections determined from
data/MC comparisons of $\Zee$ and $\Zmmg$ events, as described in the previous section.
The background is evaluated by fitting the $\mgg$ distribution in data, without reference to the MC simulation.
Thus the likelihood to be evaluated in a signal-plus-background fit is

\begin{equation}
\Lk=\Lk(\text{data}|s(p,\mgg)+f(\mgg)),
\end{equation}

where $p$ comprises those parameters of the signal, such as $\mH$ or the signal strength, that are allowed to vary
in the fit, $s(p,\mgg)$ is the parametric signal model, and $f(\mgg)$ the background fit function.

The chosen test statistic, used to determine how signal- or background-like the data are,
is based on the profile likelihood ratio.
Systematic uncertainties are incorporated into the analysis via
nuisance parameters and treated according to the frequentist paradigm.
A description of the general methodology can be found in Refs.~\cite{LHC-HCG-Report, Chatrchyan:2012tx}.
Unless stated otherwise, the results presented here are obtained using asymptotic formulae~\cite{Cowan:2010st},
including updates introduced in the \textsc{RooStats} package~\cite{RooStats}.

It is important that the choice of background fit function does not
bias the estimate of background obtained from the fit for any signal mass hypothesis, $\mH$, in
the range of the search.

A change has been made with respect to the method used to obtain
previous results, which is described in Ref.~\cite{Chatrchyan:2013lba}.
Previously, a single fit function was chosen for each class after a study of the
potential bias on the estimated background.
The potential bias using the chosen function was required to be negligible.
The number of degrees of freedom of the fit was increased until the
bias became at least five times smaller than the statistical uncertainty in the number of fitted events in a mass window
corresponding to the full width at half maximum of
the corresponding signal model, for any mass in the range $110\leq\mH\leq150\GeV$.

For the results reported in this paper a method, the discrete profiling method, has been developed~\cite{discrete-profiling} to
treat the uncertainty associated with the choice of the
function used to fit the background, in a similar way to systematic
uncertainties associated with the measurements.
The choice of the function used to fit the background, in any
particular event class, is included as a discrete nuisance parameter in the likelihood
function used to extract the result.
All reasonable families of functions should be considered,
although in practice it is found that the choice needs to be
made between functions in the same families as were previously considered:
exponentials, power-law functions, polynomials in the Bernstein basis, and Laurent series.
When performing either a background-only fit, or a signal-plus-background fit, by minimizing the value of twice the
negative logarithm of the likelihood all functions in these families are tried, with a penalty term
added to account for the number of free parameters in the fitting function.

The penalized likelihood function, $\Lt_f$, for a single fixed background fitting function, $f$, is defined as
\begin{equation}
-2\ln\Lt_f=-2\ln\Lk_f+kN_{f},
\end{equation}

where $\Lk_f$ is the unpenalized likelihood function, $N_{f}$ is the number of free parameters in $f$, and $k$ is a
constant.
When measuring a quantity, $p$, the likelihood ratio, $q(p)$, is used:

\begin{equation}
q(p)=-2\ln\frac{\Lt(\text{data}|p,\hat{\theta}_p,\hat{f}_p)}{\Lt(\text{data}|\hat{p},\hat{\theta},\hat{f})},
\label{eq:q}
\end{equation}

where the numerator represents the maximum of $\Lt$ given $p$, achieved for the
best-fit values of the nuisance parameters, $\theta=\hat{\theta}_p$, and a particular background function, $f=\hat{f}_p$.
The denominator corresponds to the global maximum of $\Lt$, where $p=\hat{p}$, $\theta=\hat{\theta}$, and $f=\hat{f}$.
Choosing the functional form of the background that maximizes $\Lt$ for any particular
value of $p$ yields confidence intervals on $p$ that  can only be wider than those
obtained using the single fixed functional form from the global best fit, $f=\hat{f}$.

Two values of $k$, which sets the magnitude of the penalty for increasing the number of free parameters in the fit,
have been tested in detail.
The values of $k=1$ and $k=2$ can be justified, respectively, by the $\chi^{2}\ p$-value and the
Akaike information criterion~\cite{AIC}.
It is found in tests made with pseudo-experiments that with a value of $k=1$
the method gives consistently good coverage and negligible bias.

In order to test coverage and bias we generate pseudo-data.
To do that we need first to fit the data, thus facing a problem similar to, but not to be confused with,
the original problem of choosing the background fit function to model the background in the analysis.
The method used to generate pseudo-data is as follows.
For each event class in turn, functions from each of the families used in the discrete profiling method, and listed above, are fit to the data.
In each family, the number of degrees of freedom (number of exponentials, number of terms in the series, degree of the polynomial, etc.)
is increased until the $\chi^2$ between N+1 degrees of freedom and N degrees of freedom for the fit to data shows no significant improvement
($p\text{-value}<0.05$ obtained from the F-distribution~\cite{F-dist}).
At that point the function with N degrees of freedom is retained as representative of that family of functions.
For each event class, the fits to the data with the retained representative functions for that class, are used to generate
pseudo-background distributions.

The discrete profiling method is applied to pseudo-experiments in which
signals having a range of strengths, from half to twice that of the SM, are added to the pseudo-background.
The tests have demonstrated that the discrete profiling method provides good coverage
of the uncertainty associated with the choice of the function, for all
the functions considered as generators of background, and provides an estimate of the
signal strength with negligible bias.
The criterion used for this is similar and approximately equivalent to that used previously~\cite{Chatrchyan:2013lba},
the median of the distribution of the pull on the signal strength,
$(\mu_\text{meas.}-\mu_\text{true})/\sigma_{\mu_\text{meas.}}$, should be less than 0.14.
This value is chosen because satisfaction of this criterion ensures that any underestimation of the uncertainty in the
signal strength is less than 1\%.

\begin{figure*}[htbp]
   \centering
     \includegraphics[width=\cmsFigWidthOne]{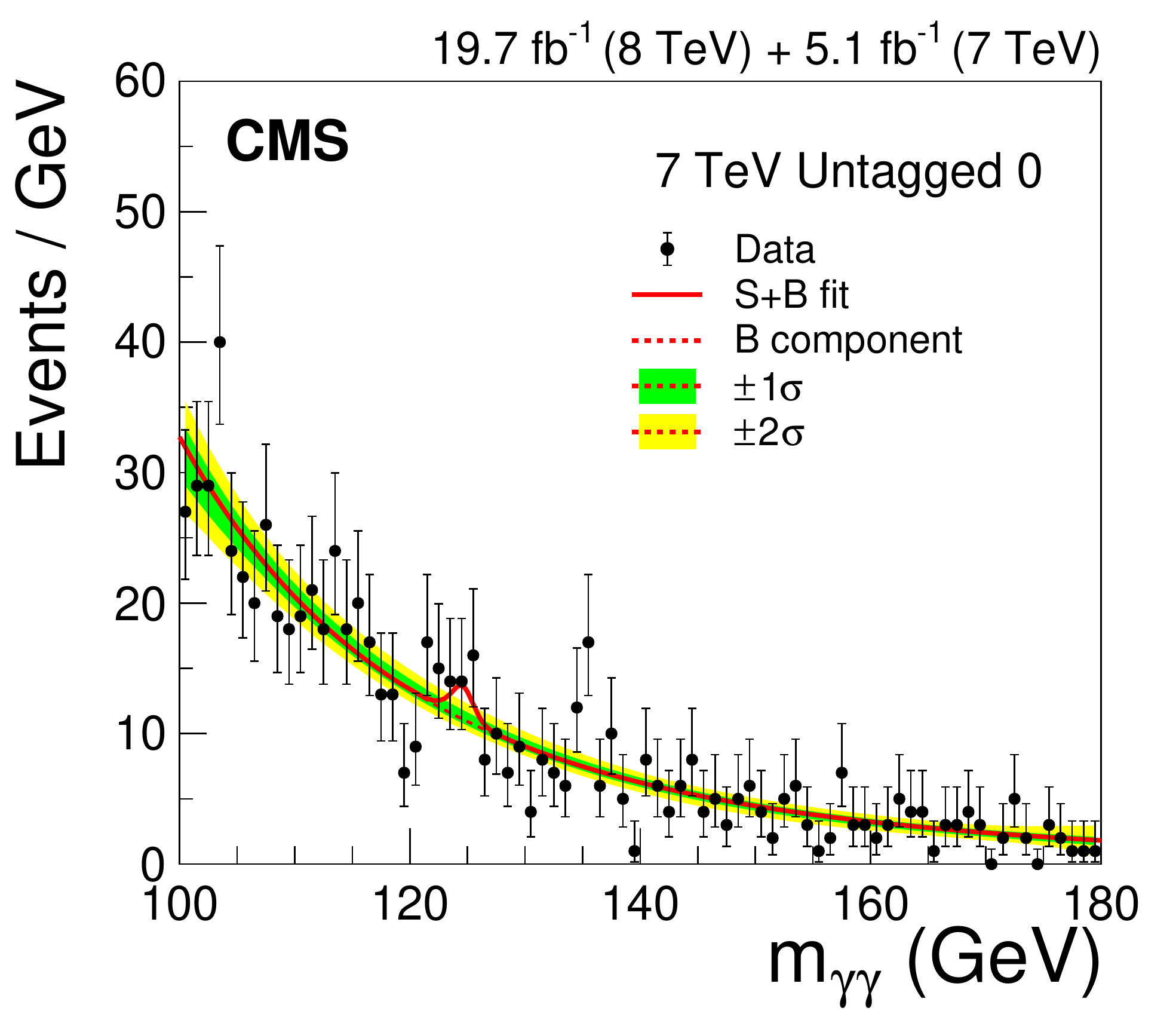}
     \includegraphics[width=\cmsFigWidthOne]{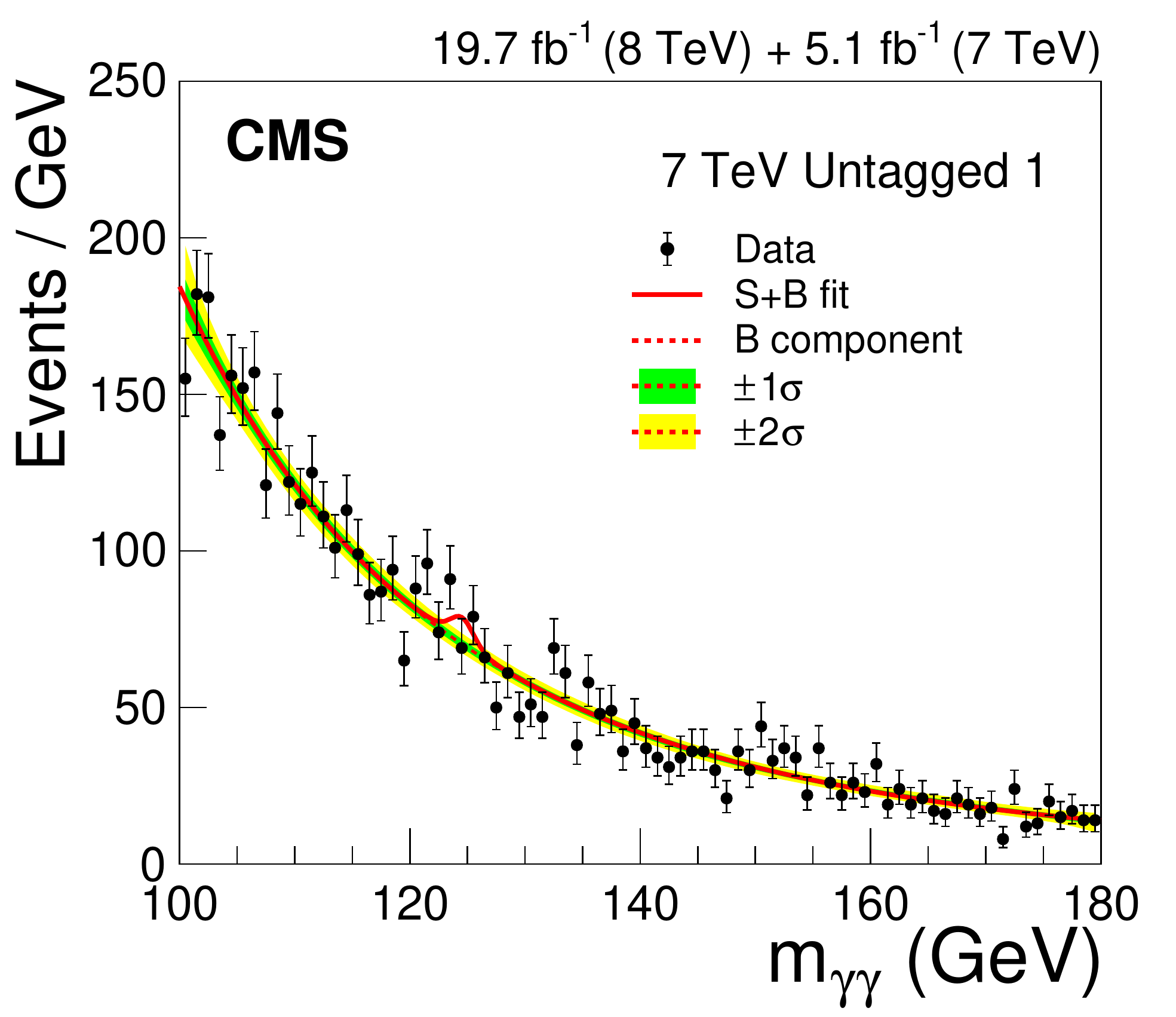}\\
     \includegraphics[width=\cmsFigWidthOne]{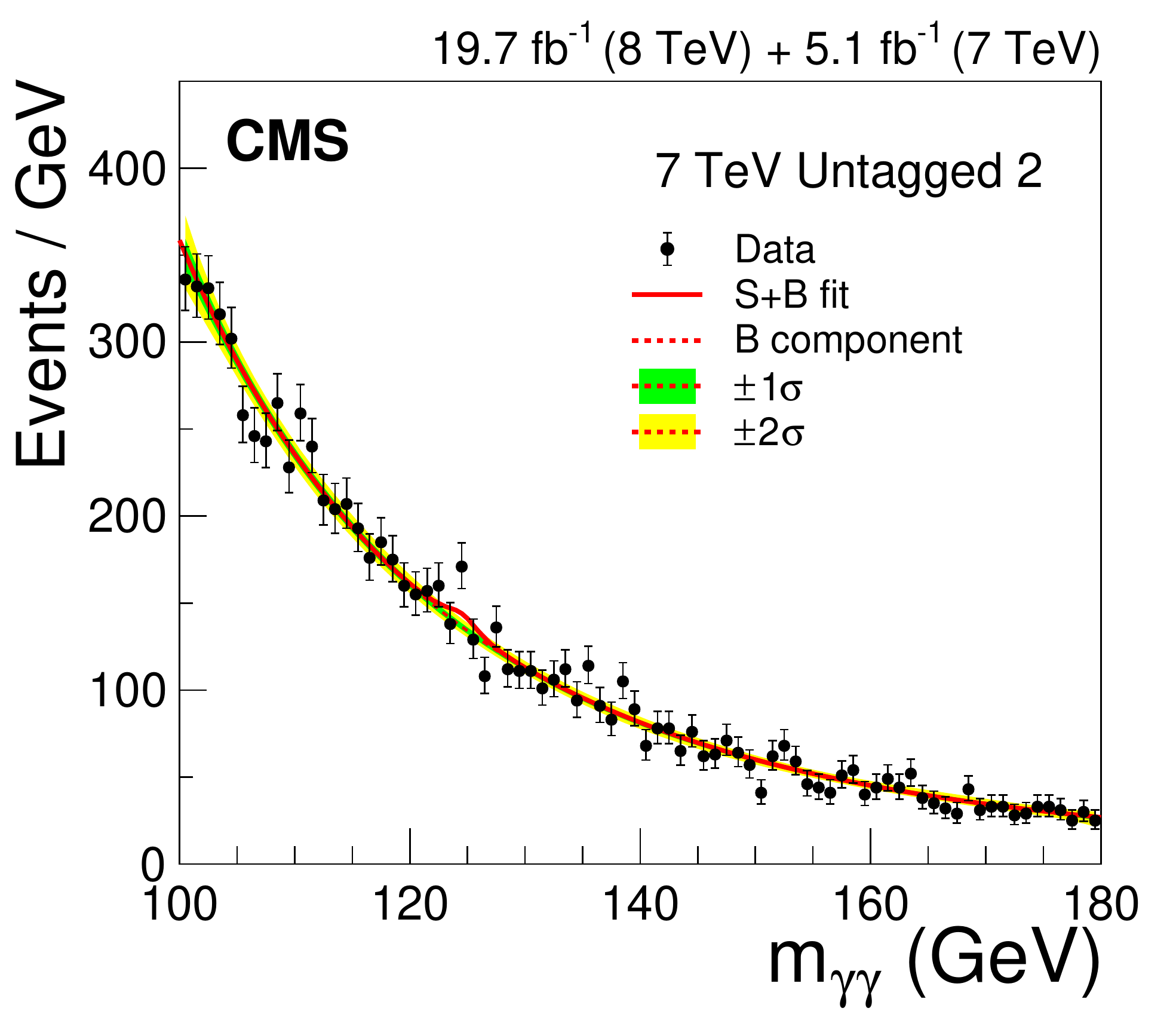}
     \includegraphics[width=\cmsFigWidthOne]{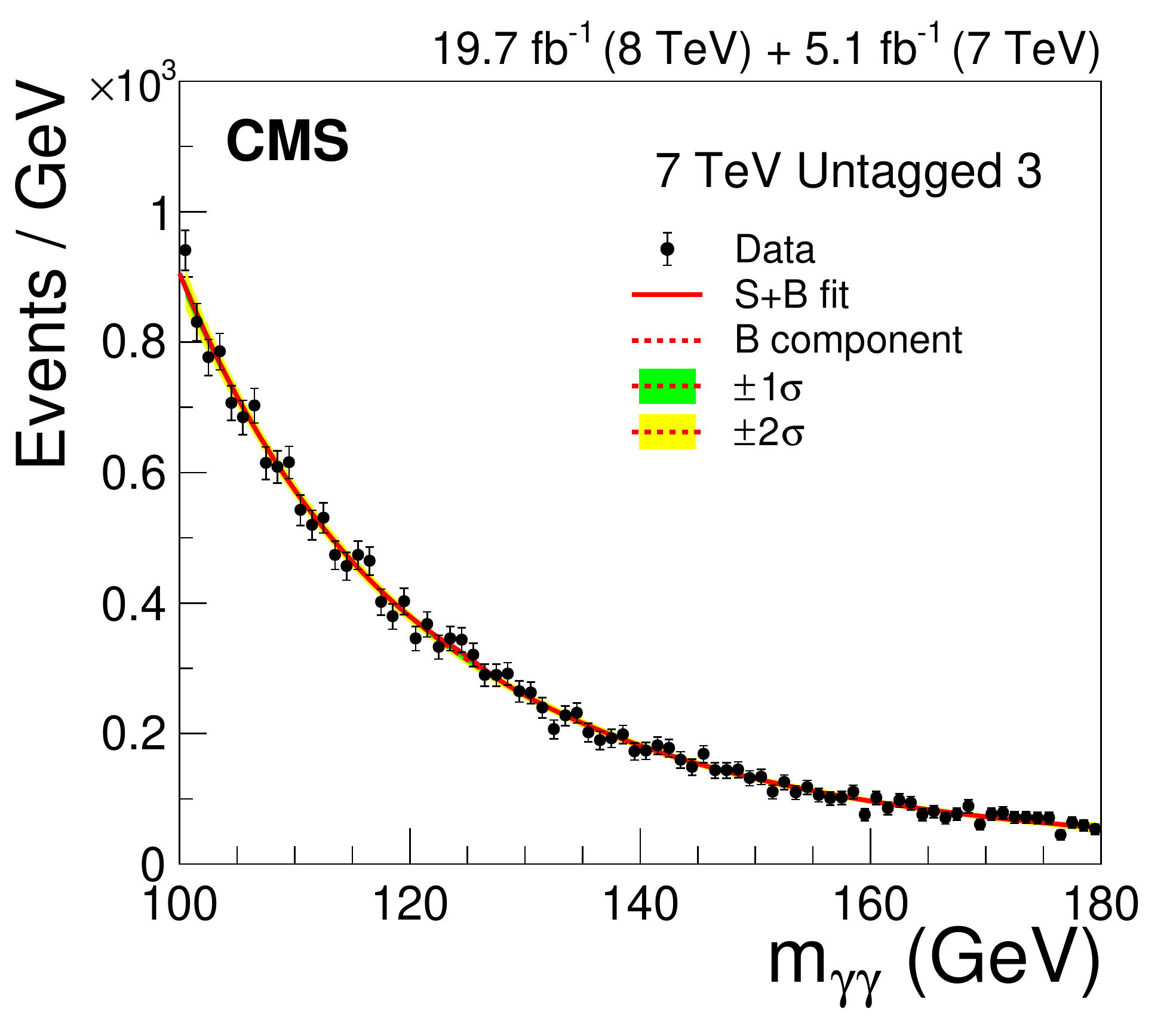}\\
     \caption{\label{fig:mgg-7-untagged} Events in the four untagged
              classes of the 7\TeV dataset, binned as a function of $\mgg$,
       together with the result of a fit of the signal-plus-background model.
The $1\sigma$ and $2\sigma$ uncertainty bands shown for the background component of the fit include the uncertainty due to the choice of function and the uncertainty in the fitted parameters. These bands do not contain the Poisson uncertainty that must be included when the full uncertainty in the number of background events in any given mass range is estimated.
}

\end{figure*}

\begin{figure*}[htbp]
   \centering
      \includegraphics[width=\cmsFigWidthOne]{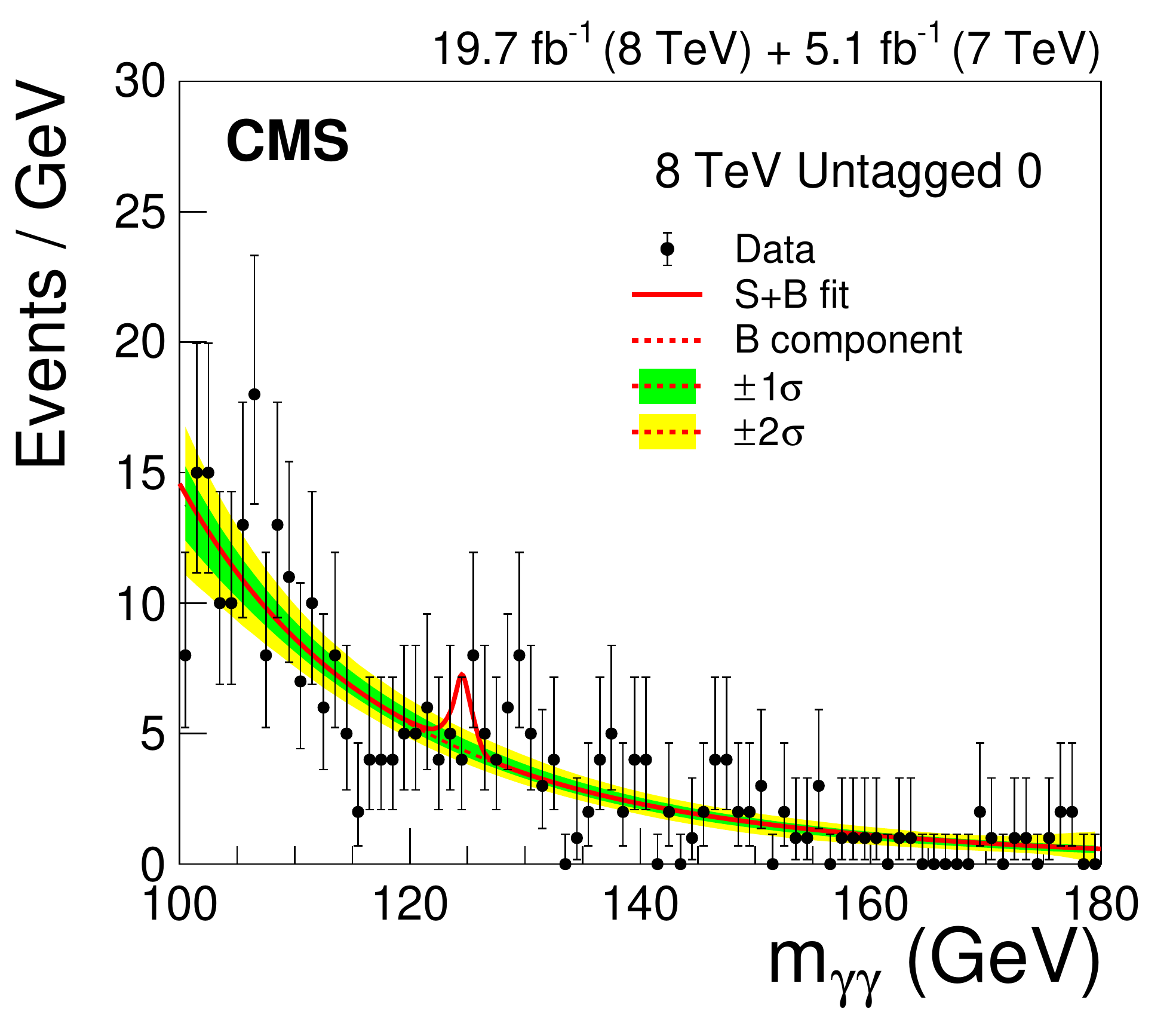}
      \includegraphics[width=\cmsFigWidthOne]{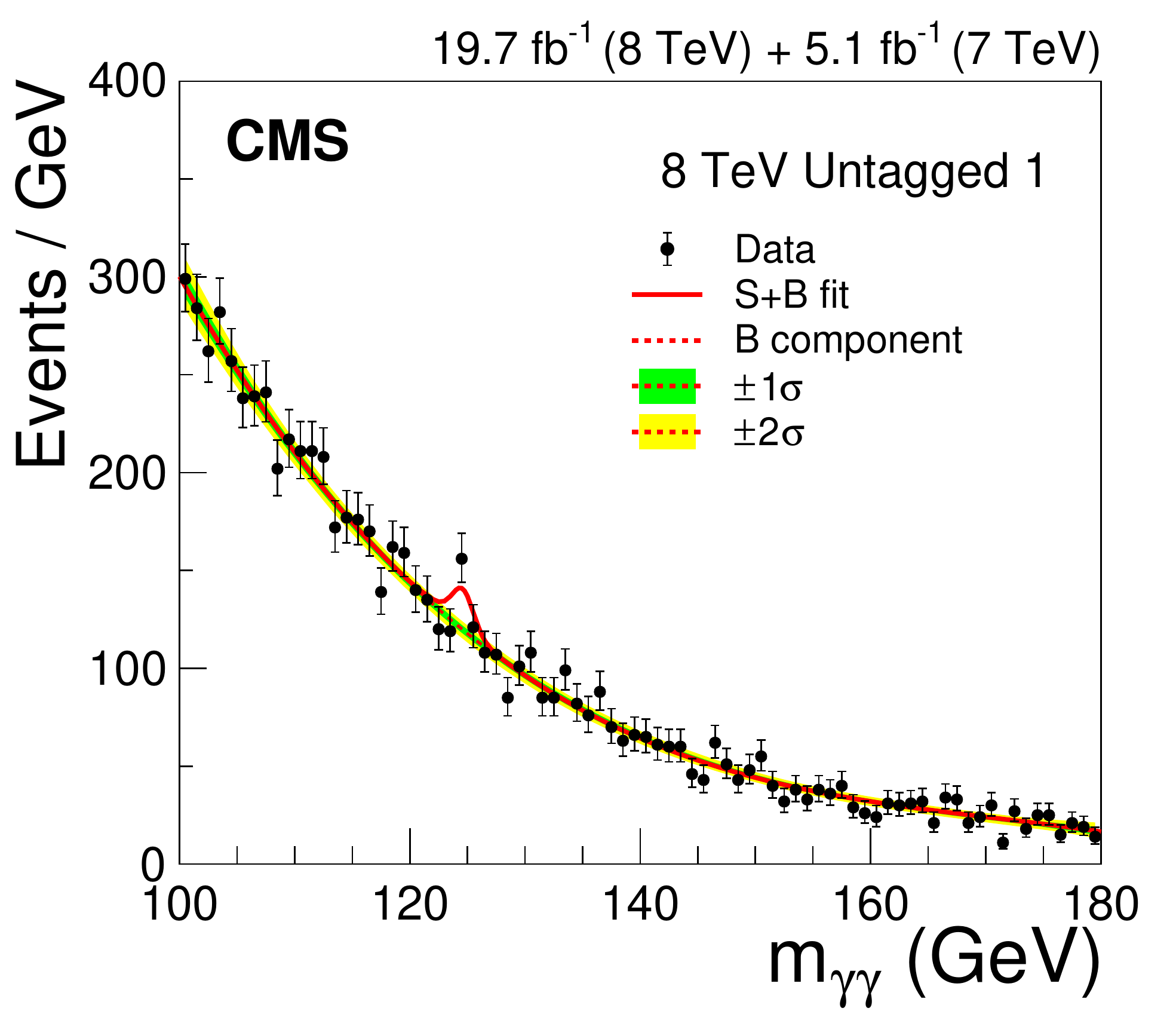}\\
      \includegraphics[width=\cmsFigWidthOne]{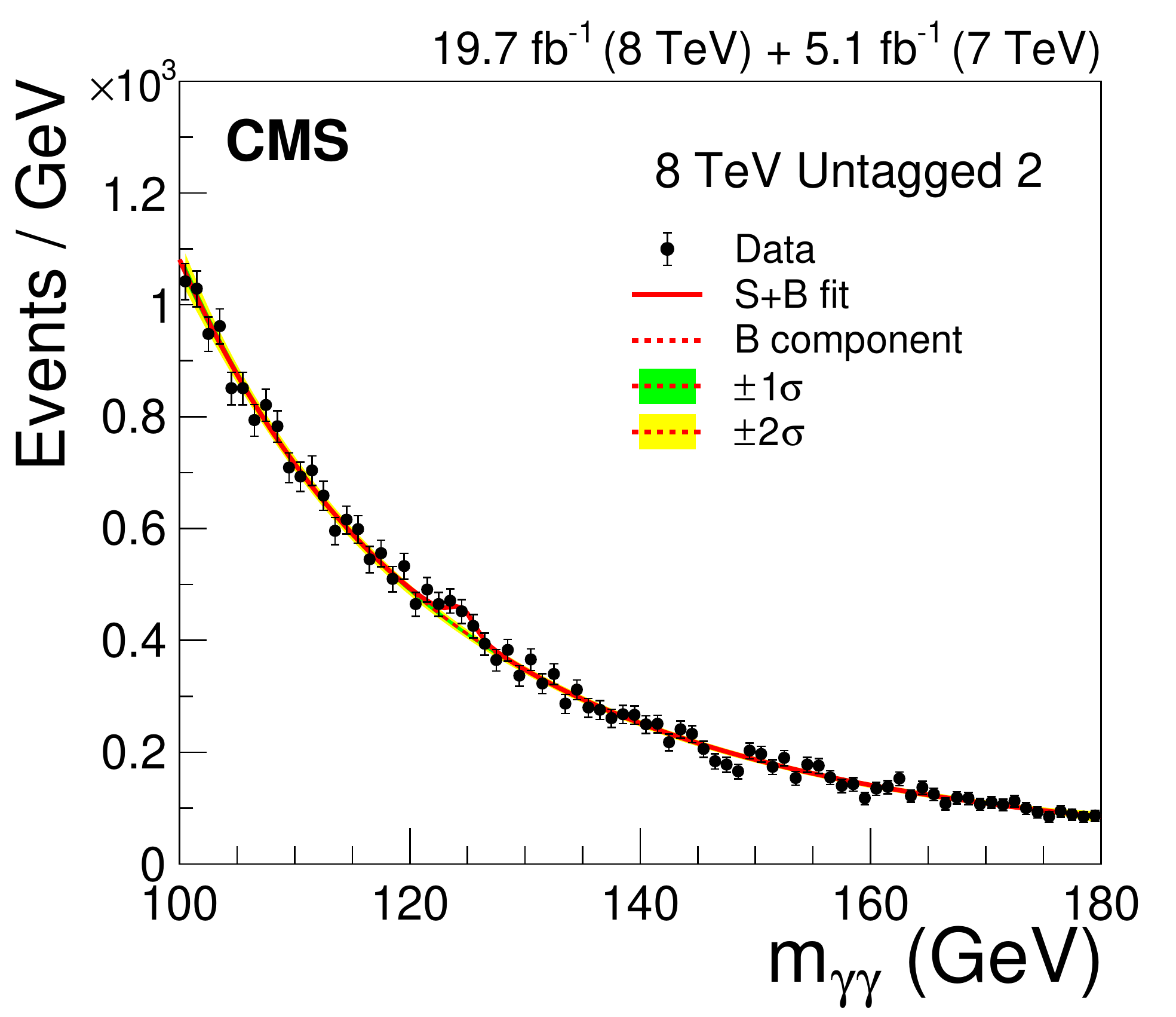}
      \includegraphics[width=\cmsFigWidthOne]{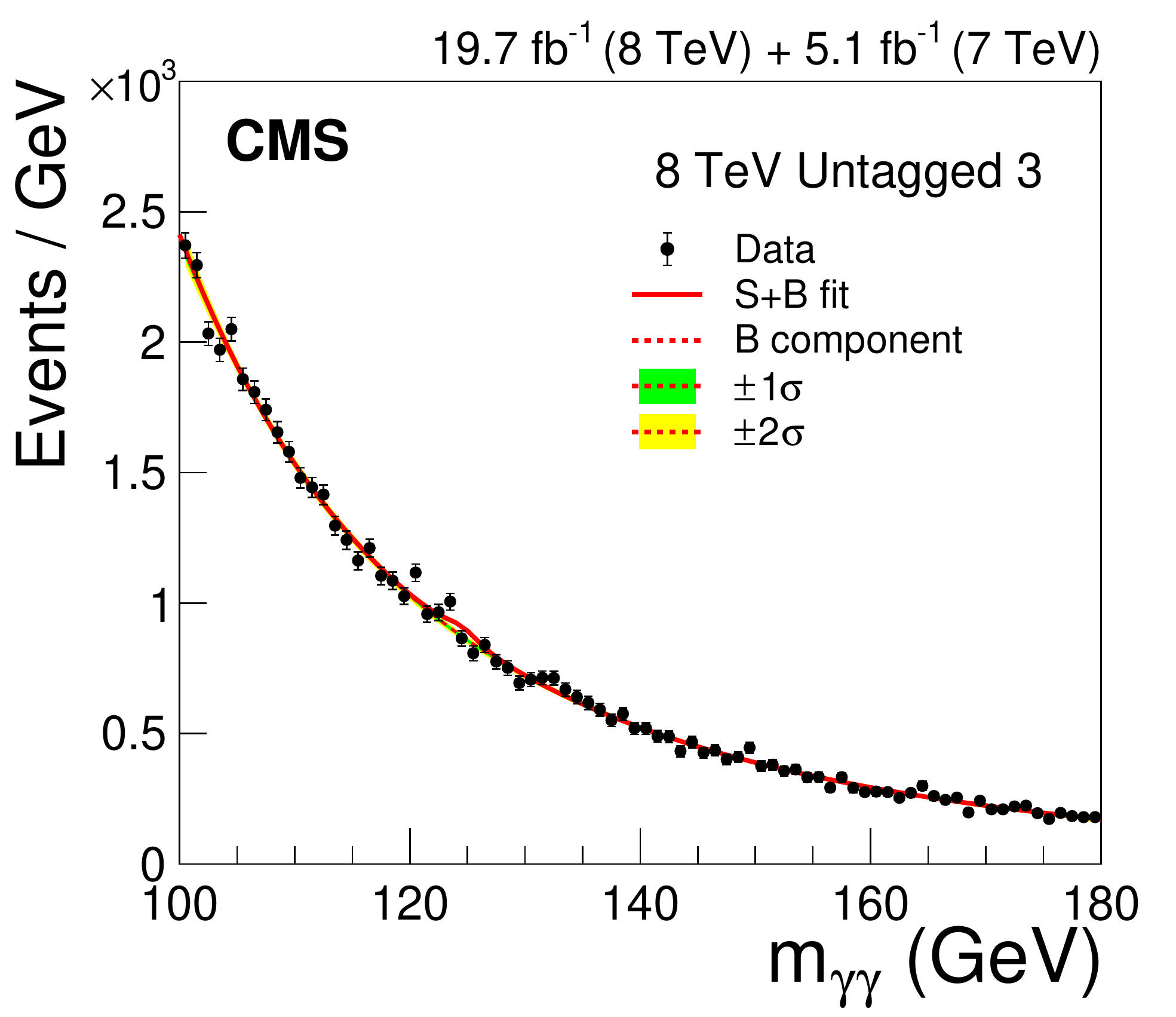}\\
      \includegraphics[width=\cmsFigWidthOne]{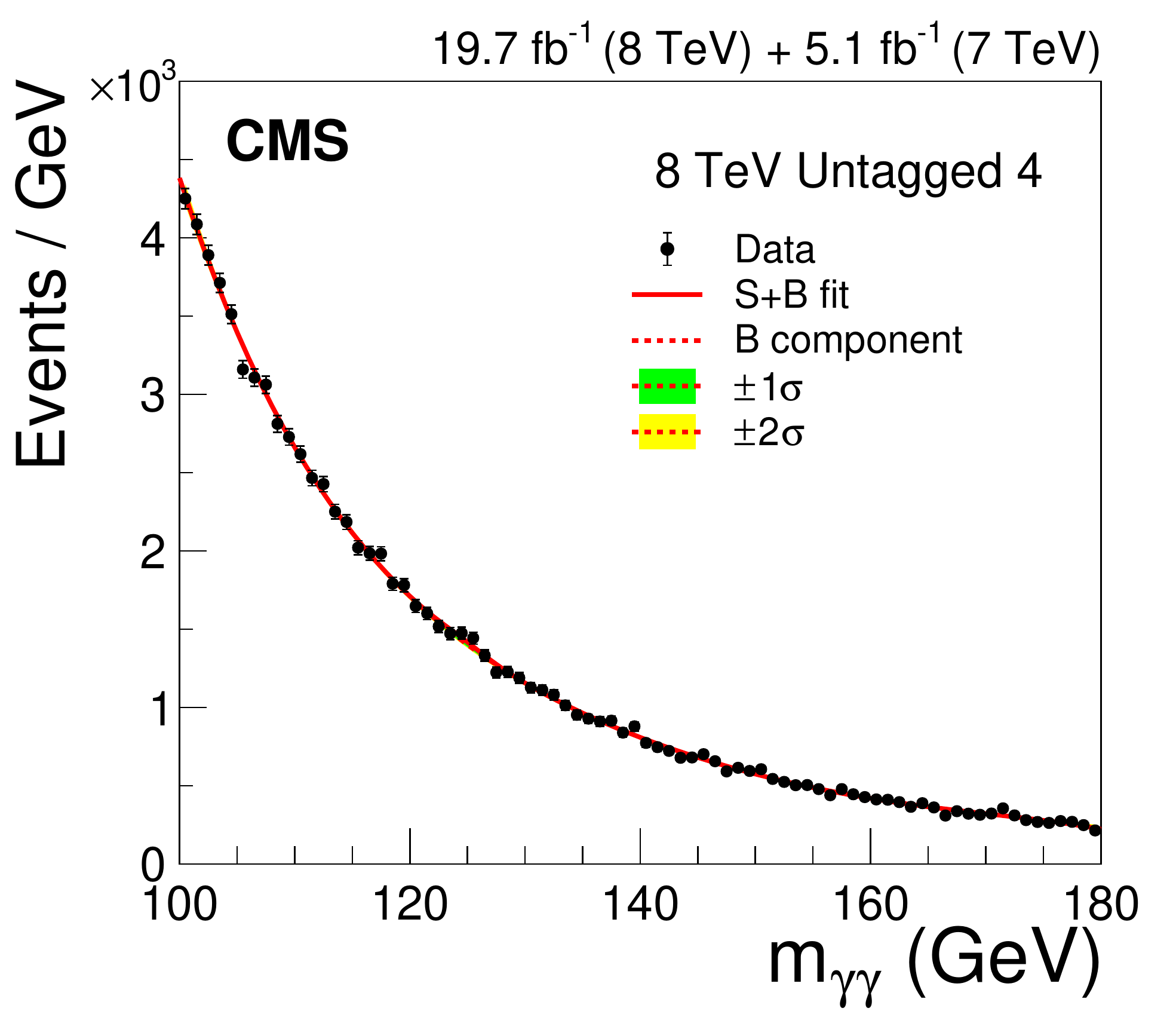}
     \caption{\label{fig:mgg-8-untagged} Events in the five untagged
              classes of the 8\TeV dataset, binned as a function of $\mgg$,
       together with the result of a fit of the signal-plus-background model.
The $1\sigma$ and $2\sigma$ uncertainty bands shown for the background component of the fit include the uncertainty due to the choice of function and the uncertainty in the fitted parameters. These bands do not contain the Poisson uncertainty that must be included when the full uncertainty in the number of background events in any given mass range is estimated.
}

\end{figure*}

\begin{figure*}[htb]
   \centering
     \includegraphics[width=\cmsFigWidthOne]{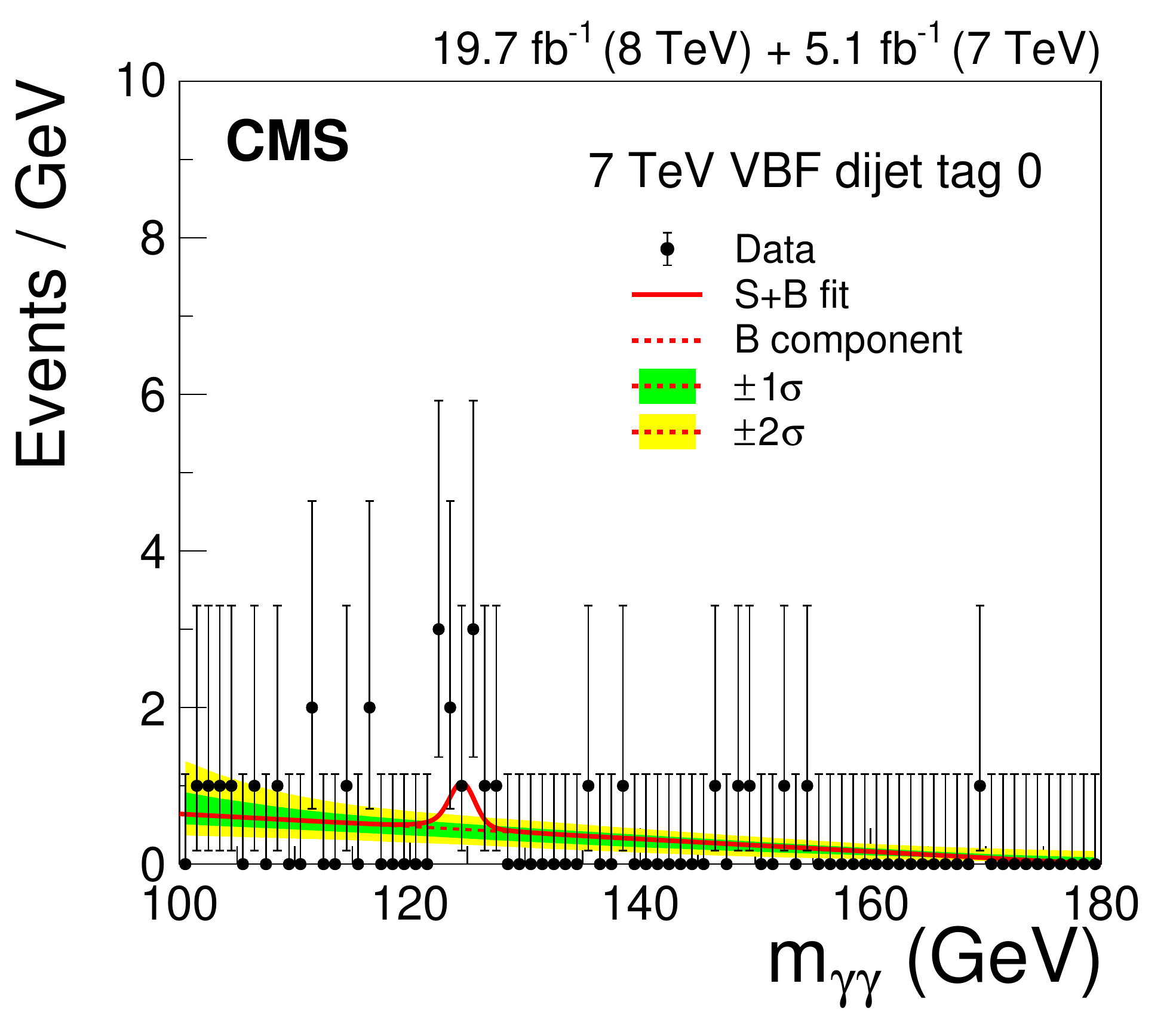}
     \includegraphics[width=\cmsFigWidthOne]{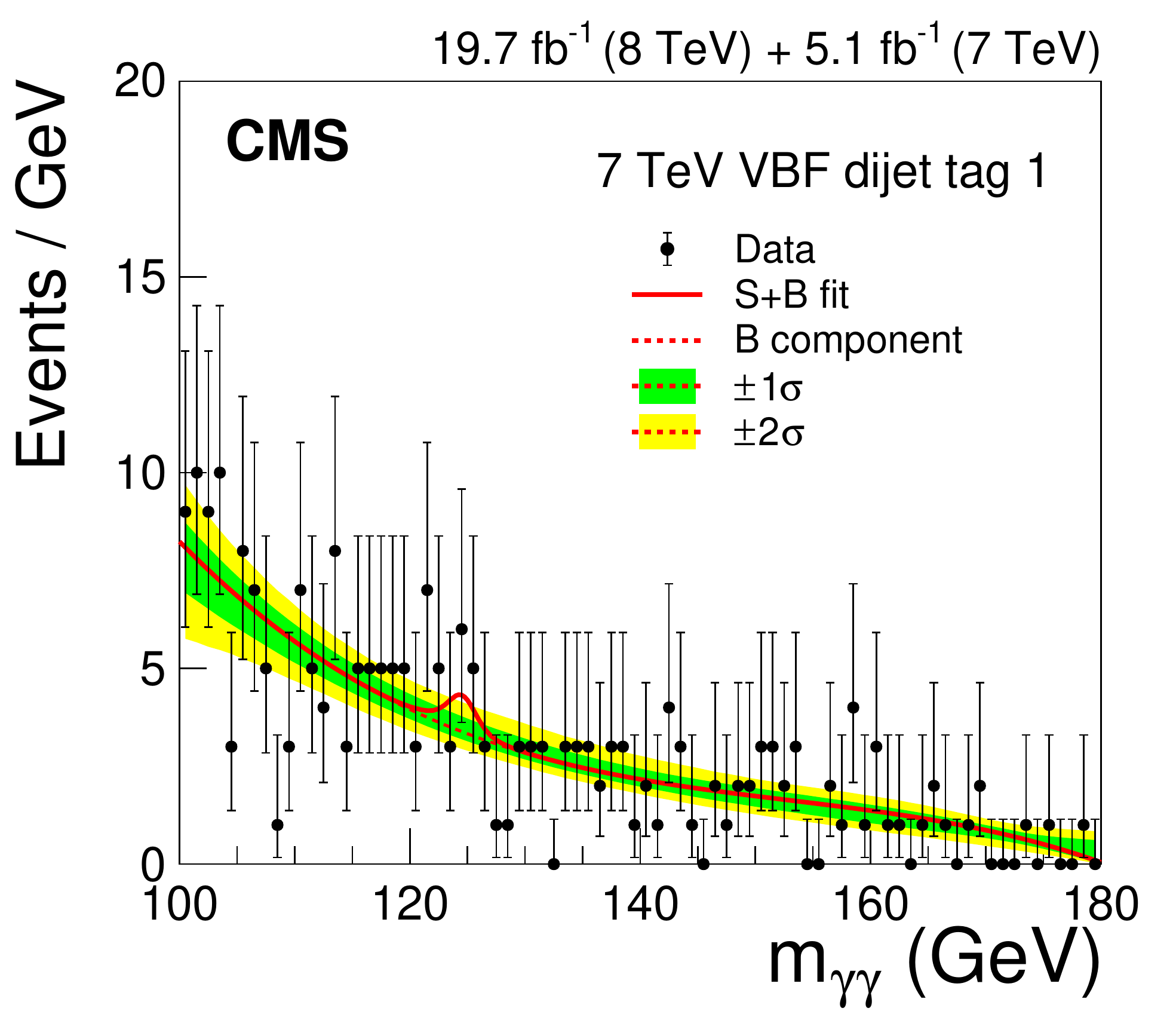}\\
     \caption{\label{fig:mgg-7-vbf} Events in the two VBF dijet-tagged
              classes of the 7\TeV dataset, binned as a function of $\mgg$,
       together with the result of a fit of the signal-plus-background model.
The $1\sigma$ and $2\sigma$ uncertainty bands shown for the background component of the fit include the uncertainty due to the choice of function and the uncertainty in the fitted parameters. These bands do not contain the Poisson uncertainty that must be included when the full uncertainty in the number of background events in any given mass range is estimated.
}
\end{figure*}

\begin{figure*}[htb]
   \centering
     \includegraphics[width=\cmsFigWidthOne]{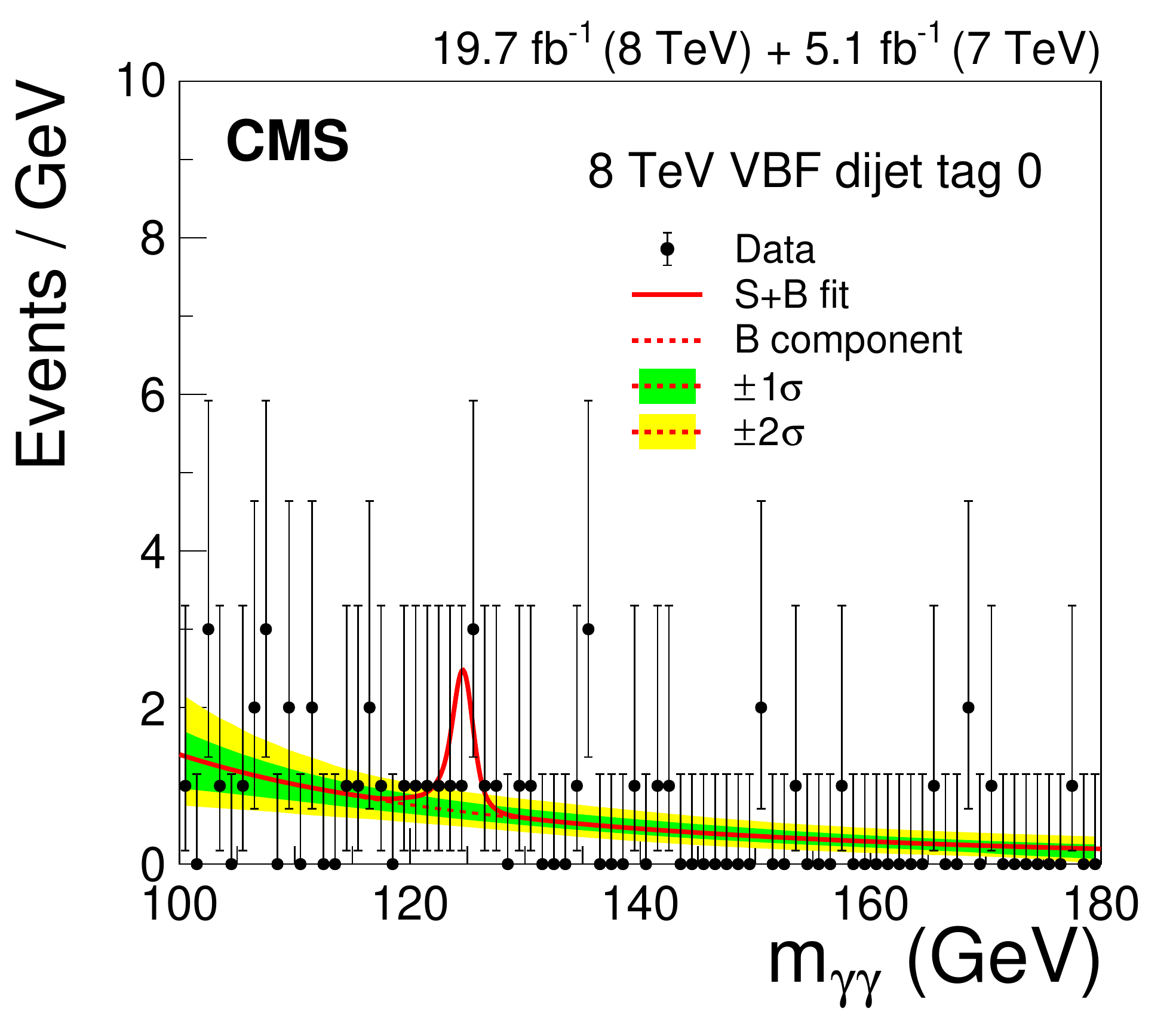}
     \includegraphics[width=\cmsFigWidthOne]{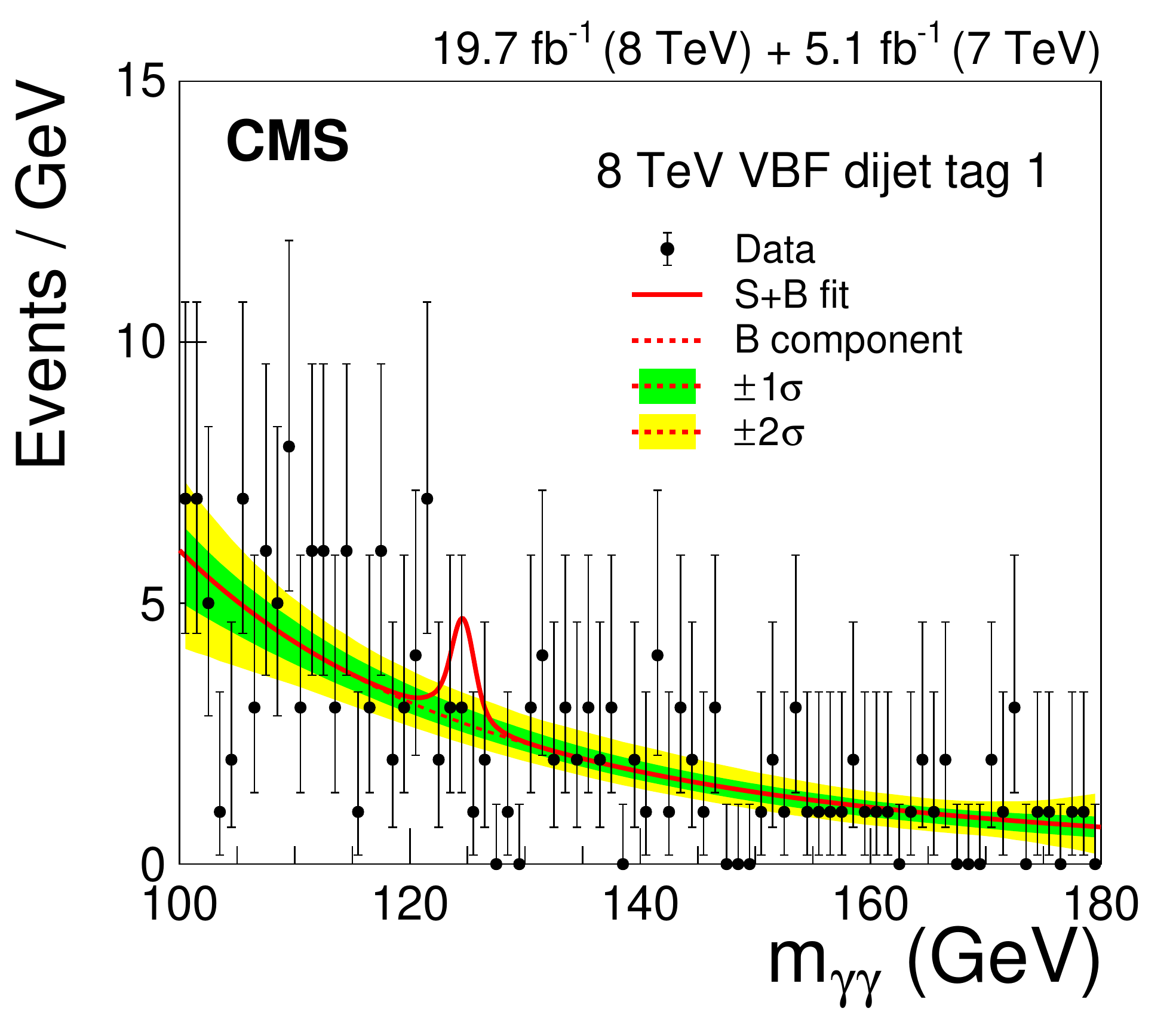}\\
     \includegraphics[width=\cmsFigWidthOne]{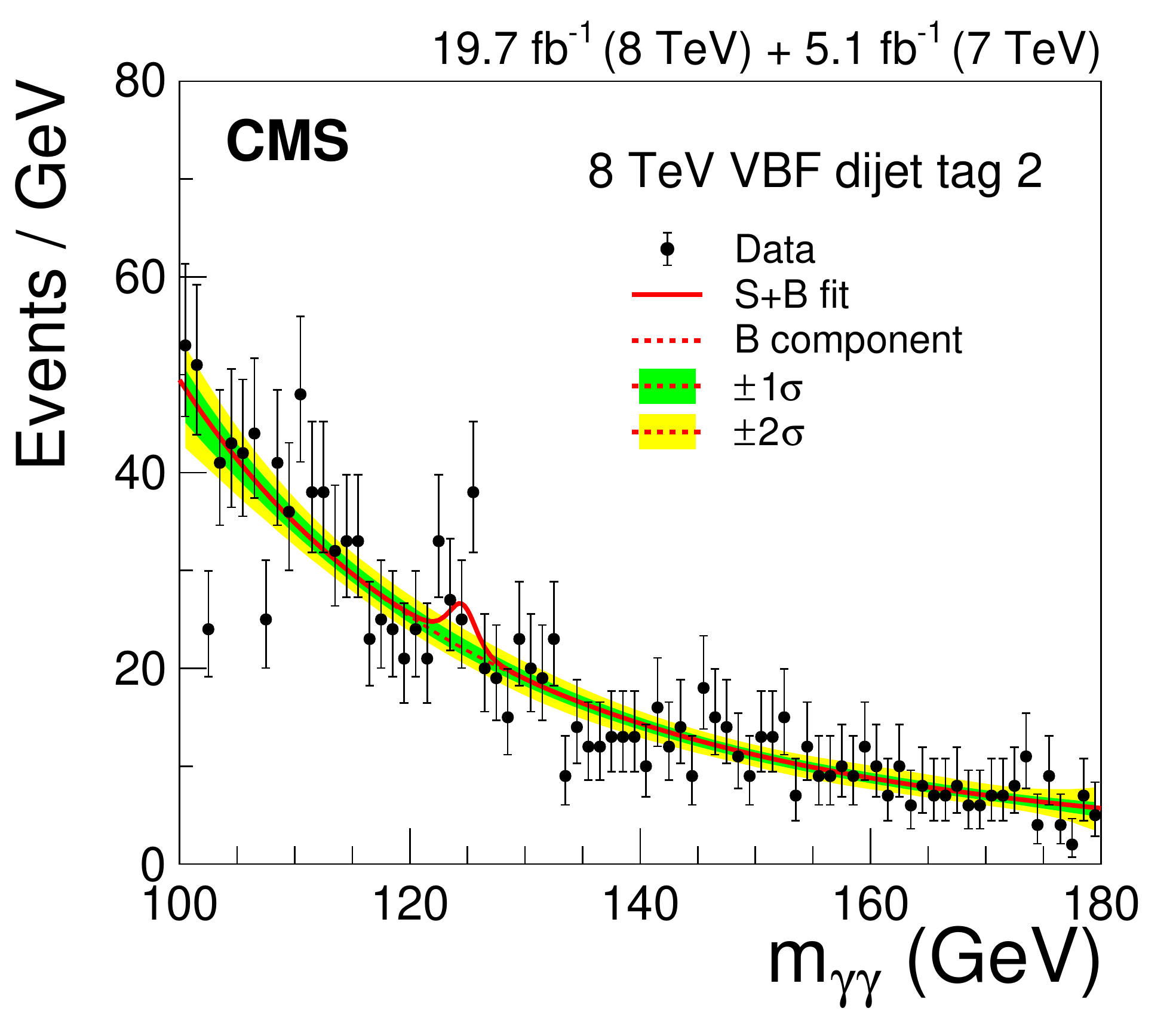}
     \caption{\label{fig:mgg-8-vbf} Events in the three VBF dijet-tagged
              classes of the 8\TeV dataset, binned as a function of $\mgg$,
       together with the result of a fit of the signal-plus-background model.
The $1\sigma$ and $2\sigma$ uncertainty bands shown for the background component of the fit include the uncertainty due to the choice of function and the uncertainty in the fitted parameters. These bands do not contain the Poisson uncertainty that must be included when the full uncertainty in the number of background events in any given mass range is estimated.
}

\end{figure*}

\begin{figure*}[htbp]
   \centering
     \includegraphics[width=\cmsFigWidthOne]{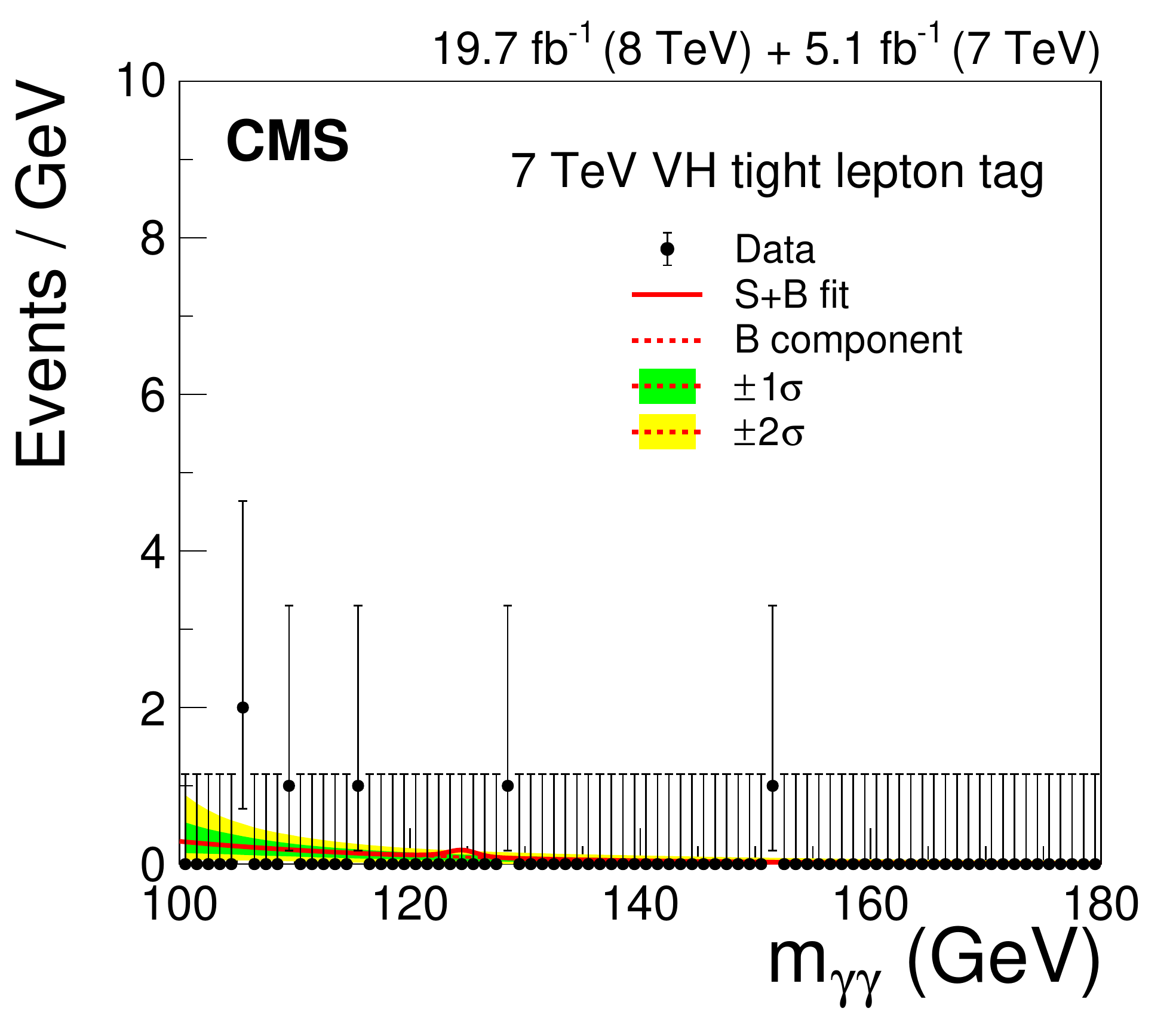}
     \includegraphics[width=\cmsFigWidthOne]{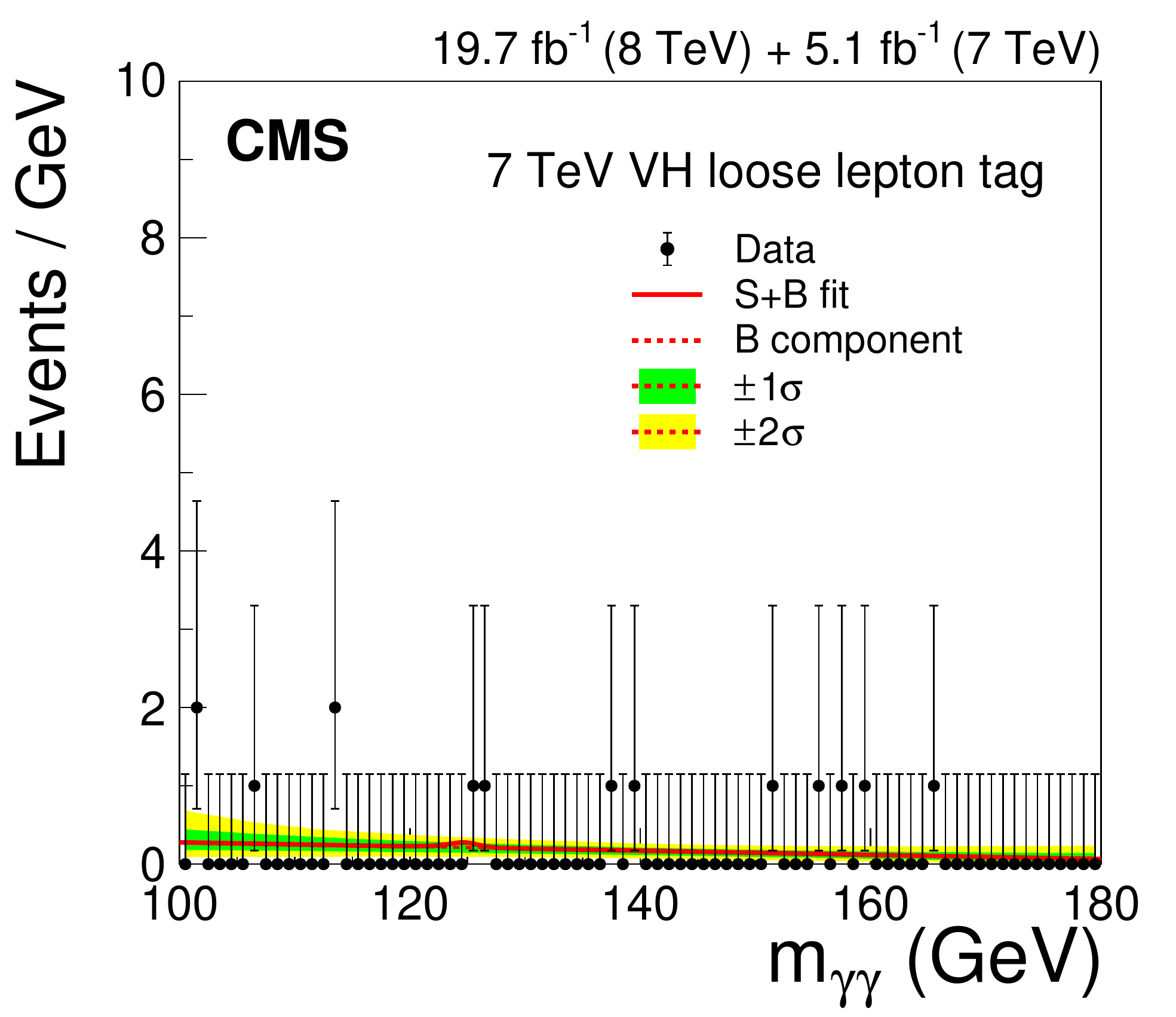}\\
     \includegraphics[width=\cmsFigWidthOne]{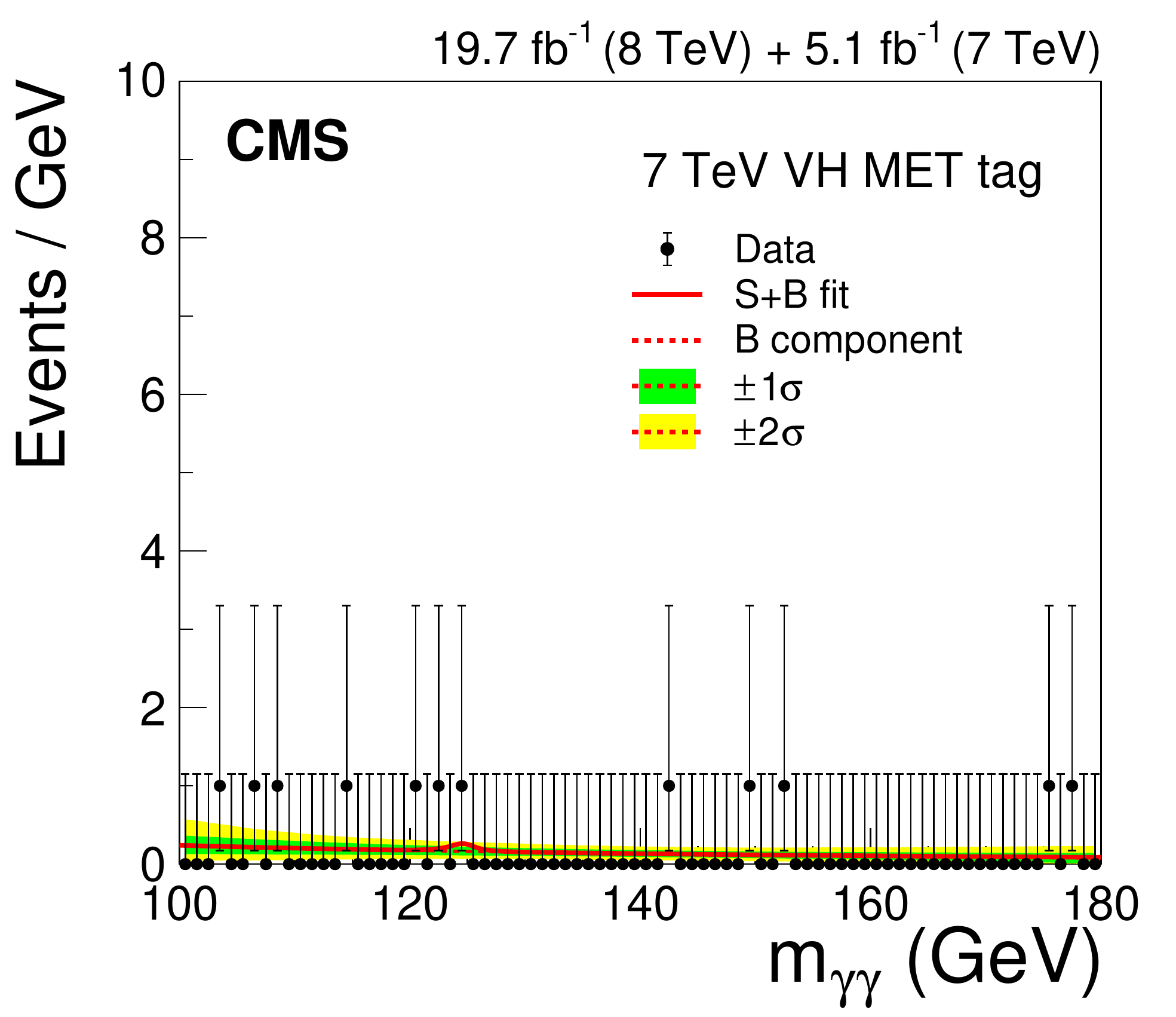}
     \includegraphics[width=\cmsFigWidthOne]{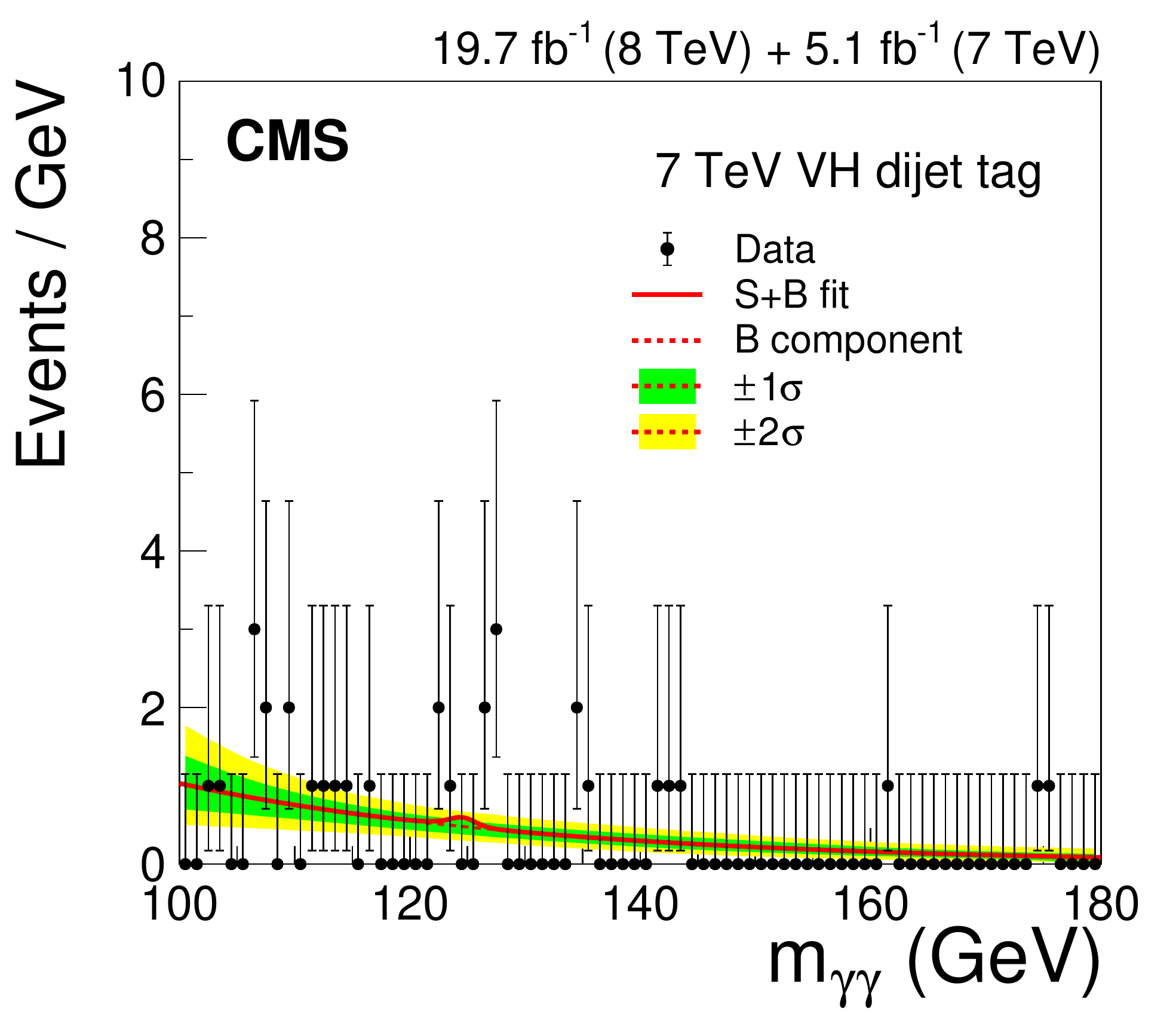}
     \caption{\label{fig:mgg-7-vh} Events in the VH-tagged
              classes of the 7\TeV dataset, binned as a function of $\mgg$,
       together with the result of a fit of the signal-plus-background model.
The $1\sigma$ and $2\sigma$ uncertainty bands shown for the background component of the fit include the uncertainty due to the choice of function and the uncertainty in the fitted parameters. These bands do not contain the Poisson uncertainty that must be included when the full uncertainty in the number of background events in any given mass range is estimated.
}
\end{figure*}

\begin{figure*}[htbp]
   \centering
     \includegraphics[width=\cmsFigWidthOne]{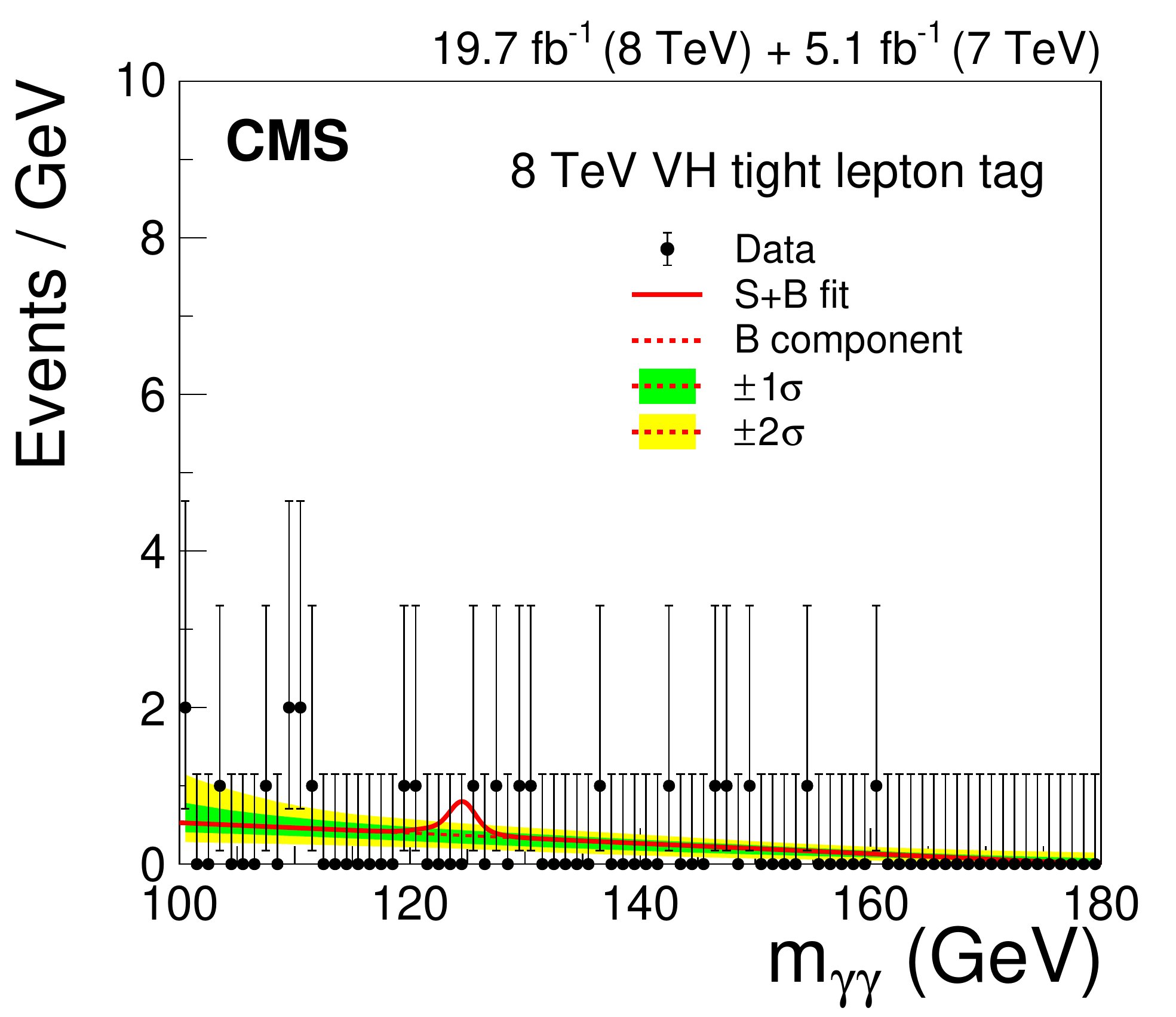}
     \includegraphics[width=\cmsFigWidthOne]{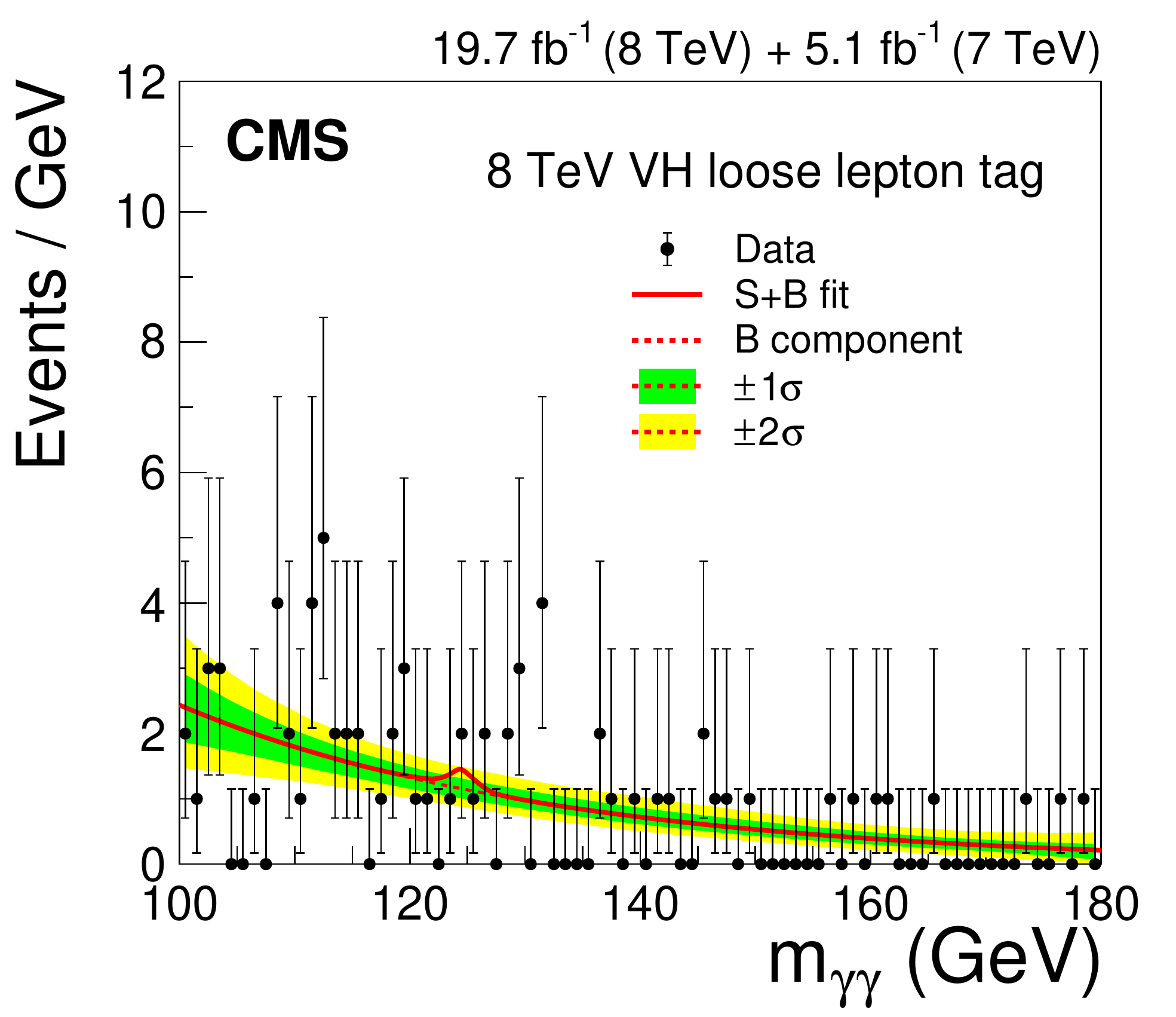}\\
     \includegraphics[width=\cmsFigWidthOne]{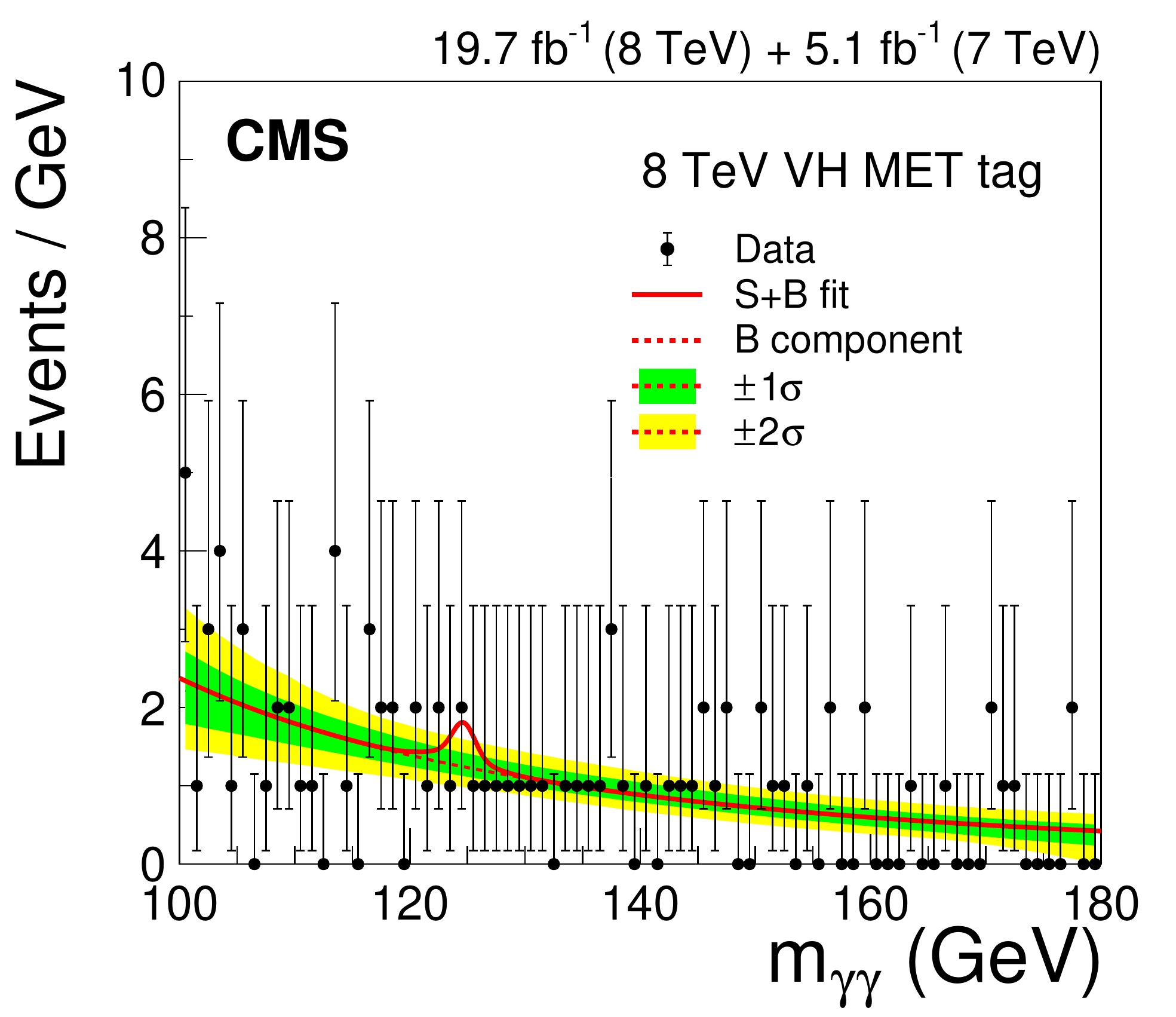}
     \includegraphics[width=\cmsFigWidthOne]{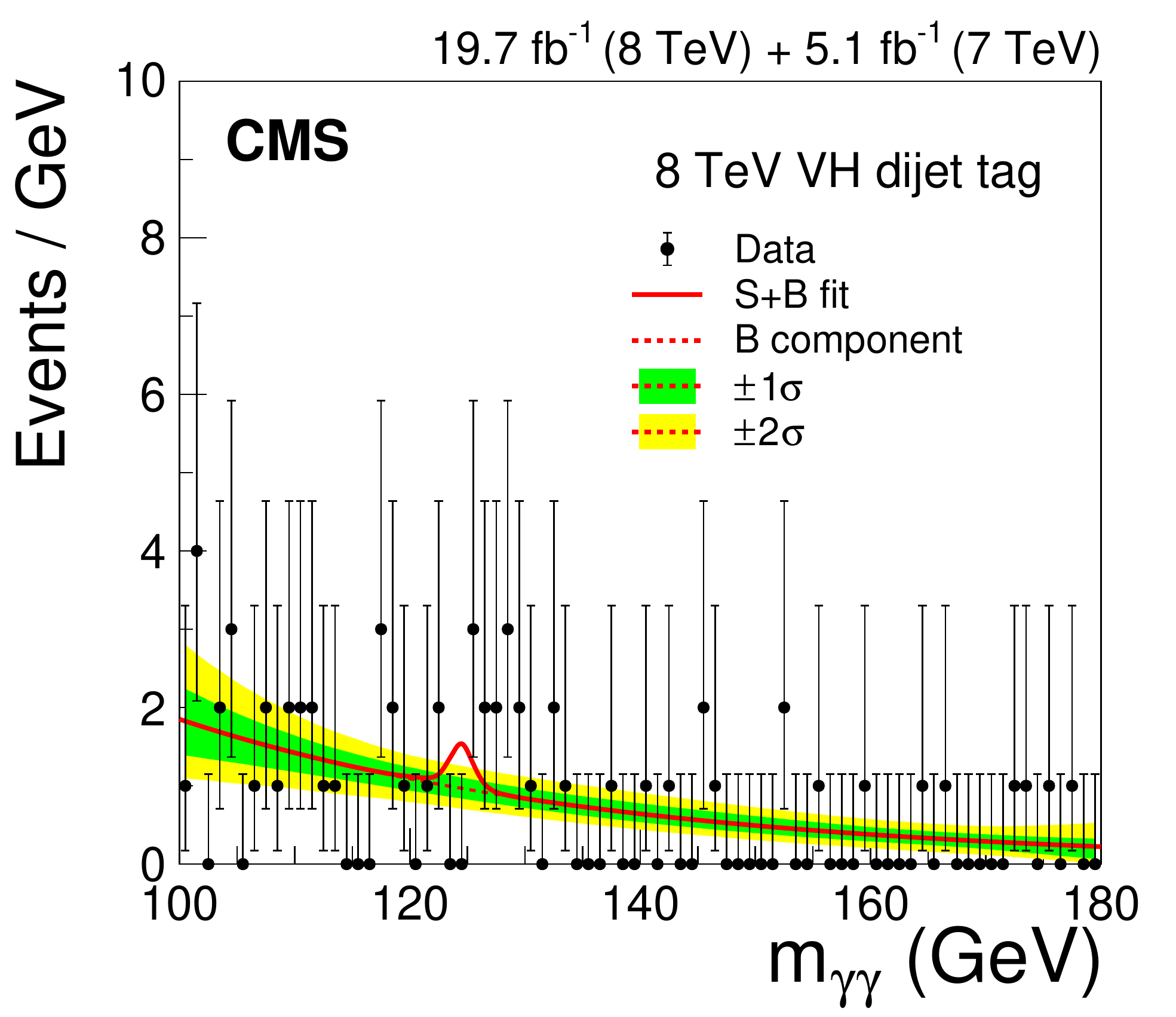}
     \caption{\label{fig:mgg-8-vh} Events in the VH-tagged
              classes of the 8\TeV dataset, binned as a function of $\mgg$,
       together with the result of a fit of the signal-plus-background model.
The $1\sigma$ and $2\sigma$ uncertainty bands shown for the background component of the fit are computed from the fit
uncertainty in the background yield in bins corresponding to those used to display the data.
These bands do not contain the Poisson uncertainty that must be
included when the full uncertainty in the number of background events in any given mass range is estimated.
}
\end{figure*}

\begin{figure}[htbp]
   \centering
     \includegraphics[width=\cmsFigWidthOne]{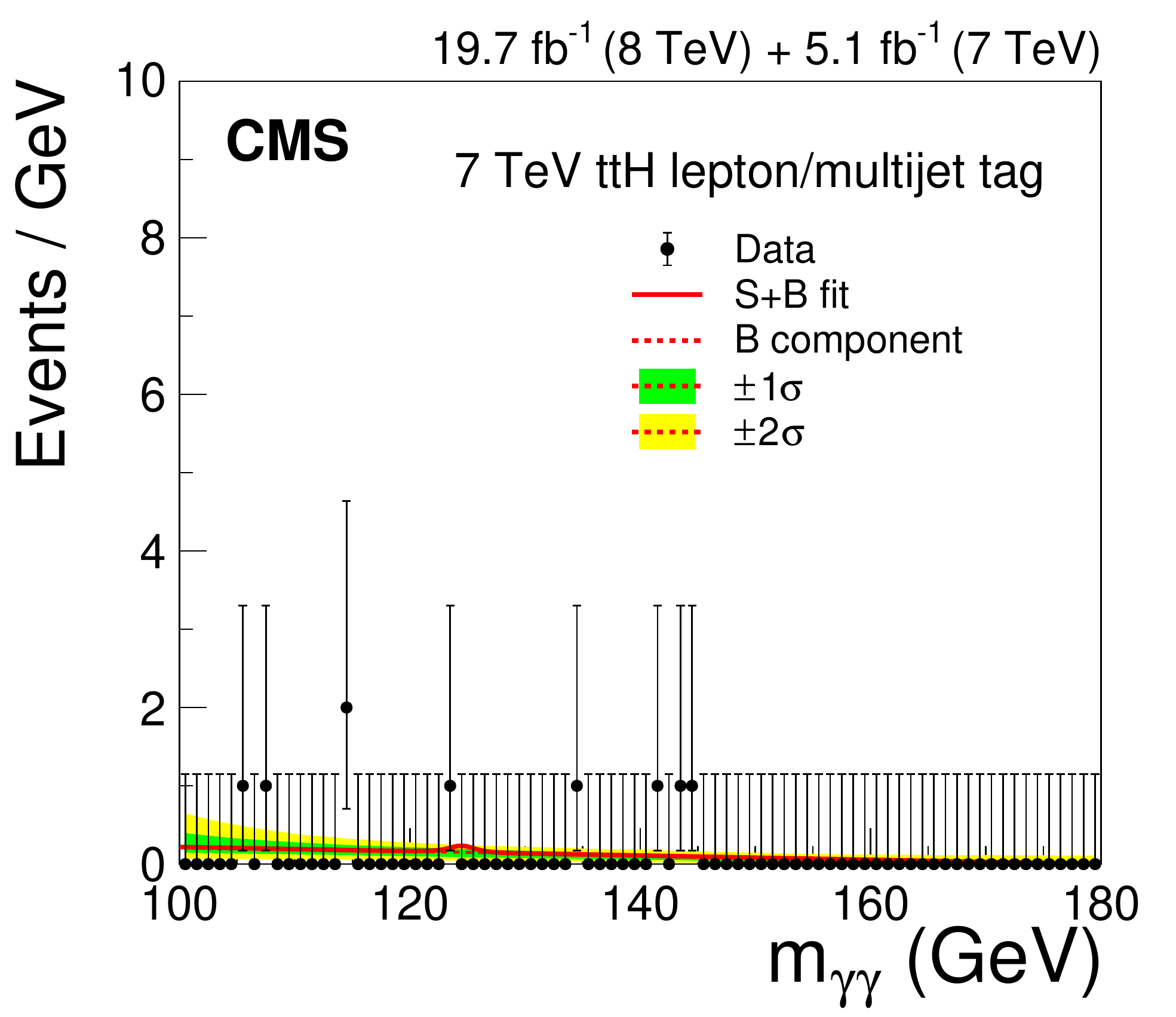}
     \caption{\label{fig:mgg-7-tth} Events in the $\ttH$-tagged
              class of the 7\TeV dataset, binned as a function of $\mgg$,
       together with the result of a fit of the signal-plus-background model for $\mH=124.7\GeV$.
The $1\sigma$ and $2\sigma$ uncertainty bands shown for the background component of the fit include the uncertainty due to the choice of function and the uncertainty in the fitted parameters. These bands do not contain the Poisson uncertainty that must be included when the full uncertainty in the number of background events in any given mass range is estimated.
}
\end{figure}

\begin{figure}[htbp]
   \centering
     \includegraphics[width=\cmsFigWidthOne]{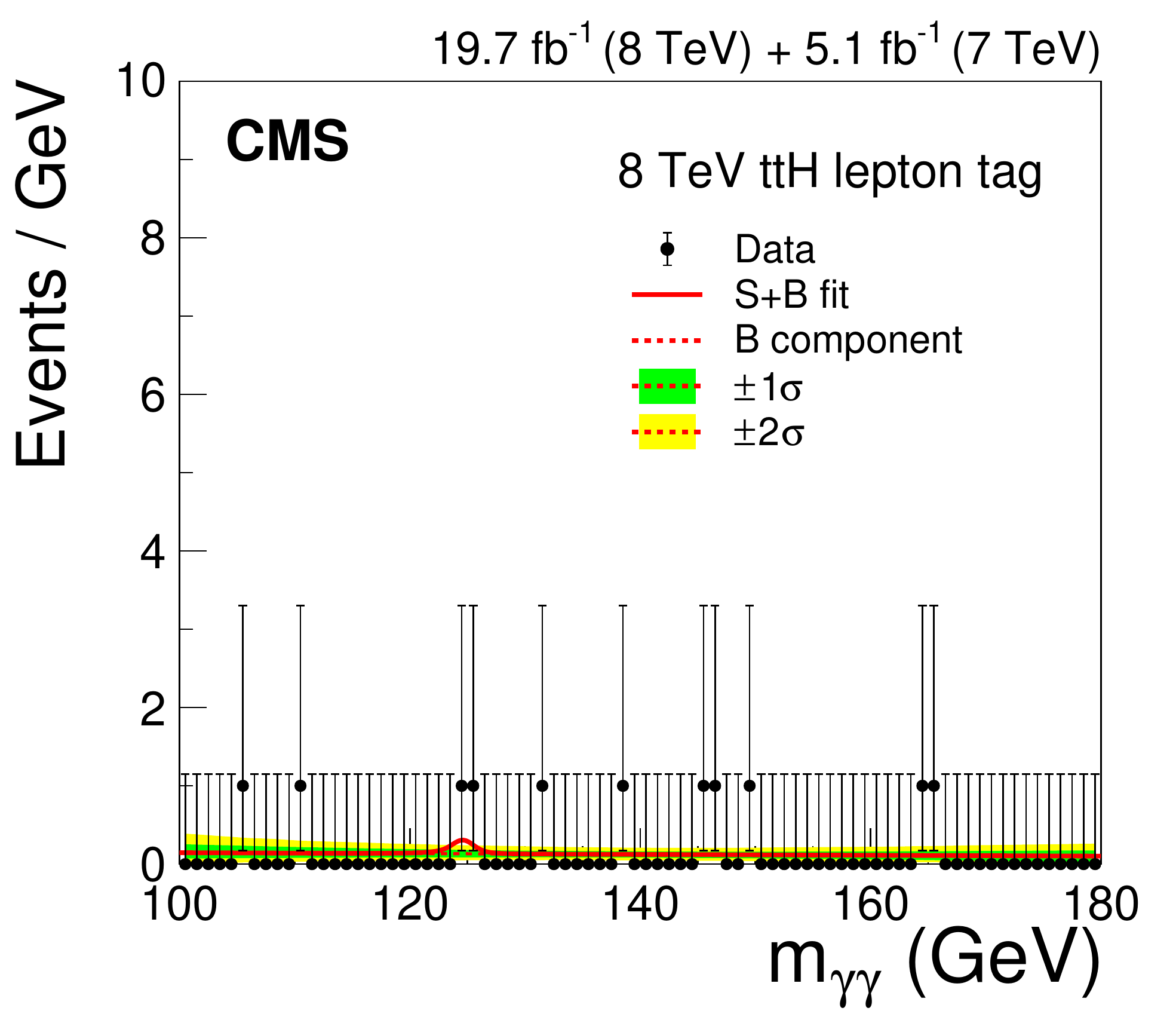}
     \includegraphics[width=\cmsFigWidthOne]{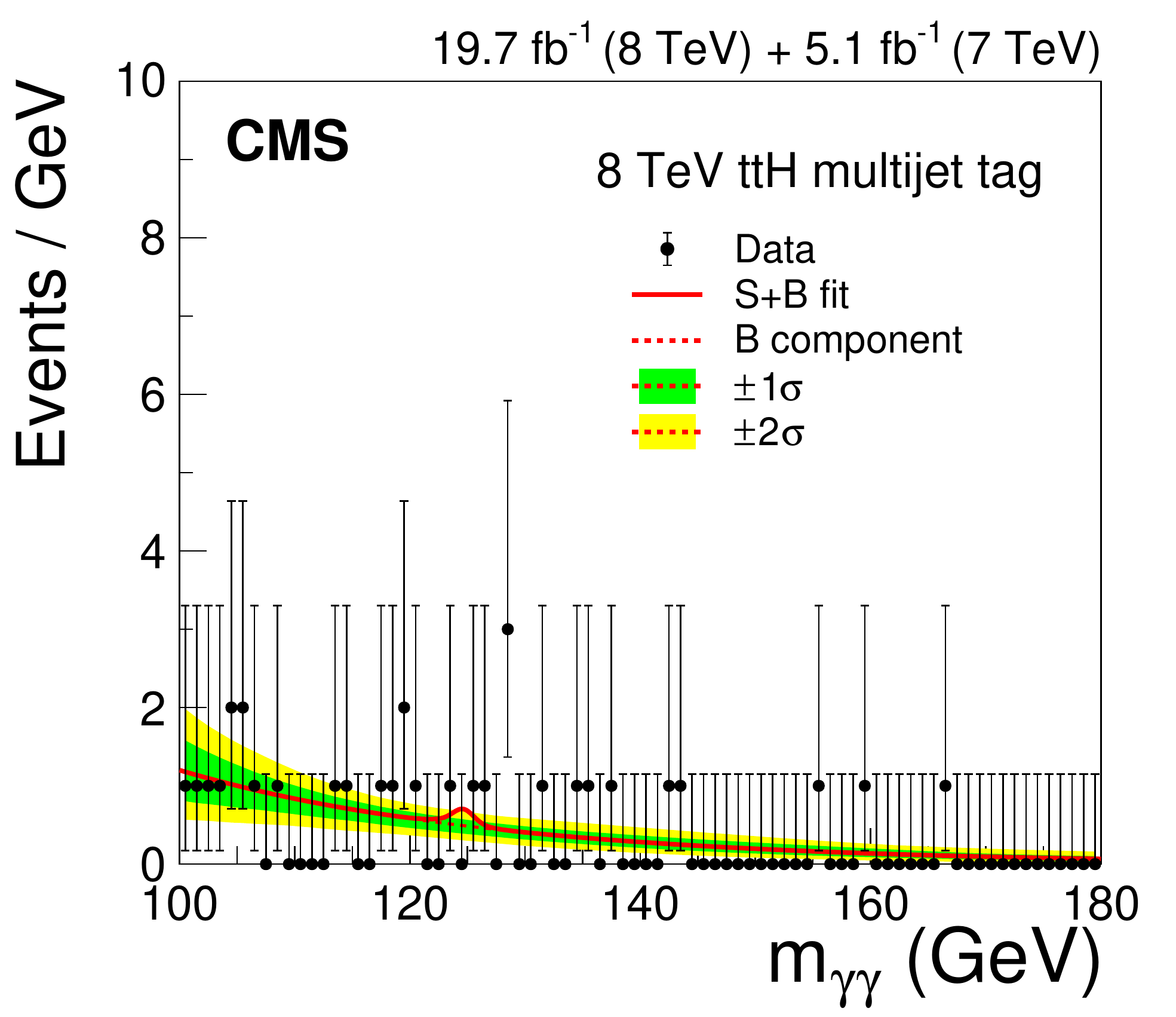}
     \caption{\label{fig:mgg-8-tth} Events in the two $\ttH$-tagged
              classes of the 8\TeV dataset, binned as a function of $\mgg$,
       together with the result of a fit of the signal-plus-background model.
The $1\sigma$ and $2\sigma$ uncertainty bands shown for the background component of the fit are computed from the fit
uncertainty in the background yield in bins corresponding to those used to display the data.
These bands do not contain the Poisson uncertainty that must be
included when the full uncertainty in the number of background events in any given mass range is estimated.
}
\end{figure}

The \mgg distributions in the 25 event classes in the 7 and 8\TeV data samples,
together with the results of a simultaneous fit of the signal-plus-background model,
are shown in Figs.~\ref{fig:mgg-7-untagged}--\ref{fig:mgg-8-tth}.
The \mgg distribution of the combined event classes is shown in Section~\ref{sec:fitresults}.
The distributions are labeled with the $\sqrt{s}$ and integrated luminosity of the combined datasets,
reflecting the fact that the signal-plus-background fit is a simultaneous fit to the 25 event classes.
Data points are drawn for all bins, including those in which there are no events.
The error bars are calculated using the Garwood procedure~\cite{garwood1936fiducial} to provide correct coverage of the Poisson uncertainty.
The $1\sigma$ and $2\sigma$ uncertainty bands shown for the background component of the fit include
the uncertainty due to the choice of function and the uncertainty in the fitted parameters,
and are computed from the variation in pseudo-experiments on the fitted background yield in bins corresponding to those used to display the data.
These bands do not contain the Poisson uncertainty that must be included when the full uncertainty in the number of background events in any given mass range is estimated.
The fit is performed on the data from all event class distributions simultaneously, with a
single overall value of the signal strength free to vary in the fit.

\clearpage
\section{Systematic uncertainties}
\label{sec:systematics}

The uncertainty related to the background modelling, and how it is
handled, has been discussed in the previous section.
The systematic uncertainties related to the signal model are
described below.
A useful measure of the relative importance of the various systematic uncertainties can be obtained by tabulating
their contributions to the total uncertainty in the final results for the best-fit signal strength and the best-fit mass.
This is done in Tables~\ref{tab:sys-mu} and \ref{tab:sys-mass} in Section~\ref{sec:fitresults} where the results of the analysis are discussed.

The systematic uncertainties assigned to all events are
\begin{itemize}

\item \textit{PDF, and theory uncertainties:}
the theory systematic uncertainties in the production cross section and the diphoton branching fraction follow
the recommendations of the LHC Higgs Cross Section Working Group~\cite{LHCHiggsCrossSectionWorkingGroup1,LHCHiggsCrossSectionWorkingGroup3}.
As can be seen in Table~\ref{tab:sys-mu}, these uncertainties make up the largest contribution to the uncertainty in the signal strength,
and are dominated by the uncertainty in the ggH process cross section,
coming from both uncertainties due to the missing higher orders and uncertainties related to the parton distribution functions.
The effect of these theory uncertainties on the overall acceptance and on the classification of the accepted events is included
by varying the $\pt$ and rapidity distributions of the simulated Higgs boson events
as they are changed by the theory uncertainties.

\item \textit{Integrated luminosity:}
the luminosity uncertainty is estimated as described in Refs.~\cite{CMS-PAS-SMP-12-008,CMS-PAS-LUM-13-001},
and amounts to a 2.2\% (2.6\%) uncertainty in the signal yield in the 7 (8)\TeV datasets, respectively.

\item \textit{Vertex finding efficiency:}
the uncertainty in the vertex finding efficiency is taken from the
uncertainty in the measurement of the corresponding data/MC scale factor
obtained using $\Zmm$ events.
We assign an additional 1\% uncertainty in the vertex finding efficiency, related to the amount of activity
resulting in charged particle tracks in signal events, which is derived
by varying the \PYTHIA underlying event tunes in ggH events.
Since the vertex-finding efficiency varies considerably with $\ptgg$, there is an uncertainty in the overall efficiency coming from the
uncertainty in the signal \PT\ distribution, leading to a further
uncertainty of 0.2\% to be added to the uncertainty in the data/MC scale factor for both the 7 and 8\TeV datasets.

\item \textit{Trigger efficiency:}
the uncertainty in the trigger efficiency is extracted from $\Zee$ events using a tag-and-probe technique.
Rescaling is used to take into account the difference in the \RNINE\ distributions of electrons and photons.
The uncertainty value obtained is slightly less than 1\%, but an uncertainty of 1\% has been assigned.

\end{itemize}

The systematic uncertainties related to individual photons are
\begin{itemize}

\item \textit{Photon energy scale uncertainty resulting from electron/photon differences:}
an important source of uncertainty in the energy scale of photons is the imperfect modelling of the difference between
electrons and photons by the MC simulation, the most important cause of which is an imperfect description of the material between the
interaction point and the ECAL.
Studies of electron bremsstrahlung, photon conversion vertices, and the multiple scattering of pions suggest
a deficit of material in the simulation.
Although the deficit is almost certainly in specific structures and localized regions --- and this hypothesis is supported by the studies ---
the data/MC discrepancies are slightly smaller than what would be caused by a 10\% uniform deficit of material
in the region $\abs{\eta}<1.0$ and a 20\% uniform deficit for $\abs{\eta}>1.0$.
The resulting uncertainty in the energy scale has been assessed using simulated samples in which the tracker material is increased
uniformly by 10 and 20\%, and an uncertainty, with differing magnitude in eight bins
($\eta$: three barrel and one endcap, and $\RNINE$: two bins) is assigned to photon energies.
The systematic uncertainty in the energy scale ranges from 0.03\% in the central ECAL barrel up to 0.3\% in the outer endcap.
Two nuisance parameters, one for $\abs{\eta}<1.0$ and one for the remainder of the $\eta$ range used in the analysis,
are introduced to model this uncertainty, which is fully correlated between the 7 and 8\TeV datasets.

Another difference between data and simulation, relevant to electron-photon differences, is the modelling of the varying
fraction of scintillation light reaching the photodetector as a function of the longitudinal depth in the crystal at which it was emitted.
Ensuring adequate uniformity was a major accomplishment in the lead tungstate crystal development that was achieved by depolishing one face of
each barrel crystal, but an uncertainty in the degree of uniformity achieved remains~\cite{Paramatti,Auffray}.
In addition, the uniformity is modified by the radiation-induced loss of transparency of the crystals.
The effect of the uncertainty, including the effect of radiation-induced transparency loss, has been simulated.
It results in a difference in the energy scale between electrons and unconverted photons which is not present in the standard simulation.
The magnitude of the uncertainty in the photon energy scale is 0.04\% for photons with $\RNINE>0.94$ and 0.06\% for those with
$\RNINE<0.94$, but the signs of the energy shifts are opposed, and the two anti-correlated uncertainties result in an
uncertainty about 0.015\% in the mass scale.

A further small uncertainty is added to account for imperfect electromagnetic shower simulation by $\GEANT4$~version~9.4.p03.
A simulation made with an improved shower description, using the Seltzer--Berger model for the bremsstrahlung energy spectrum~\cite{Seltzer:1974zz},
changes the energy scale for both electrons and photons.
The much smaller changes in the \textit{difference} between the electron and photon energy scales, although mostly consistent with zero,
are interpreted as a limitation on our knowledge of the correct simulation of the showers, leading to a further uncertainty of 0.05\%.

\item \textit{Energy scale nonlinearity:}
possible differences between MC simulation and data in the extrapolation from shower energies typical of electrons from $\Zee$ decays,
to those typical of photons from $\Hgg$ decays, have been investigated with $\Zee$ data samples by binning the events according to the scalar sum
of the $\ET$ of the two electron showers, and by studying electron showers in $\Wenu$ events in which the electron $\pt$ is also
measured by the tracker.
The effect of the differential nonlinearity in the measurement of photon energies has an effect of up to 0.1\% on
the diphoton mass scale for diphoton masses close to $\mgg=125\GeV$.
In the best untagged event class, in which the diphoton transverse momentum is particularly high, the effect is up to 0.2\%.
The uncertainties are not completely correlated between the 7 and 8\TeV datasets, since
the energy response regression (Section~\ref{sec:photonE}), which would be strongly implicated in any nonlinearity,
uses independent sets of regression weights for the two datasets.
Moreover, $\ET$-dependent scale corrections have been applied at 8\TeV for barrel photons, while the corrections
at 7\TeV are not $\ET$-dependent.
Studies suggest that there may be as much as 20\% correlation between the uncertainties in the energy scale nonlinearities in the
7 and 8\TeV datasets, and this correlation is included in the implementation of the uncertainties.
This uncertainty makes a significant contribution to the uncertainty in the measured Higgs boson mass, as can be seen in Table~\ref{tab:sys-mass}.

\item \textit{Measuring and correcting the energy scale in data, and the energy resolution in simulation:}
the energy scale and resolution in data are measured with electrons from
$\Zee$ decays. The statistical uncertainties in the measurements are small,
but the methodology, which is described in Section~\ref{sec:photonE},
gives rise to a number of systematic uncertainties related to the imperfect agreement between data and MC simulation.
These are estimated and accounted for in the same eight bins
(4 bins in $\abs{\eta}$ and 2 bins in $\RNINE$) as are used to derive the scale corrections and the resolution smearings for simulated events.
The uncertainties range from 0.05\% for unconverted photons in the ECAL central barrel, to 0.1\% for converted photons in the ECAL outer endcaps.
In addition, for the barrel region, the uncertainty in the energy dependence
of the Gaussian smearing applied to the simulation, is also accounted for.
The energy dependence of the smearing is controlled by a parameter that shares the smearing between a constant term and a term proportional to $1/\sqrt{\ET}$, and the uncertainty pertains to this sharing.
Finally, there is an overall uncertainty that accounts for possible misdescription of the $\Zee$ line-shape in simulation.

\item \textit{Photon identification BDT score, and estimate of the per-photon energy resolution:}
the uncertainties in these two quantities are discussed together since they are studied in the same way, and
the dominant underlying cause of the observed differences between data and simulation
is, almost certainly, the imperfect simulation of the shower shape --- despite the fact
that no obvious differences between data and simulation can be observed when the shower shape variables are
examined individually.
The combined contribution of the uncertainties in these two quantities dominates
the experimental contribution to the systematic uncertainty in the signal strength,
and has been labeled ``shower shape modelling'' in Table~\ref{tab:sys-mu}.

The agreement between data and simulation is examined when the photon candidates are
electron showers reconstructed as photons in $\Zee$ events, photons in $\Zmmg$ events, and
leading photons in preselected diphoton events where $\mgg>160\GeV$.
It is found that among the input variables to the diphoton BDT,
only the distributions of the photon identification BDT score and the per-photon energy resolution estimate show
significant differences between data and simulation.
A variation of $\pm$0.01 on the photon identification BDT score, together with an uncertainty
in the per-photon energy resolution estimate, parameterized as a rescaling of the resolution estimate by $\pm$10\%
about its nominal value, fully covers the differences observed in all three of the above data samples.

\item \textit{Photon preselection efficiency:}
the uncertainty in the photon preselection efficiency is taken as the uncertainty in the data/MC
preselection efficiency scale factors, which are measured using $\Zee$ events with a tag-and-probe technique (see Table~\ref{tab:PhotEff}).

\end{itemize}

The effect of the single photon uncertainties is propagated to the diphoton quantities:
diphoton efficiency, diphoton mass scale, and diphoton mass resolution.
For instance, to obtain the magnitude of the mass-scale uncertainty resulting from a particular photon energy uncertainty,
which may relate only to certain photons (such as barrel photons with $\RNINE>0.94$),
the energy of photons in simulated signal events to which the uncertainty applies is shifted by the $1\,\sigma$ single photon uncertainty.
The resulting shift of the mean of the diphoton mass distribution in each event class is determined.
This shift corresponds to the effect of the single photon energy uncertainty in the diphoton mass scale and may be different for each event class.
The effect of single photon uncertainties on the diphoton selection efficiency and diphoton resolution are determined in a similar way.

{\tolerance=700
The sources of systematic uncertainty for the event classes targeting specific production modes are
\begin{itemize}

\item \textit{Uncertainties in jet requirements:}
the largest uncertainty related to the tagging of production processes comes from a theory uncertainty and concerns the probability of producing
additional jets in gluon-fusion Higgs boson production.
The Stewart--Tackmann procedure~\cite{Stewart:2011cf} recommended by the LHC Higgs Cross Section Working
Group~\cite{LHCHiggsCrossSectionWorkingGroup3} has been used to quantify the uncertainty in the yield of
ggH events in the VBF dijet-tagged classes.
The resulting uncertainty agrees comfortably with our previous estimation~\cite{Chatrchyan:2013lba} derived
by varying the underlying event tunes in ggH events produced by \PYTHIA, and that method is retained to estimate
the uncertainty associated with additional jet production in the yield of ggH events in the \ttH multijet-tagged class.
There is a further contribution to the uncertainty in the yield of ggH events in
the \ttH multijet-tagged class arising from the uncertainty in the probability of gluon splitting to $\bbbar$, which
is estimated from the discrepancy observed between data and \POWHEG simulation in the fraction of additional b-tagged jets in
samples of $\ttbar$+jets events, where the $\ttbar$ pair is identified by the presence of two charged leptons in the final state.
Additionally, since few events from the simulated signal samples of ggH are selected for the \ttH multijet-tagged class,
there is a contribution due to the limited sample size.
For the VBF dijet-tagged classes, the VH dijet-tagged class, and the \ttH multijet-tagged class there is an uncertainty in the effect of
the algorithm used to reject jets from pileup (in the 8\TeV dataset only).
Further small contributions are due to the uncertainties in the jet energy scale and resolution corrections.

\item \textit{Lepton identification efficiency:}
for both electrons and muons, the uncertainty in the identification efficiency is computed by varying the data/simulation efficiency scale
factor by its uncertainty.
The resulting differences in the selection efficiency for the event classes tagged by leptons,
range from 0.2\% to 0.5\% depending on the event category,
and are taken as systematic uncertainties.

\item \textit{$\MET$ selection efficiency:}
systematic uncertainties due to \MET\ reconstruction are estimated both in signal events in which
real \MET\ is expected (such as in $\PW(\ell\nu)\PH$ production) and in the other Higgs production mechanisms.
For WH events the uncertainty is estimated by applying or not the \MET\ corrections 
and taking the difference in efficiency of 2.6\% as a systematic uncertainty.
For the other processes, ggH, VBF, and $\ttH$, what is uncertain is the fraction of
events in the tail of the \MET\ distribution.
This is evaluated by comparing diphoton data and simulated events in control samples enriched
in $\Pgg$+jet events, which have a similar \MET\ distribution to the Higgs signal events.
The systematic uncertainty amounts to 4\%.

\item \textit{b-tagging efficiency:}
the uncertainty in the b-tagging efficiency used in the selection for the $\ttH$-tagged classes,
is evaluated by varying the measured b-tagging efficiency scale factors between data and simulation within their uncertainty.
The resulting uncertainty in the signal yield is 1.3\% in the lepton-tagged class and 1.1\%  in the multijet-tagged class.
\end{itemize}
\par}

\section{Alternative analyses}
\label{sec:crosscheck}

Three alternative analyses are performed using particular variations of methodology, which
help to provide verification of different aspects of the analysis described in the previous
sections.

\subsection{Cut-based analysis}
\label{sec:cut-based}
The first of these, the ``cut-based'' analysis described in Ref.~\cite{Chatrchyan:2013lba}, does not use multivariate
techniques for selection or classification of events.
Photon identification is performed by dividing photons into four
mutually exclusive categories depending on whether the photon is in
the barrel or endcap, and on whether or not it has $\RNINE>0.94$.
The identification selection requirements are then
particular to the category, and use a subset of the
discriminating variables that are used in the multivariate photon
identification described in Section~\ref{sec:photonID}.

Four mutually exclusive diphoton event classes are constructed by splitting the
events according to the same categorization criteria as is used for single photons in the
photon identification.
Subsequently these four classes are each split according to the transverse momentum of the diphoton system.
The four event classes are
\begin{enumerate}\setcounter{enumi}{-1}
\item Both photons are in the barrel and have $\RNINE>0.94$.
\item Both photons are in the barrel and at least one of them fails the requirement of $\RNINE>0.94$.
\item At least one photon is in the endcap and both photons have $\RNINE>0.94$.
\item At least one photon is in the endcap and at least one of them fails the requirement $\RNINE>0.94$.
\end{enumerate}
Photons with a high value of the $\RNINE$ variable are predominantly
unconverted and have a better energy resolution than those with a lower value,
and photon candidates with a high value of $\RNINE$ are also less likely to arise from misidentification of jet fragments.
Similarly, photons in the barrel have both better energy resolution
and are more likely to be signal photons.
Thus, the classification serves a similar purpose to the one
using the BDT event classifier: events with good diphoton mass resolution,
resulting from photons with good energy resolution,
and with better signal-to-background ratio are grouped together.
Each of the four event classes is then split into two according to the transverse momentum of the diphoton system.
Since the \ptgg spectrum resulting from Higgs bosons produced by the VBF, VH, or \ttH processes is significantly harder
than that of the diphoton background, this separation improves the sensitivity of the analysis by increasing
the expected signal-to-background ratio in the high-\ptgg event classes.
The magnitude of the improvement in sensitivity is about 5\%, and has a very weak dependence on the
precise value of the \ptgg threshold chosen.
To avoid modification of the shape of the invariant mass spectrum by the threshold, the classification uses
the ratio $\ptgg/\mgg$, with a threshold value of 0.32, corresponding to $\ptgg=40\GeV$ at $\mgg=125\GeV$.

Event classes tagged by signatures of VBF, VH, and \ttH production are also included in the cut-based analysis.
The event classes tagged for VH and \ttH production are defined in exactly the same way as described
in Section~\ref{sec:exclusive-tags}, with the exception that the minimum requirements on the diphoton BDT scores
are replaced by the cut-based photon identification requirements.
A dijet tag is defined to select signal events produced by the VBF process by requiring a pair of
jets satisfying requirements on the same variables as are used by the main analysis in the dijet BDT
described in Section~\ref{sec:VBF}.
These selection requirements are listed in Table~\ref{tab:vbf-selection}.
The tagged events are subdivided into two classes depending on whether they additionally satisfy tighter
requirements on the \pt of the second jet and the dijet mass, $\ptjb>30\GeV, \mjj>500\GeV$.

Signal and background models are constructed in the same way as in the main analysis and are fitted
to the $\mgg$ distributions.
Since this analysis does not use multivariate techniques for event selection or for event classification,
it provides some degree of cross-checking on their use in the main analysis.

\begin{table}[htbp]
\caption{Selection requirements for the VBF dijet tag in the cut-based and dijet 2D analyses.
               The variables are defined in Section~\ref{sec:VBF}
}
\centering
\begin{tabular}{ l l }
\hline
Variable  &  Requirement \\
\hline

 $\ptga/\mgg$  & $>$0.5 \\
 $\ptgb$      & $>$25\GeV     \\
 $\ptja$    & $>$30\GeV     \\
 $\ptjb$   & $>$20\GeV     \\
 $\abs{\Delta\eta_\text{jj}}$ & $>$3       \\
 $\abs{\eta_{\gamma\gamma}-(\eta_\text{j1}+\eta_\text{j2})/2}$ & $<$2.5       \\
 $\mjj$           & $>$250\GeV   \\
 $ \Dggjj$ & $>$2.6 \\
\hline
\end{tabular}
\label{tab:vbf-selection}
\end{table}
\subsection{Sideband background model analysis}
The second alternative analysis approach, the ``sideband background model'' analysis described in Ref.~\cite{Chatrchyan:2013lba},
uses the same multivariate techniques as the standard analysis to select the events,
but employs a very different procedure to model the background.
For any given mass hypothesis, \mH, a signal region is defined as
the ${\pm}2\%$ range centred on $\mH$.
A contiguous set of sidebands is defined in the mass distribution on either side of the signal region, from
which the background is extracted.
Each sideband is defined to have the same width of ${\pm}2\%$ relative to the diphoton mass
that corresponds to its centre.
A total of eight sidebands are defined, four on either side of the signal region.
Six sidebands are used to obtain the background estimate, with a sideband on either
side of the signal region left unused in order to avoid signal contamination.

The result is extracted by counting events in the signal region, in bins
that are defined using two-dimensional (2D) distributions of
the diphoton BDT score and the diphoton mass in the form $\Delta m/\mH$, where
$\Delta m=\mgg-\mH$ and $\mH$ is the Higgs boson mass hypothesis.
The distributions, for simulated signal and background events, are in the form of histograms, and after applying a smoothing algorithm
to them, seven event bins are defined for the untagged events by defining regions
ranked by signal-to-background ratio in the 2D plane.
For the tagged events, the event bins correspond to the tagged classes described
in Section~\ref{sec:exclusive-tags}.

The overall normalization of the background model is obtained from a parametric fit to the inclusive mass spectrum, with
the signal region excluded from the fit, and it is easy to account for the small uncertainty associated with the choice of function in this single fit.
The number of events in each event bin is obtained from the data in each of the six sidebands.
It is assumed that, for any sideband, the fraction of events in each bin is a linear function of the invariant mass
of the sideband central mass, and that there is negligible signal contamination in the sidebands.
These assumptions have been verified within the assigned systematic uncertainties.
The sideband background analysis does not rely on a parametric fit to the \mgg distribution
to model the background shape in the signal region, and thus provides a valuable cross-check of the background modelling used in the main analysis.

\subsection{Dijet 2D analysis}
The third alternative analysis, the ``dijet 2D'' analysis, uses a different method for extracting the signal produced by the VBF production process.
The dijet invariant mass, $\mjj$, of the pair of jets that accompany the production of a Higgs boson by the VBF mechanism,
tends to be larger than that of pairs of jets found in either background events or in events produced by the ggH process.
The analysis takes advantage of this by extracting the VBF signal in a parametric 2D fit of signal and background in the ($\mgg$, $\mjj$) plane.
The initial selection of events for the analysis makes a requirement on the photon identification BDT score (Section~\ref{sec:photonID}).
Dijet-tagged events are required to satisfy the same requirements as for the VBF dijet tag in the cut-based analysis, shown
in Table~\ref{tab:vbf-selection}.
The invariant mass of the dijet pair is required to satisfy $\mjj>250\GeV$, and the selected events in the 7 and 8\TeV datasets are divided
in two and four event classes, respectively, based solely on the estimated diphoton mass resolution.
The remaining events, not selected for the VBF dijet-tagged classes, are classified in the same way as in the main analysis.
The 2D fit is applied to the events in the dijet-tagged classes using parametric 2D signal and background models.
The signal in the other event classes is extracted using a one-dimensional fit to the $\mgg$ distribution, as in the main analysis.
This analysis provides an alternative approach to extracting the VBF signal, which provides most of the sensitivity in the measurement
of vector-boson-initiated production.
\section{Results}
\label{sec:fitresults}

\begin{figure}[htbp]
   \centering
     \includegraphics[width=\cmsFigWidth]{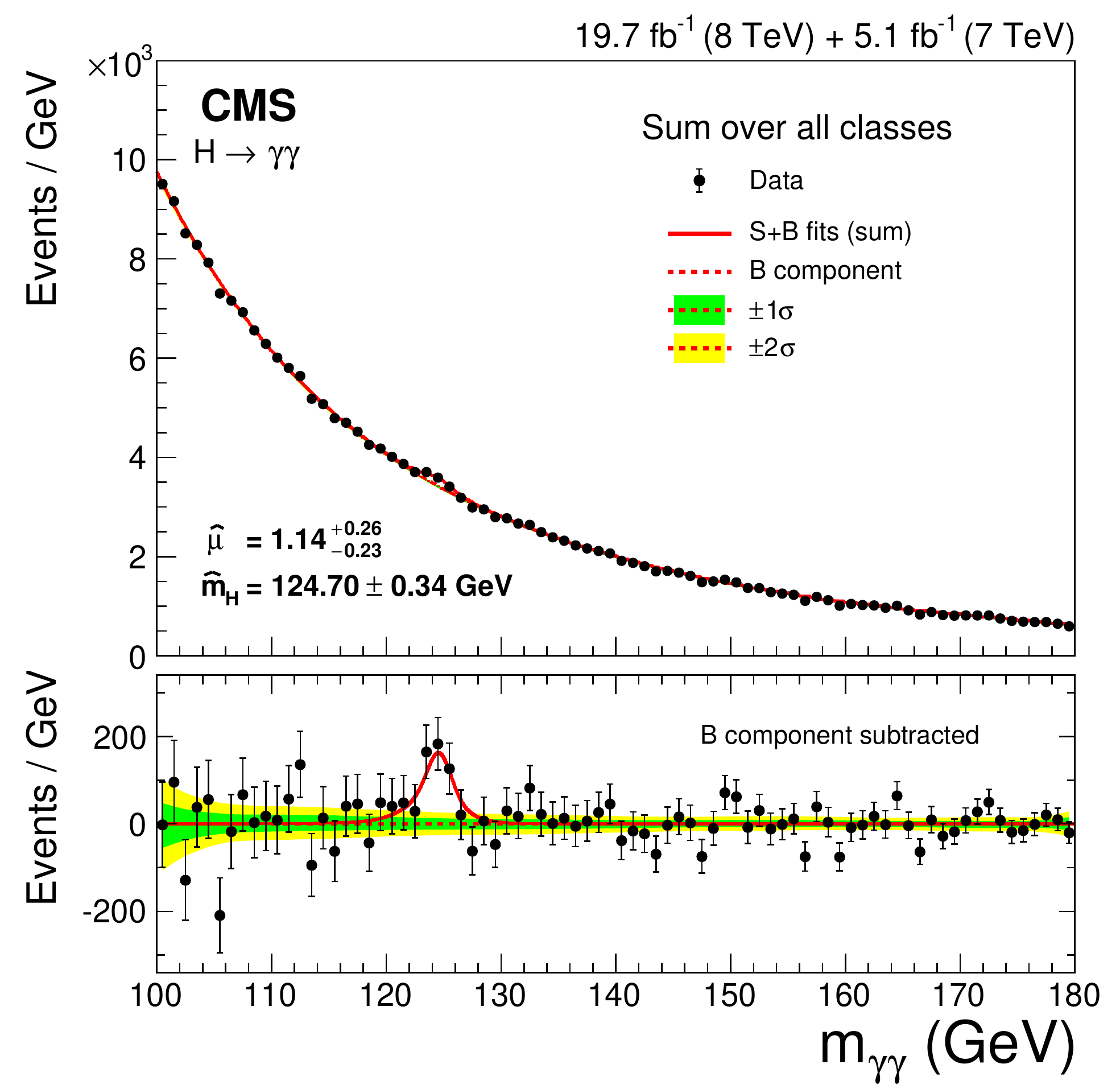} 
     \caption{\label{fig:mgg-all}Sum of the 25 signal-plus-background model
       fits to the event classes in both the 7 and 8\TeV datasets,
       together with the data binned as a function of \mgg.
The $1\sigma$ and $2\sigma$ uncertainty bands shown for the background component of the fit are computed from the fit
uncertainty in the background yield in bins corresponding to those used to display the data.
These bands do not contain the Poisson uncertainty that must be
included when the full uncertainty in the number of background events in any given mass range is estimated.
The lower plot shows the residual data after subtracting the fitted background component.
}
\end{figure}

Figure~\ref{fig:mgg-all} shows the \mgg distribution of the combined data in the 7 and 8\TeV samples,
together with the sum of the signal-plus-background fits to the 25 event classes which results in a
best-fit mass $\mH=124.7\GeV$.
The uncertainty bands shown on the background component of the fit include the uncertainty due to the choice of function and the uncertainty in the fitted parameters.
These bands do not contain the Poisson uncertainty which must be included when the full uncertainty
in the number of background events in any given mass range is estimated.
The excess of events over the background expectation visible near $\mgg=125\GeV$ can be seen more clearly
after subtraction of the background component, shown in the lower plot.

\subsection{Significance of the signal and its strength}

\begin{figure}[htbp]
  \centering
     \includegraphics[width=\cmsFigWidth]{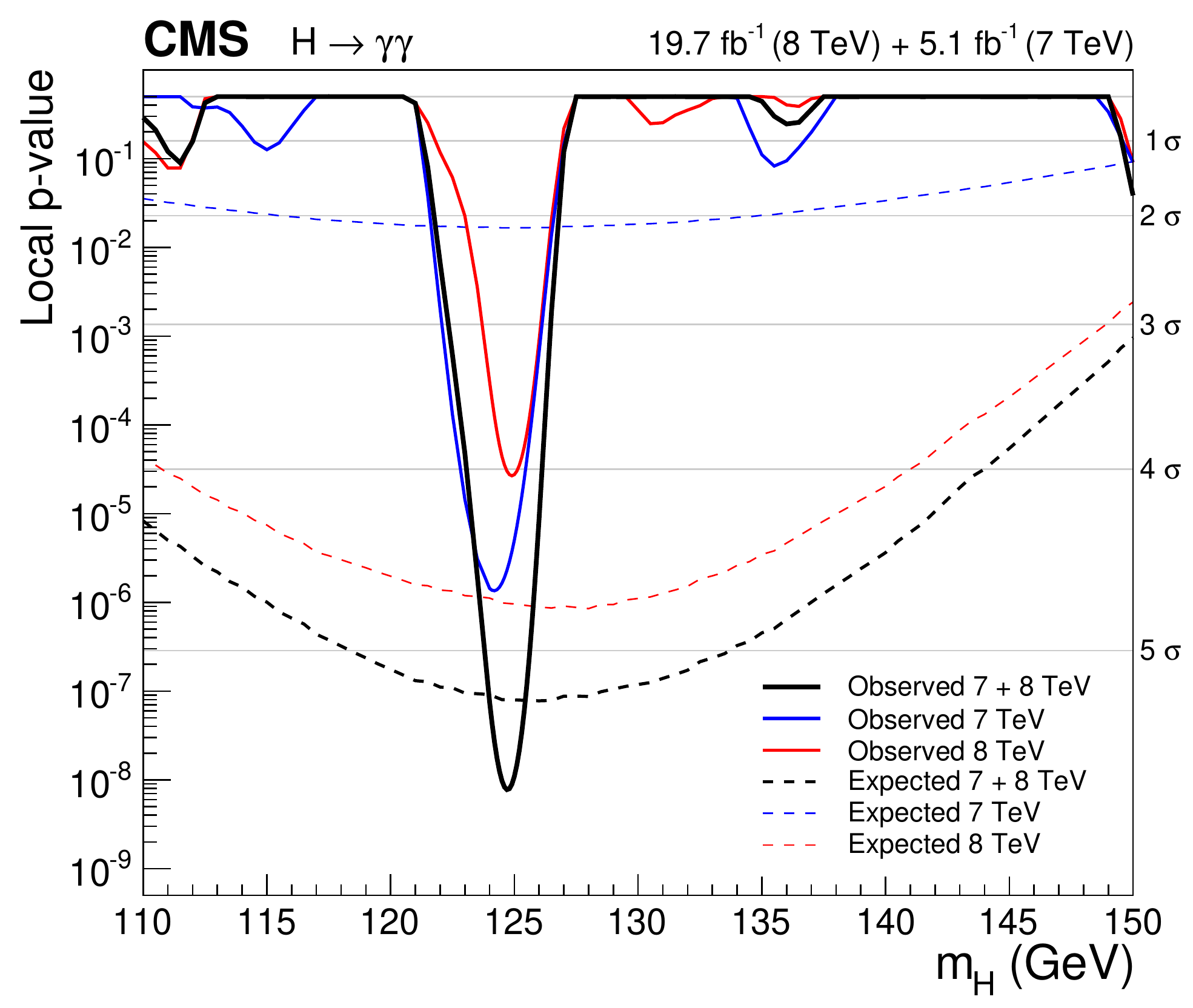} 
    \caption{Local $p$-values as a function of $\mH$ for the 7\TeV,
      8\TeV, and the combined dataset.
The values of the expected significance, calculated using the background expectation obtained from
the signal-plus-background fit, are shown as dashed lines.
    }
    \label{fig:p-value}
\end{figure}

The local $p$-value quantifies the probability for the background to produce a fluctuation as large, or larger,
than the apparent signal observed, within a specified search range and uncorrected for the ``look-elsewhere effect"~\cite{LEE}.
Figure~\ref{fig:p-value} shows the local $p$-value,
in the mass range $110<\mH<150$\GeV, calculated separately for the 7 and 8\TeV datasets as well as their combination.
Lines indicating the $p$-values expected for a SM Higgs boson, for the three cases, are also shown.
The values of expected significance have been calculated using the background expectation obtained from
the signal-plus-background fit, the so-called \textit{post-fit} expectation.
The post-fit model corresponds to the parametric bootstrap described in the statistics literature~\cite{Efron1979,Lee2005},
and includes information gained in the fit regarding the values of all parameters, including the best-fit mass.

\begin{figure}
  \centering
      \includegraphics[width=\cmsFigWidth]{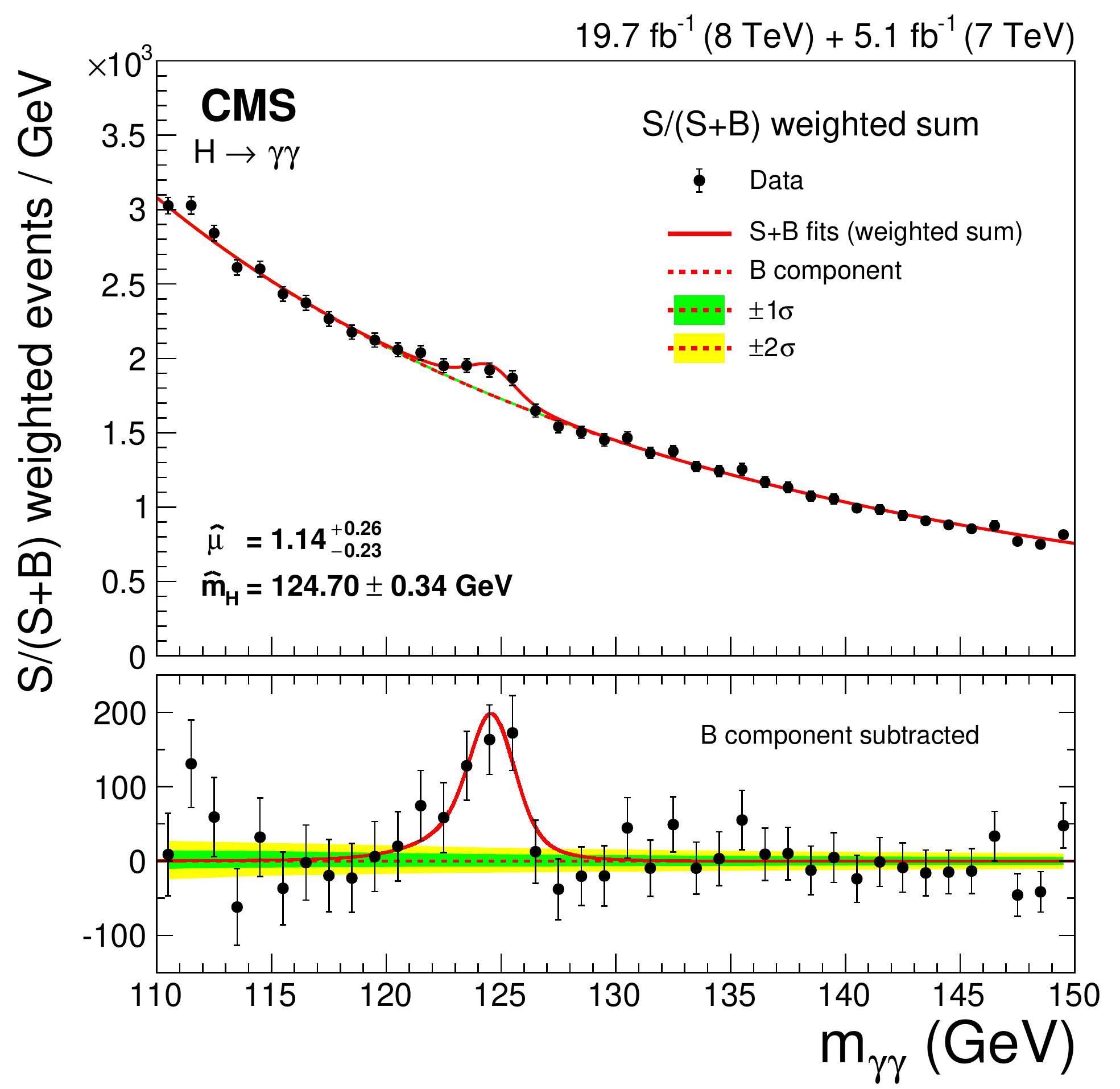} 
    \caption{Diphoton mass spectrum weighted by the ratio $S/(S+B)$ in each event class,
    together with the background subtracted weighted mass spectrum.
}
    \label{fig:mgg-weighted}
\end{figure}

The significance of the minimum of the local $p$-value, at 124.7\GeV, is
5.7\,$\sigma$ where a local significance of 5.2\,$\sigma$ is expected from the SM Higgs boson.
To better visualize the excess of events, with respect to the background expectation, and its significance,
the diphoton mass spectrum is plotted with each event used in the analysis weighted by a factor depending on the category in which
it falls.
The weight is proportional to $S/(S+B)$, where $S$ and $B$ are the numbers of
expected signal and background events, respectively, counted in a mass window corresponding
to ${\pm}1\sigma_\text{eff}$ and centred on $\mgg=124.7\GeV$.
The background is calculated from the signal-plus-background fit.
The motivation for this choice of weights is explained in Ref.~\cite{Barlow:1986ek}.
The weighted data, the weighted signal model, and the weighted background model are
normalized such that the integral of the weighted signal model matches
the number of signal events obtained from the best fit.
The resulting distribution, and the corresponding background subtracted spectrum, are shown in Fig.~\ref{fig:mgg-weighted}.

\begin{figure}
  \centering
      \includegraphics[width=\cmsFigWidthOne]{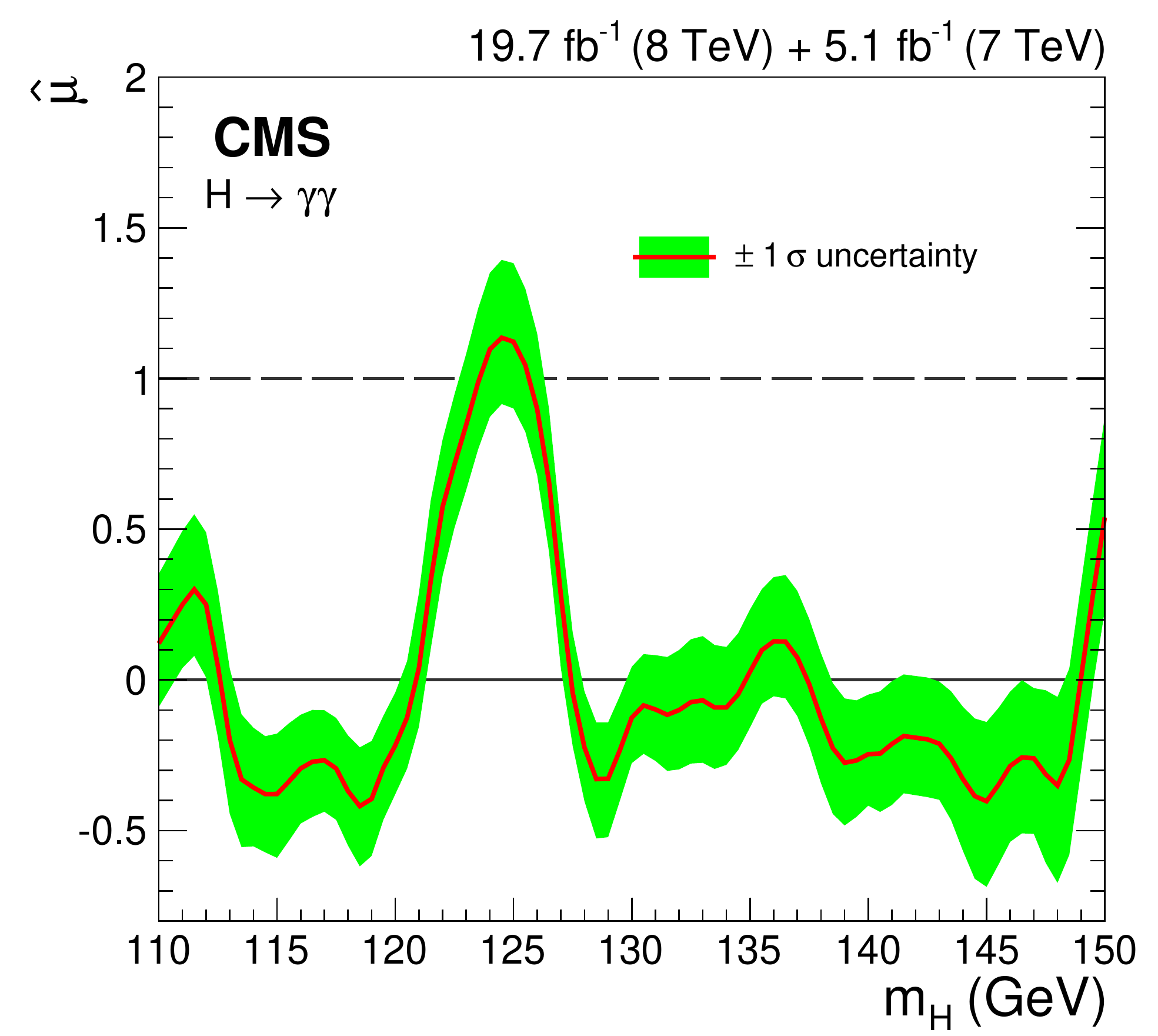} 
      \includegraphics[width=\cmsFigWidthOne]{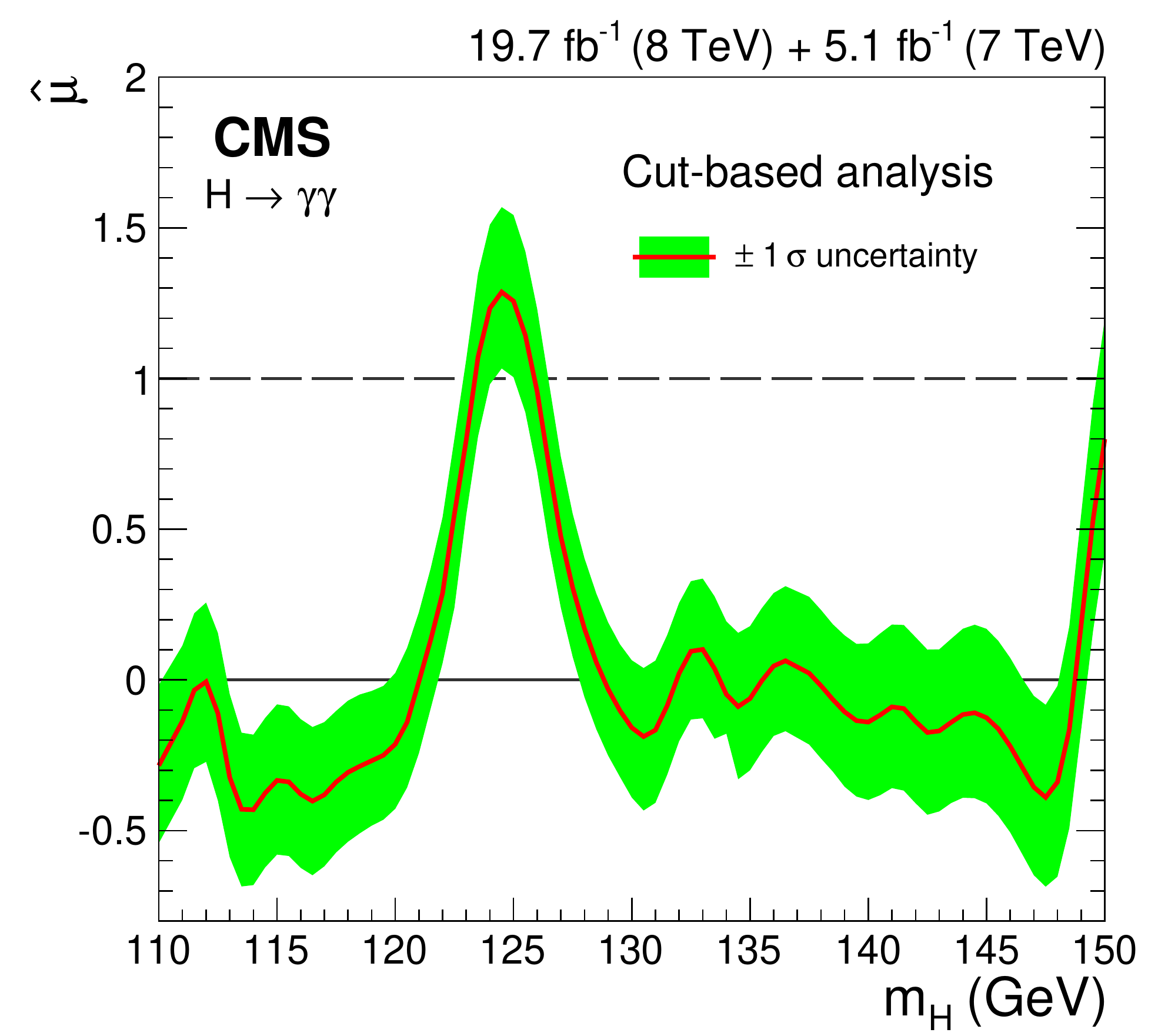}   
    \caption{Best-fit signal strength, $\muhat$, shown as a function of
      the mass hypothesis, \mH. The results are shown for
      the standard analysis (\cmsLeft), and for the cut-based cross-check
      analysis (\cmsRight).
}
    \label{fig:mu-vs-mh}
\end{figure}

The signal strength is quantified by $\mu=\musm$, where \musm denotes the production cross section
times the relevant branching fractions, relative to the SM expectation.
In Fig.~\ref{fig:mu-vs-mh} the combined best-fit signal strength, $\muhat$, is shown as a function of the Higgs boson
mass hypothesis, both for the standard analysis (\cmsLeft) and for the cut-based analysis (\cmsRight).
The two analyses agree well across the entire mass range.
In addition to the signal around 125\GeV, both analyses see a small upward fluctuation at 150\GeV,
which is found to have a maximum local significance of just over $2\,\sigma$ at $\mH=151\GeV$---slightly beyond the mass range of our analysis.

The best-fit signal strength for the main analysis, when the value of $\mH$ is treated as an unconstrained parameter in the fit,
is $\muhat=1.14^{+0.26}_{-0.23}$, with the corresponding best-fit mass being $\mHhat=124.7\GeV$.
The expected uncertainties in the best-fit signal strength, at this mass, are +0.24 and $-0.22$.
The values of the best-fit signal strength, derived separately for the 7 and
8\TeV datasets, are listed in Table~\ref{tab:breakdown}.
For the cut-based analysis the corresponding value is $\muhat=1.29^{+0.29}_{-0.26}$ at $\mHhat=124.6\GeV$,
and for the sideband background model analysis the value measured is $\muhat=1.06^{+0.26}_{-0.23}$ at $\mHhat=124.7\GeV$.
These values are shown in Table~\ref{tab:exp-mu-alternative} together with the expected uncertainty, and the
corresponding values for the main analysis.

\begin{table}
    \topcaption{\label{tab:breakdown}Values of the best-fit signal
      strength, $\muhat$, when \mH\ is treated as an unconstrained parameter, for the 7\TeV, 8\TeV, and combined datasets.
      The corresponding best-fit value of $\mH$, $\mHhat$, is also given.
      }
  \centering
   \renewcommand{\arraystretch}{1.2}
    \begin{tabular}{lcc}
      & $\muhat$  & $\mHhat$ (\GeVns)       \\  \hline
      7\TeV  & $2.22^{+0.62}_{-0.55}$ & 124.2  \\
      8\TeV  & $0.90^{+0.26}_{-0.23}$ & 124.9  \\ \hline
      Combined & $1.14^{+0.26}_{-0.23}$ & 124.7  \\
    \end{tabular}
\end{table}

\begin{table}
    \topcaption{\label{tab:exp-mu-alternative} Expected and observed best-fit values of the signal
      strength for a SM Higgs boson signal in the alternative analyses, together with their uncertainties,
       indicating the expected uncertainty in the measurement at the best-fit values of $\mH$,
       and the best-fit values obtained from the data.
      The corresponding values for the main analysis are shown for comparison.
      }
  \centering
   \renewcommand{\arraystretch}{1.2} 
    \begin{tabular}{lcc}
     & Expected & Observed \\  \hline
      Main analysis & $1.00^{+0.24}_{-0.22}$ & $1.14^{+0.26}_{-0.23}$ \\
      Cut-based analysis & $1.00^{+0.26}_{-0.24}$ & $1.29^{+0.29}_{-0.26}$ \\
      Sideband bkg. model analysis & $1.00^{+0.25}_{-0.22}$ & $1.06^{+0.26}_{-0.23}$ \\
    \end{tabular}
\end{table}

The uncertainty in the signal strength may be  separated into statistical and systematic contributions, with the latter
further divided into those having, or not, a theoretical origin:
$\muhat=1.14\pm0.21\stat\ ^{+0.09}_{-0.05}\syst\ ^{+0.13}_{-0.09}\thy$,
where the statistical contribution includes all uncertainties in the background modelling.
The separation of contributions can be taken further and Table~\ref{tab:sys-mu} lists
a finer breakdown of the contributions to the systematic uncertainty, where the contributions of the 81
nuisance parameters in the analysis are grouped according to their physical origin, as relevant to the signal strength uncertainty.

\begin{table}[htbp]
\topcaption{Magnitude of the uncertainty in the best fit signal strength, $\muhat$, induced by the systematic uncertainties in the signal model.
    To obtain the values, the quadratic subtraction, needed to remove the statistical uncertainty, is made for
    the positive and negative uncertainties separately. The values quoted are the average magnitudes of the positive and negative uncertainties.
    The statistical uncertainty includes all uncertainties in the background modelling.
}
\centering
\begin{tabular}{ l c }
\multirow{2}{*}{Source of uncertainty} & Uncertainty \\
 &  {in $\muhat$} \\
\hline
PDF and theory & 0.11\\
Shower shape modelling (Section~\ref{sec:systematics}) & 0.06 \\
Energy scale and resolution & 0.02\\
Other & 0.04 \\
\hline
All syst. uncert. in the signal model & 0.13 \\
 \multicolumn{1}{r}{Statistical} & \multicolumn{1}{r}{0.21} \\
\hline
 \multicolumn{1}{r}{Total} & \multicolumn{1}{r}{0.25} \\
\end{tabular}
\label{tab:sys-mu}
\end{table}

\begin{figure}
  \centering
    \includegraphics[width=\cmsFigWidth]{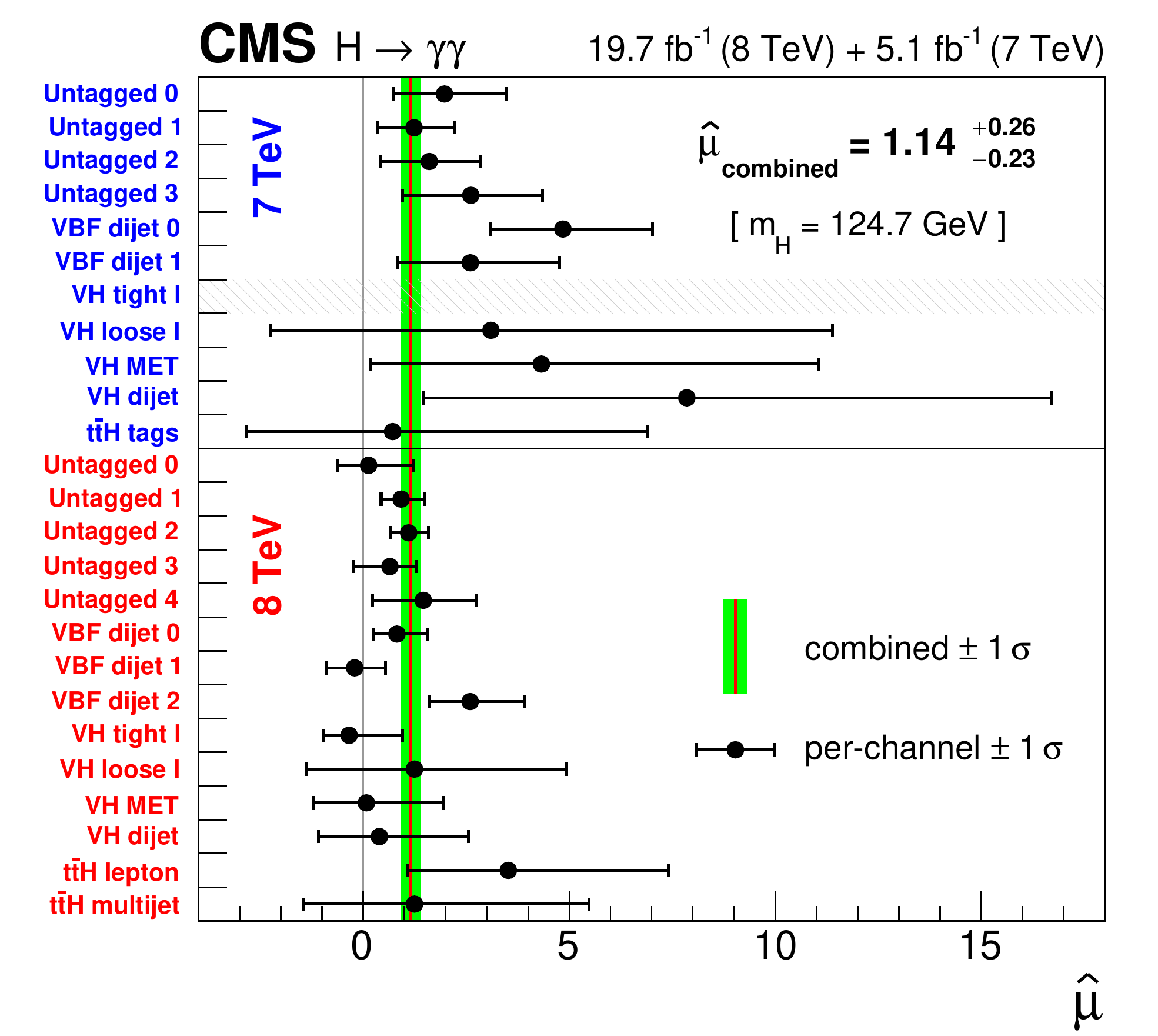} 
    \caption{Values of $\muhat$ measured individually for all event classes in the 7 and 8\TeV datasets,
    fixing $\mH=124.7\GeV$.
    The horizontal bars indicate ${\pm}1\,\sigma$ uncertainties in the values, and the
    vertical line and band indicate the best-fit signal strength in the combined fit to the data and its uncertainty.
  }
    \label{fig:compatibility}
\end{figure}

In Fig.~\ref{fig:compatibility} the best-fit signal strength, $\muhat$, is shown
for each event class in the combined 7 and 8\TeV datasets, fixing $\mH=124.7\GeV$ in the fits.
The horizontal bars indicate ${\pm}1\,\sigma$ uncertainties in the values, and the
vertical line and band indicate the best-fit signal strength in the combined fit to the data and its uncertainty.
The signal-plus-background fit for the VH tight-lepton tagged class in the 7\TeV dataset, when done alone,
does not converge because in this class and in the region of $\mgg$ where the signal is expected there are no events in the data.
No value for the signal strength in this class is shown in the figure.
The $\chi^2$ probability of the values for the 24 remaining classes being compatible
with the overall best-fit signal strength is 74\%.

\subsection{Mass measurement}

\begin{figure}
  \centering
    \includegraphics[width=0.49\textwidth]{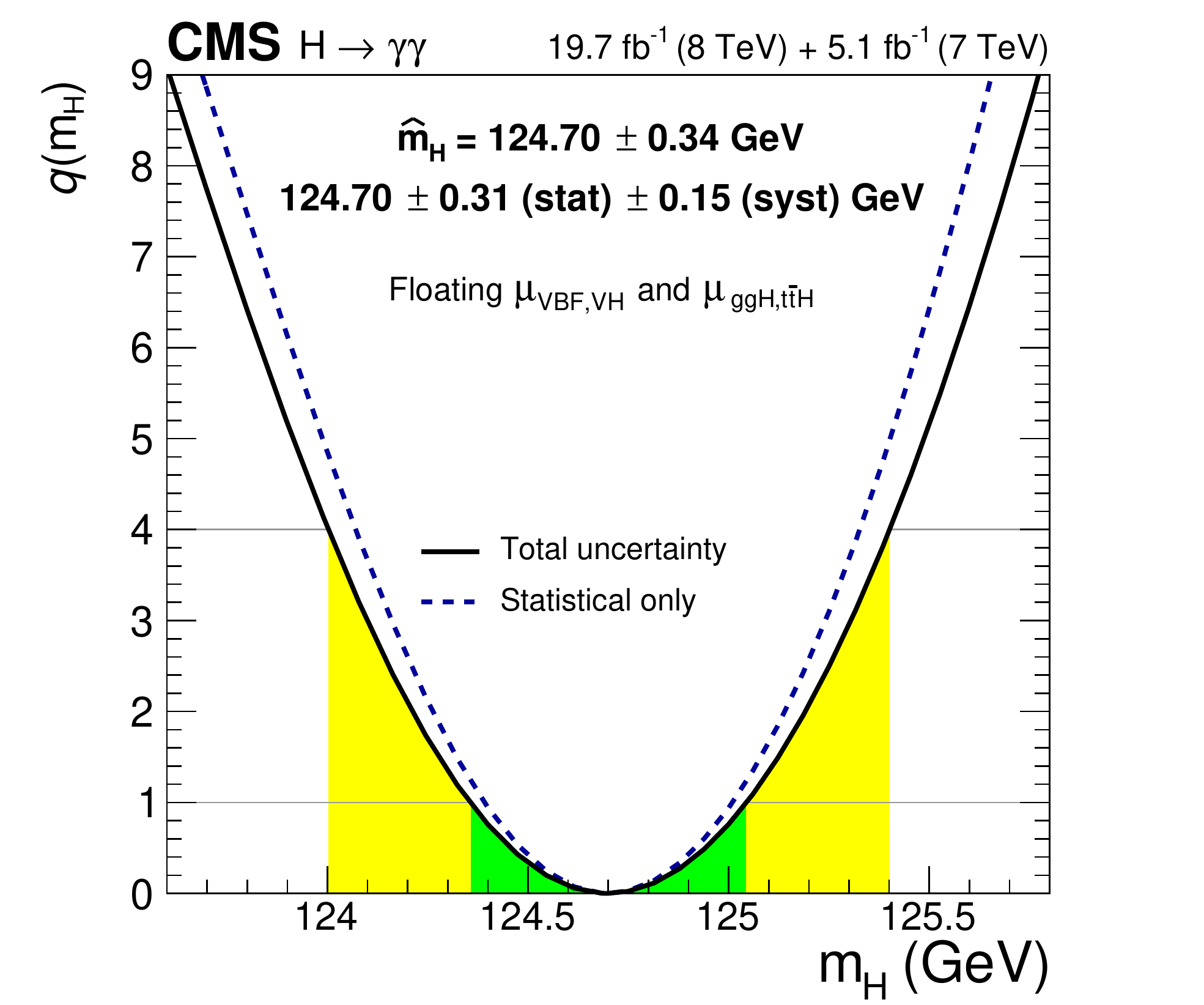}  
    \includegraphics[width=0.49\textwidth]{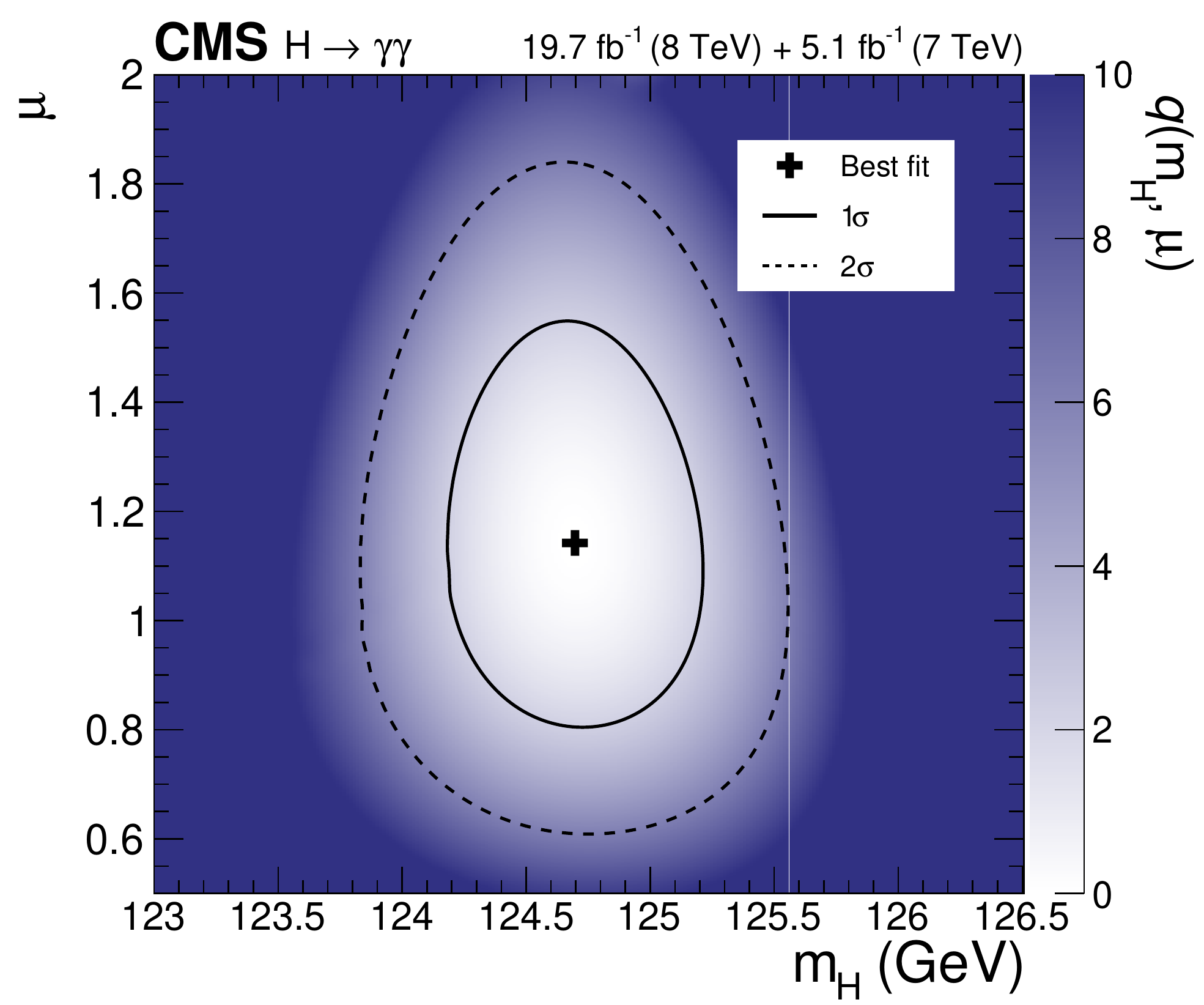}  
    \caption{(\cmsLeft) Scan of the likelihood ratio, $q$,
      as a function of the hypothesised mass when $\mu_{\Pg\Pg\PH, \ttH}$ and $\mu_\text{VBF, VH}$
      are allowed to vary independently.
      (\cmsRight) Map of $q(\mH,\mu)$ showing the $1\,\sigma$ and $2\,\sigma$ regions,
      and the best-fit point $(\mHhat, \muhat)=(124.70\GeV, 1.14)$.
}
    \label{fig:mu-vs-mh-likelihood}
\end{figure}

The four main Higgs boson production mechanisms can be associated with either
fermion couplings (ggH and \ttH) or vector boson couplings (VBF and VH).
To make the measurement of the mass of the observed resonance less model dependent
the signal strengths of the production processes involving the Higgs boson coupling to fermions and the
production processes involving the coupling to vector bosons, are allowed to vary independently.
The two signal strength modifiers are denoted $\mu_{\Pg\Pg\PH, \ttH}$ and $\mu_\text{VBF, VH}$.
Figure~\ref{fig:mu-vs-mh-likelihood} (\cmsLeft) shows the resulting scan of the
negative-log-likelihood ratio, $q$, defined in Equation~\ref{eq:q}, as a function of the mass hypothesis,
where $\mu_{\Pg\Pg\PH, \ttH}$ and $\mu_\text{VBF, VH}$ are treated as unconstrained parameters in the fit,
giving the mass of the observed boson as $124.70\pm0.34\GeV$.

Figure~\ref{fig:mu-vs-mh-likelihood} (\cmsRight) shows a map of the value of $q$ in a two-dimensional
scan of the ($\mH$, $\mu$) plane.
Here only a single signal strength modifier is allowed to vary, thus requiring
$\mu=\mu_{\Pg\Pg\PH, \ttH}=\mu_\text{VBF, VH}$,
and the mass measured is unchanged.
If the mass is measured in the 7~and 8\TeV datasets separately the values are found to differ by
less than $1\sigma$.
The uncertainty in the measured mass can be separated into statistical and systematic
contributions: $\mHhat=124.70\pm0.31\stat\pm0.15\syst\GeV$.
Systematic uncertainties from theory play a negligible role.
However, the effect of interference between ggH and the continuum diphoton background produced via quark loops
has not been taken into account.
This interference is expected to result in a downward shift of the observed mass~\cite{Martin:2012xc,Dixon:2013haa}.
Taking the parameterization given in Ref.~\cite{Dixon:2013haa} we expect a shift
of less than 20\MeV in our analysis.

The calibration of the energy scale is achieved using $\Zee$ events as a reference,
as described in Section~\ref{sec:photonE}.
Systematic uncertainties related to individual photons as described in Section~\ref{sec:systematics} are propagated to the signal model, where they result in uncertainties in the signal peak position and width.
The three main sources of systematic uncertainty in the energy scale that contribute to the uncertainty in the measured mass are shown in Table~\ref{tab:sys-mass},
where the contributions of the 81 nuisance parameters in the analysis are grouped according to their physical origin, as relevant to the mass uncertainty.
The largest contributions are due to the possible imperfect simulation of
(i) differences in detector response to electrons and photons arising from a number of factors that have been discussed
in Section~\ref{sec:systematics}, and
(ii) the energy scale nonlinearity in the extrapolation from the $\cPZ$-boson mass to the Higgs boson mass.
A further contribution comes from the uncertainties in the setting of the energy scale itself, that is, in the procedure
and methodology of using measurement of the invariant mass in \Zee\ events in which the electron showers are reconstructed as photons.
Other sources of systematic uncertainty contribute little.

Additional possible sources of uncertainty that have been investigated and found to be negligible are
a possible bias related to the choice of background parameterization, which has been studied using pseudo-experiments
where the effect is found to be less than 10\MeV; the effect of the switch of preamplifier when very large signals,
$E\gtrsim200\GeV$ in the barrel and $\ET\gtrsim80\GeV$ in the endcaps, are digitized using a preamplifier with lower gain;
and the effect of imperfect simulation of the effect of signals from interactions in previous bunch crossings.

\begin{table*}[htbp]
\topcaption{Magnitude of the uncertainty in the best fit mass induced by the systematic uncertainties in the signal model.
These numbers have been obtained by quadratic subtraction of the statistical uncertainty.
The statistical uncertainty includes all uncertainties in the background modelling.
}
\centering
\begin{tabular}{ l c }
\multirow{2}{*}{{Source of uncertainty}} & {Uncertainty in} \\
 & {$\widehat{m}_{\PH}$} {(\GeVns)}\\
\hline
Imperfect simulation of electron-photon differences & 0.10 \\
Linearity of the energy scale & 0.10 \\
Energy scale calibration and resolution & 0.05 \\
Other & 0.04 \\
\hline
All systematic uncertainties in the signal model & 0.15 \\ 
 \multicolumn{1}{r}{Statistical} & \multicolumn{1}{r}{0.31} \\
\hline
 \multicolumn{1}{r}{{Total}} & \multicolumn{1}{r}{{0.34}} \\
\end{tabular}\label{tab:sys-mass}
\end{table*}

\subsection{Production mechanisms and coupling modifiers}

\begin{figure}[hbtp]
  \centering
    \includegraphics[width=\cmsFigWidth]{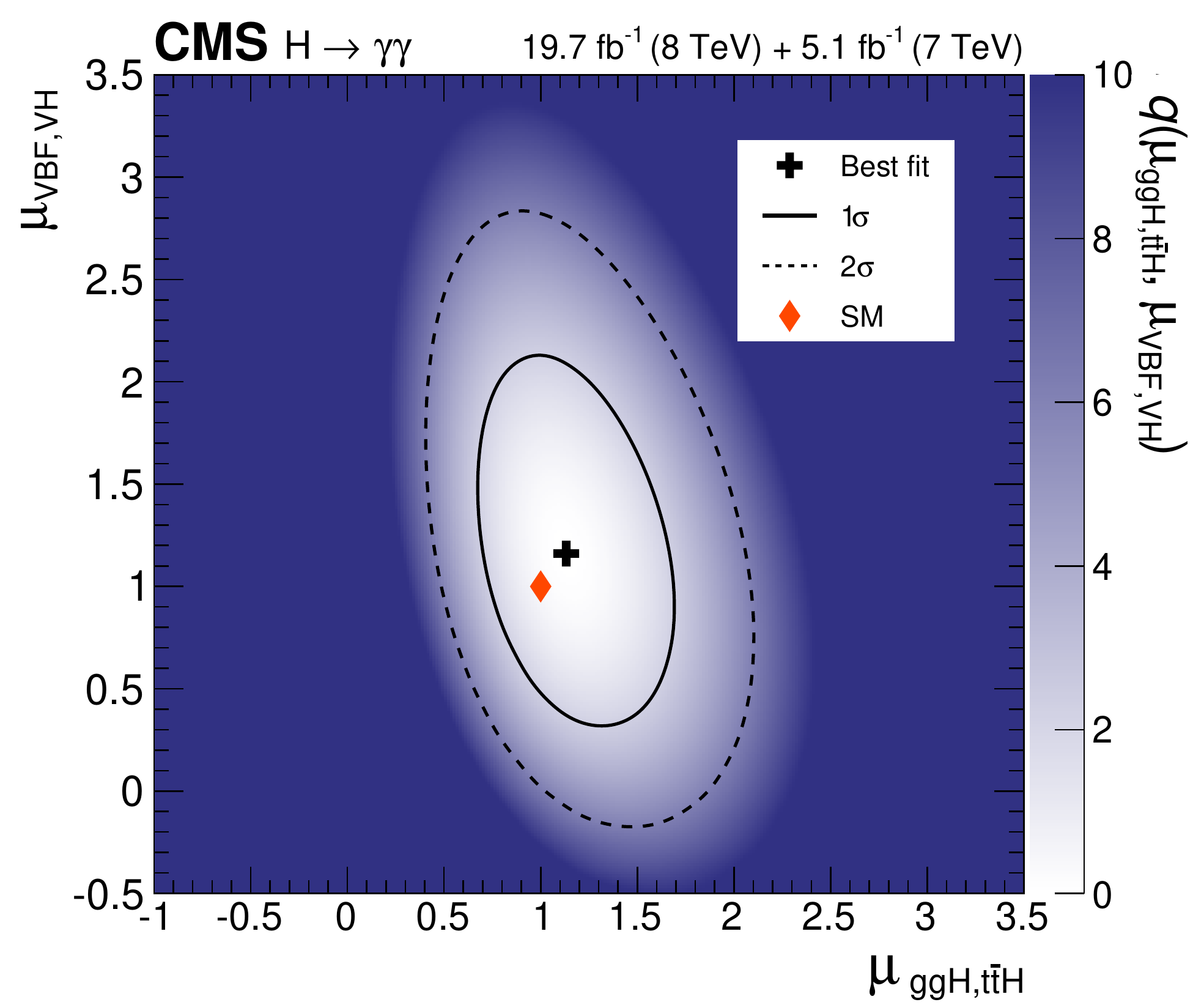}  
    \caption{Map of the likelihood ratio $q(\mu_{\Pg\Pg\PH, \ttH},\mu_\text{VBF, VH})$
       with \mH\ treated as an unconstrained parameter.
       The $1\,\sigma$ and $2\,\sigma$ uncertainty contours are shown.
       The cross indicates the best-fit values,
       ($\muhat_{\Pg\Pg\PH, \ttH}, \muhat_\text{VBF, VH})=(1.13, 1.16)$,
        and the diamond represents the SM expectation.
}
    \label{fig:rvrf-2d}
\end{figure}

Figure~\ref{fig:rvrf-2d} shows the $1\,\sigma$ and $2\,\sigma$ contours,
computed as the variations around the likelihood maximum,
for the signal strength modifiers $\mu_{\Pg\Pg\PH, \ttH}$ and $\mu_\text{VBF, VH}$.
The best-fit values of these signal strength modifiers, when they are both allowed to vary,
and \mH\ is treated as an unconstrained parameter in the fit,
are found to be $\muhat_{\Pg\Pg\PH, \ttH}=1.13^{+0.37}_{-0.31}$ and $\muhat_\text{VBF, VH}=1.16^{+0.63}_{-0.58}$.
These numbers are tabulated in Table~\ref{tab:exp-mu-unc}, together with the
expected uncertainty in each signal strength modifier.

\begin{table}[hbtp]
    \topcaption{\label{tab:exp-mu-unc} Expected and observed best-fit values of the signal
      strength modifiers $\mu_{\Pg\Pg\PH, \ttH}$ and $\mu_\text{VBF, VH}$ for a SM Higgs boson signal
      together with their uncertainties,
      indicating the expected uncertainty in the measurement and the best-fit values obtained from the data.
      }
  \centering
   \renewcommand{\arraystretch}{1.2}
    \begin{tabular}{l|c|c}
     & Expected & Observed \\  \hline
      $\muhat_{\Pg\Pg\PH, \ttH}$ & $1.00^{+0.34}_{-0.30}$ & $1.13^{+0.37}_{-0.31}$ \\
      $\muhat_\text{VBF, VH}$ & $1.00^{+0.57}_{-0.51}$ & $1.16^{+0.63}_{-0.58}$ \\
    \end{tabular}
\end{table}
\renewcommand{\arraystretch}{1.0}

\begin{figure}[hbtp]
  \centering
    \includegraphics[width=\cmsFigWidth]{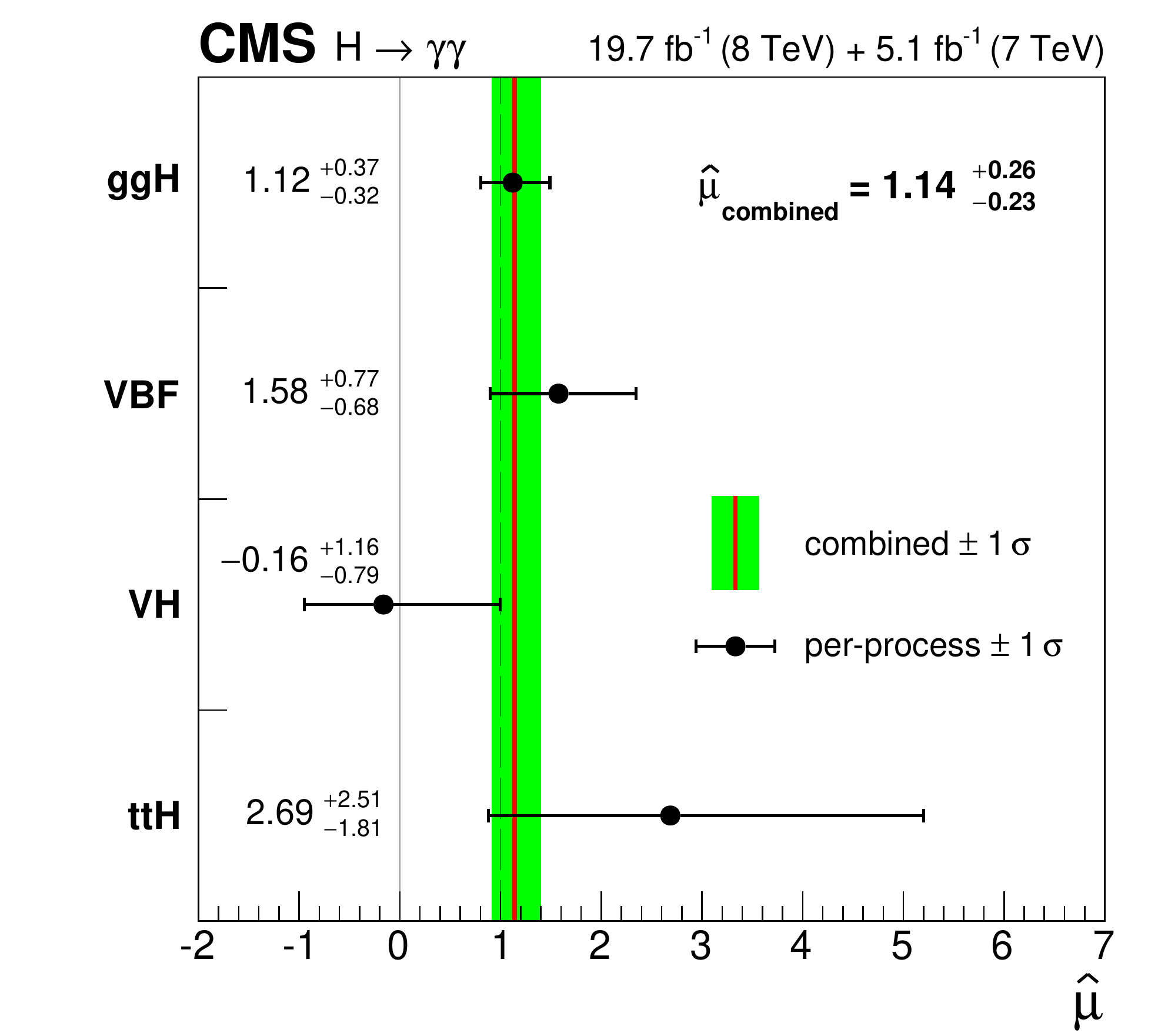} 
    \caption{Best-fit signal strength, $\muhat$, measured for each of the production processes in a combined fit
     where the signal strengths of all four processes have been allowed to vary independently in the fit.
     The signal mass, common to all four processes, is treated as an unconstrained parameter in the fit.
     The horizontal bars indicate ${\pm}1\,\sigma$ uncertainties in the values for the individual processes.
     The band corresponds to ${\pm}1\,\sigma$
     uncertainties in the value obtained from the combined fit with a single signal strength.
}
    \label{fig:proc-comp}

\end{figure}

If the signal strengths of all four production processes are allowed to vary independently in the fit,
the values of $\musm$ measured for each process are compatible with the expectations for a SM Higgs boson,
as shown in Fig.~\ref{fig:proc-comp}.
The signal mass, common to all four processes, is treated as an unconstrained parameter in the fit.
The horizontal bars indicate ${\pm}1\,\sigma$ uncertainties in the values.
For comparison, the dijet 2D analysis obtains the value $\muhat_\text{VBF}=1.6^{+0.9}_{-0.7}$, whereas the
result of the main analysis, shown in the plot, is $\muhat_\text{VBF}=1.6^{+0.8}_{-0.7}$.
Table~\ref{tab:process-breakdown} shows the four signal strengths observed, with the contributions to their uncertainties
separated into statistical and systematic components.
The systematic uncertainty has been separated, where feasible, into the contributions from
theoretical uncertainties, and other (experimental) uncertainties.

\begin{table} [hbtp]
    \topcaption{\label{tab:process-breakdown}Best-fit signal strength modifiers for the four production processes.
    The total uncertainty for each process is separated into statistical (stat) and systematic contributions.
    The systematic uncertainty has been separated, where feasible, into the contributions from
    theoretical (theo), and experimental (exp) uncertainties.
    To obtain the values, the quadratic subtraction, needed to remove the statistical uncertainty, is made for
    the positive and negative uncertainties separately. The values quoted are the average magnitudes of the positive and negative uncertainties.
      }
  \centering
   \renewcommand{\arraystretch}{1.2}
    \begin{tabular}{l c | c c c c}
      & &  \multicolumn{4}{c}{Uncertainty} \\
      \multirow{2}{*}{Process} & \multirow{2}{*}{$\hat{\mu}$} & \multirow{2}{*}{total} & \multirow{2}{*}{stat} & \multicolumn{2}{c}{systematic} \\
      & & & & theo & exp \\ \hline
      ggH & $\phantom{-}1.12^{+0.37}_{-0.32}$  & 0.34 & 0.30 & 0.13 & 0.09 \\
      VBF & $\phantom{-}1.58^{+0.77}_{-0.68}$ & 0.73 & 0.69 & 0.20 & 0.15 \\
      VH &  $-0.16^{+1.16}_{-0.79}$ & 0.97 & 0.97 & \multicolumn{2}{c}{0.08}  \\
      $\ttH$ & $\phantom{-}2.69^{+2.51}_{-1.81}$ & 2.2 & 2.1 & \multicolumn{2}{c}{0.4} \\
    \end{tabular}
\end{table}
\renewcommand{\arraystretch}{1.0}

\begin{figure}[hbtp]
  \centering
    \includegraphics[width=0.49\textwidth]{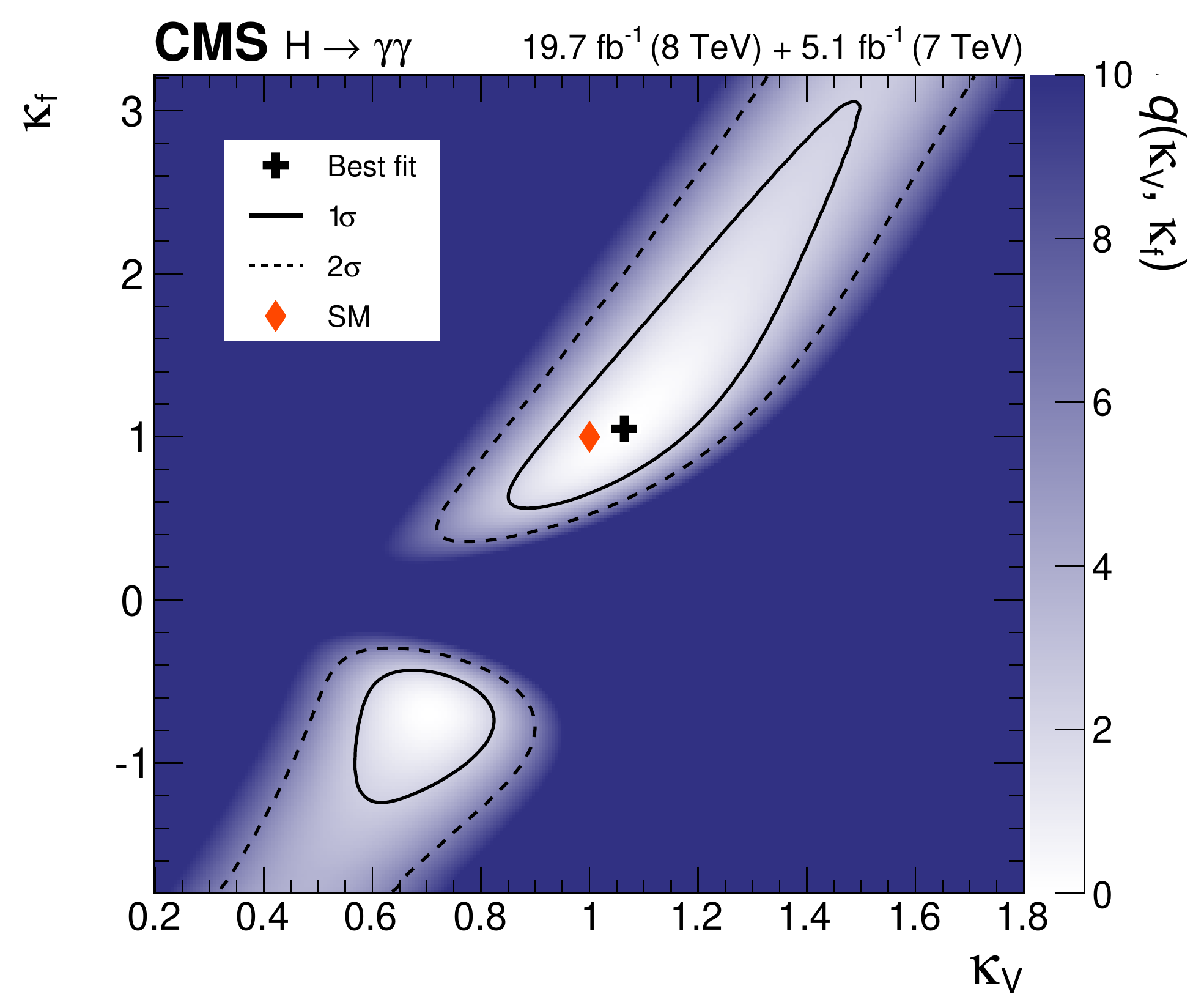} 
    \includegraphics[width=0.49\textwidth]{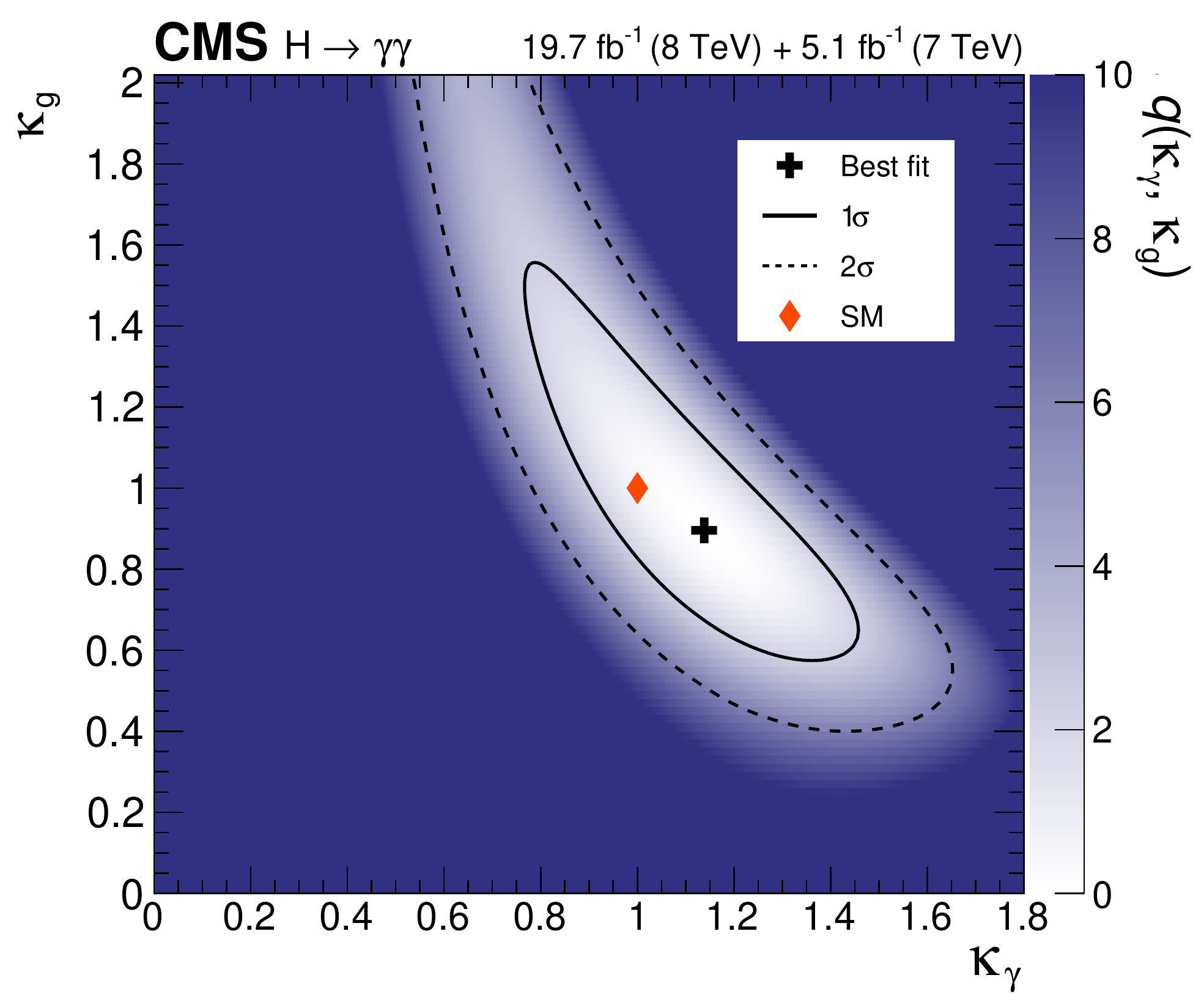} 
    \caption{Maps of the  likelihood ratio $q(\kV,\kf)$ (\cmsLeft), and $q(\kgm,\kg)$ (\cmsRight),
      showing the $1\,\sigma$ and $2\,\sigma$ uncertainty contours.
      The crosses indicate the best-fit values, and the diamonds indicate the SM expectation.
}
    \label{fig:kappas}
\end{figure}

Various parameterizations of the couplings can be used to further test the compatibility of the observed new particle with the predictions
for a SM Higgs boson~\cite{LHCHiggsCrossSectionWorkingGroup3}.
Figure~\ref{fig:kappas} shows two-dimensional likelihood scans of $\kf$ versus $\kV$ (\cmsLeft) and $\kg$ versus $\kgm$ (\cmsRight).
The variables $\kV$ and $\kf$ are, respectively, the coupling modifiers of the new particle to vector bosons and to fermions;
alternatively, $\kgm$ and $\kg$ are the effective coupling modifiers to photons and to gluons; all four variables
are expressed relative to the SM expectations.
For each scan a fixed value of $\mH=124.7\GeV$ is used, and it has been verified that
allowing $\mH$ to vary produces an indistinguishable result.
The best-fit points are $(\kV, \kf)=(1.06, 1.05)$, and $(\kgm, \kg)=(1.14, 0.90)$.

\subsection{Decay width}
It is possible to set a limit on the width of the observed signal, albeit a limit far in excess of the SM expectation of 4\MeV for $\mH=125\GeV$.
To accommodate the natural width of the Higgs boson, the Gaussian
components used in the signal model of the SM analysis, where the signal width is assumed to be negligible as compared to the detector resolution,
are replaced by an analytic convolution of a Breit--Wigner distribution (modelling a nonzero decay width) with a Gaussian distribution (modelling
the detector resolution).

\begin{figure}[hbtp]
  \centering
    \includegraphics[width=\cmsFigWidth]{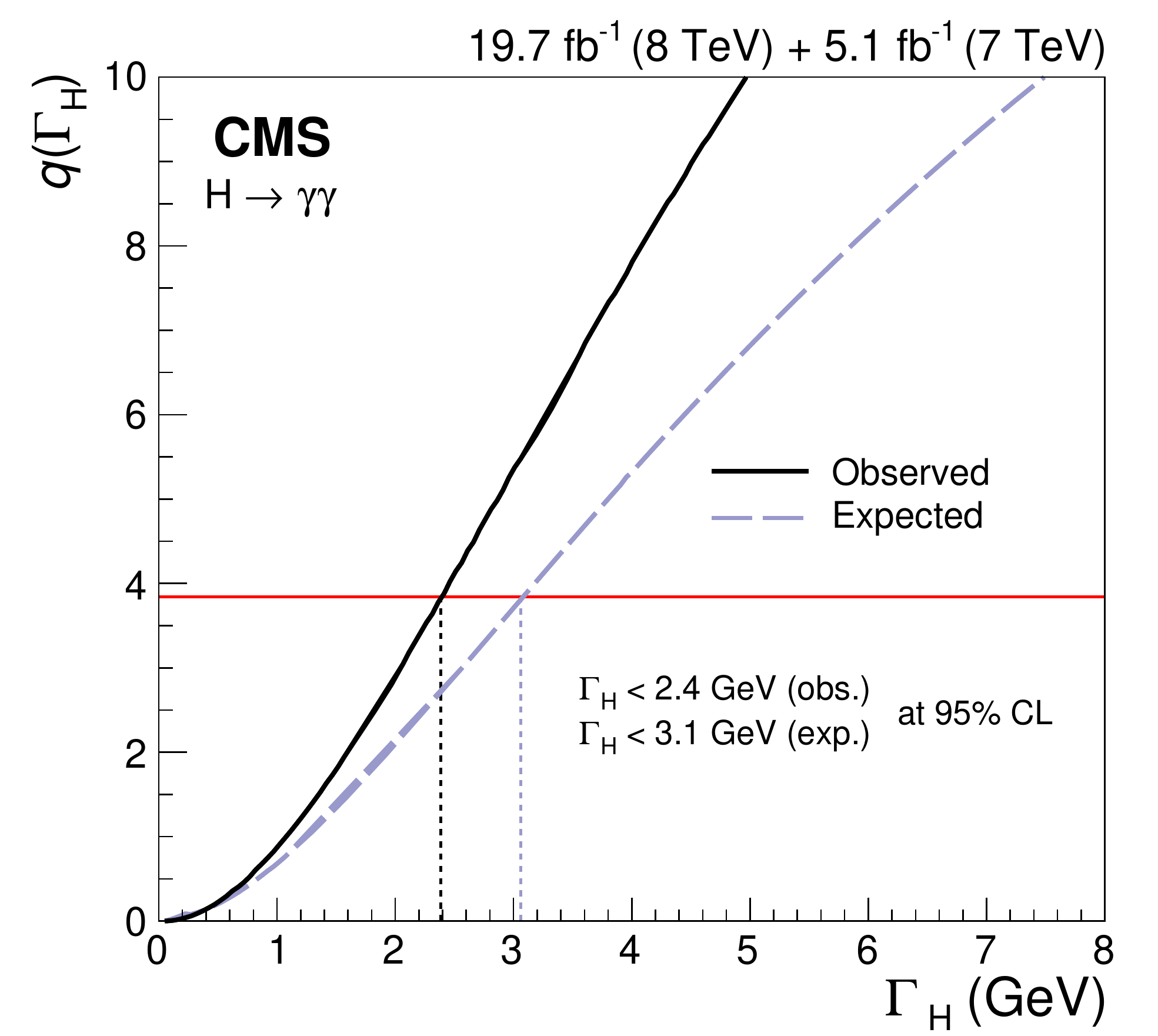} 
    \caption{Scan of the negative-log-likelihood ratio as a function of the Higgs boson decay width.
The observed (expected) upper limit on the width is found to be 2.4 (3.1)\GeV at a 95\% CL.
}
    \label{fig:width}
\end{figure}

A profile likelihood estimator is used to calculate upper limits on the width of the observed boson whilst
allowing the Higgs boson mass to vary in the fit.
Figure~\ref{fig:width} shows a scan of the negative-log-likelihood ratio as a function of the observed new particle's decay width for the combined 7 and 8\TeV dataset.
The observed (expected) upper limit on the width is found to be 2.4 (3.1)\GeV at a 95\% confidence level (CL).

\subsection{Search for additional Higgs-boson-like states}

\begin{figure}[hbtp]
  \centering
    \includegraphics[width=\cmsFigWidth]{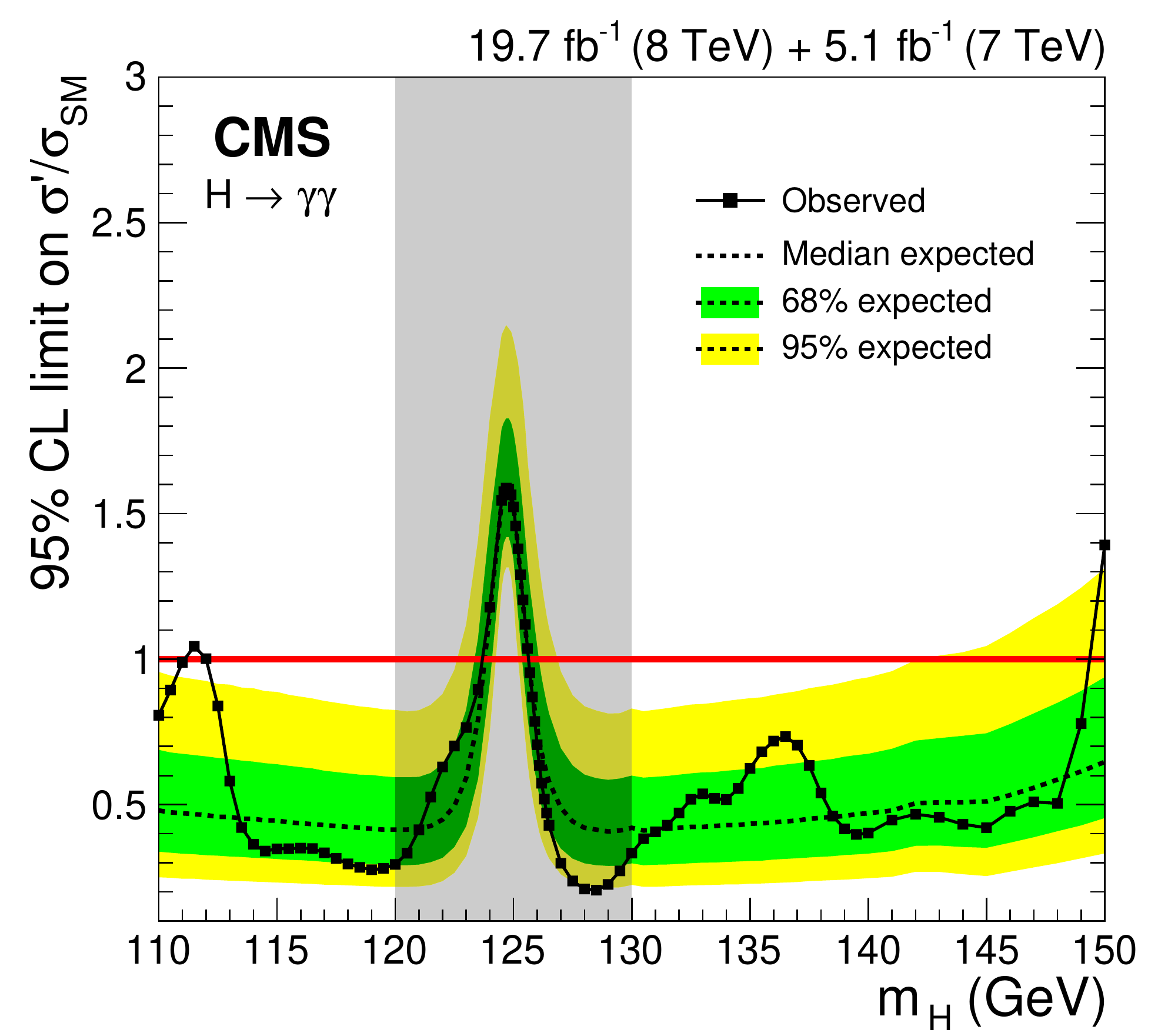} 
    \caption{Exclusion limit on the signal strength, $\musmx$, for a second Higgs-boson-like
      state with SM couplings taking the observed state at 125\GeV as
      part of the background. The shading indicates a window with a width of 10\GeV, centred at the best-fit mass,
      where the expected sensitivity to a second Higgs boson is severely degraded due to the presence of the already observed state.
}
    \label{fig:another-higgs-limit}
\end{figure}

To search for a possible additional Higgs-boson-like state, $\PH'$, in the mass range
$110\leq m_{\PH'}\leq150\GeV$, the observed signal around 125\GeV is added to the background
model and its mass and signal strength are allowed to vary in the fit.
An additional, independent signal model is introduced as a second Higgs boson, for which the exclusion
limits are calculated using the modified frequentist method and the \CLs criterion~\cite{Junk:1999kv, Read:2002hq}.
In order to set limits for the combined 7 and 8\TeV datasets it is necessary to make
an assumption about the ratio of cross sections of the new state at 7 and 8\TeV.
By expressing the limit in terms of the SM cross section times branching fraction
we implicitly assume that the ratio is that of the SM.
The resulting exclusion limit is shown in Fig.~\ref{fig:another-higgs-limit}.
Once sufficiently away from 125\GeV, the same limit is obtained as when searching for a single SM Higgs boson.
The shading indicates a window with a width of 10\GeV, centred at the best-fit mass,
where the expected sensitivity to a second Higgs boson
is severely degraded due to the presence of the already observed state.

\begin{figure}[hbtp]
  \centering
      \includegraphics[width=0.49\textwidth]{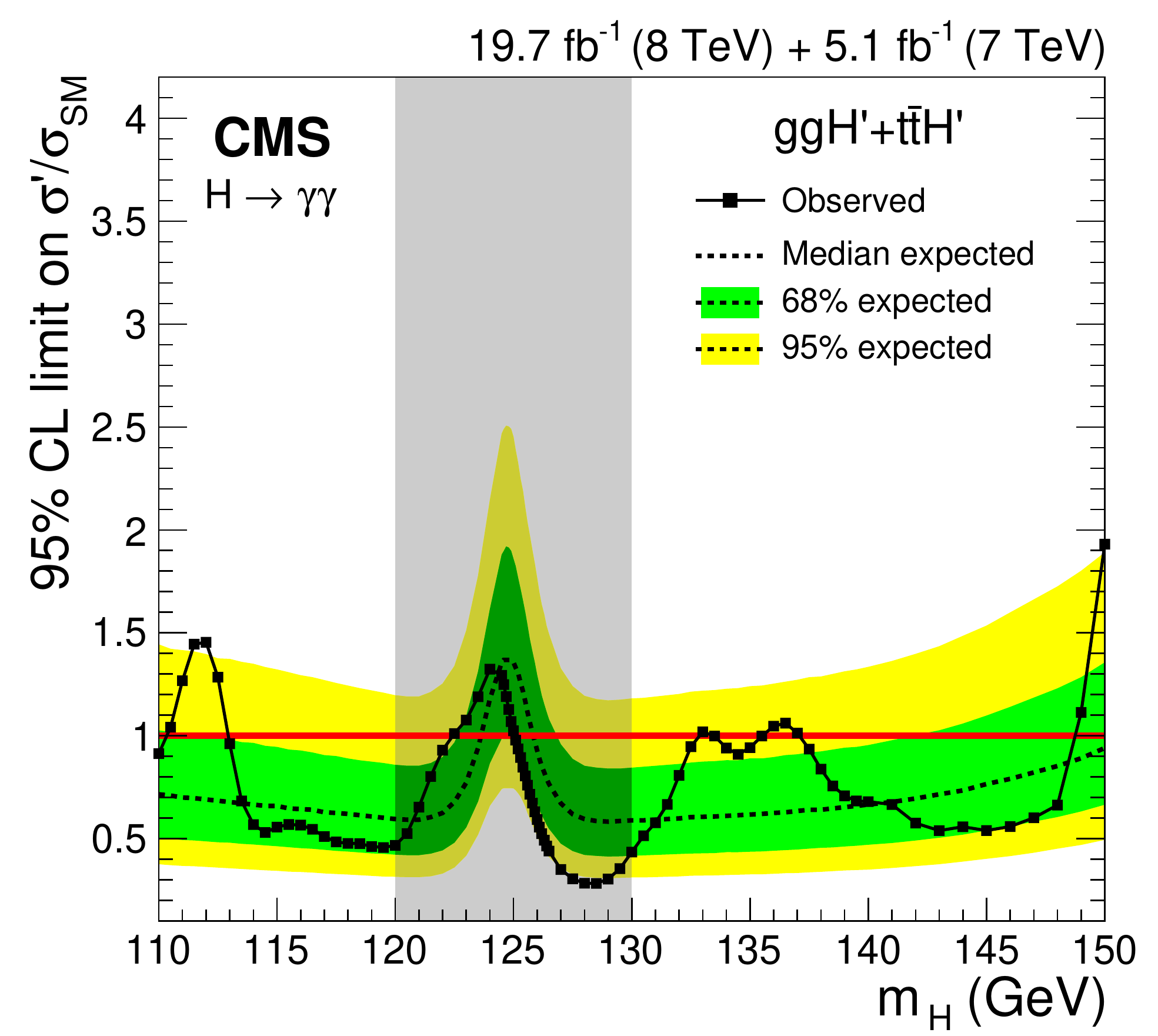} 
      \includegraphics[width=0.49\textwidth]{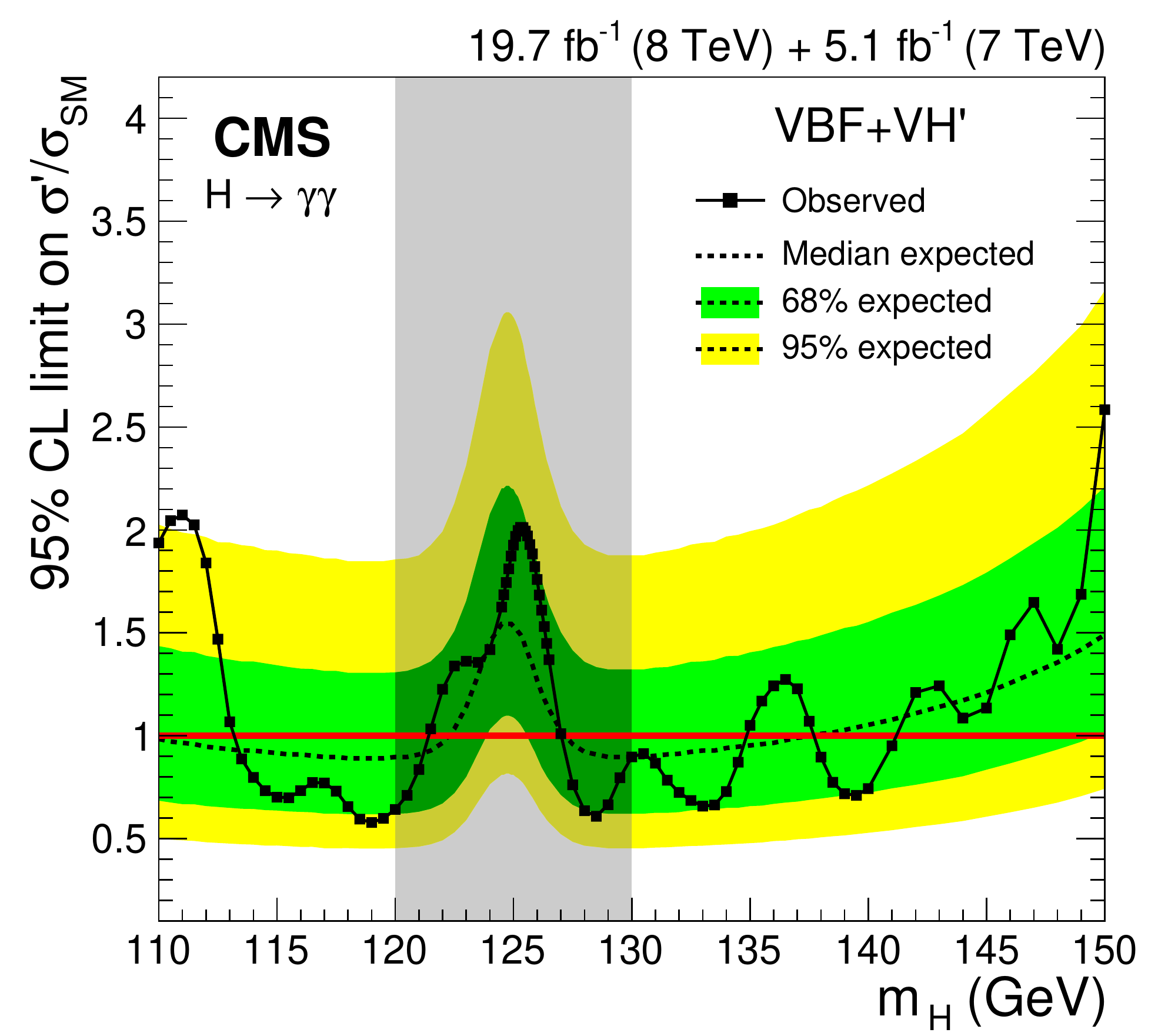} 
      \caption{Exclusion limits on $\musmx$ for a second
      Higgs-boson-like state produced with gluon-gluon fusion only (\cmsLeft) or
      VBF and VH only (\cmsRight) taking the observed state at 125\GeV as part of the background.
      The shading indicates a window with a width of 10 GeV, centred at the best-fit mass,
      where the expected sensitivity to a second Higgs boson
      is severely degraded due to the presence of the already observed state.}
      \label{fig:another-higgs-limit-proc}

\end{figure}

A further particular case of interest is when the second state couples only to
fermions, for example in the alignment limit of some two-Higgs-doublet models~\cite{Branco:2011iw}.
We also examine the case where the second state couples only to bosons at the tree level.
Figure~\ref{fig:another-higgs-limit-proc} shows the exclusion limits obtained when the
observed signal near 125\GeV is added to the background model and its mass and signal strength are allowed to vary in the fit,
and an additional state produced (\cmsLeft) only by the gluon-fusion process, or (\cmsRight) only by the VBF and VH processes.
The limits are given in terms of the SM cross section times branching fraction for those processes.
Even for the VBF and VH processes, which have lower cross sections, an additional state with SM-like signal strength
is excluded or disfavoured over much of the mass range.

\begin{figure}[h!]
  \centering
      \includegraphics[width=\cmsFigWidth]{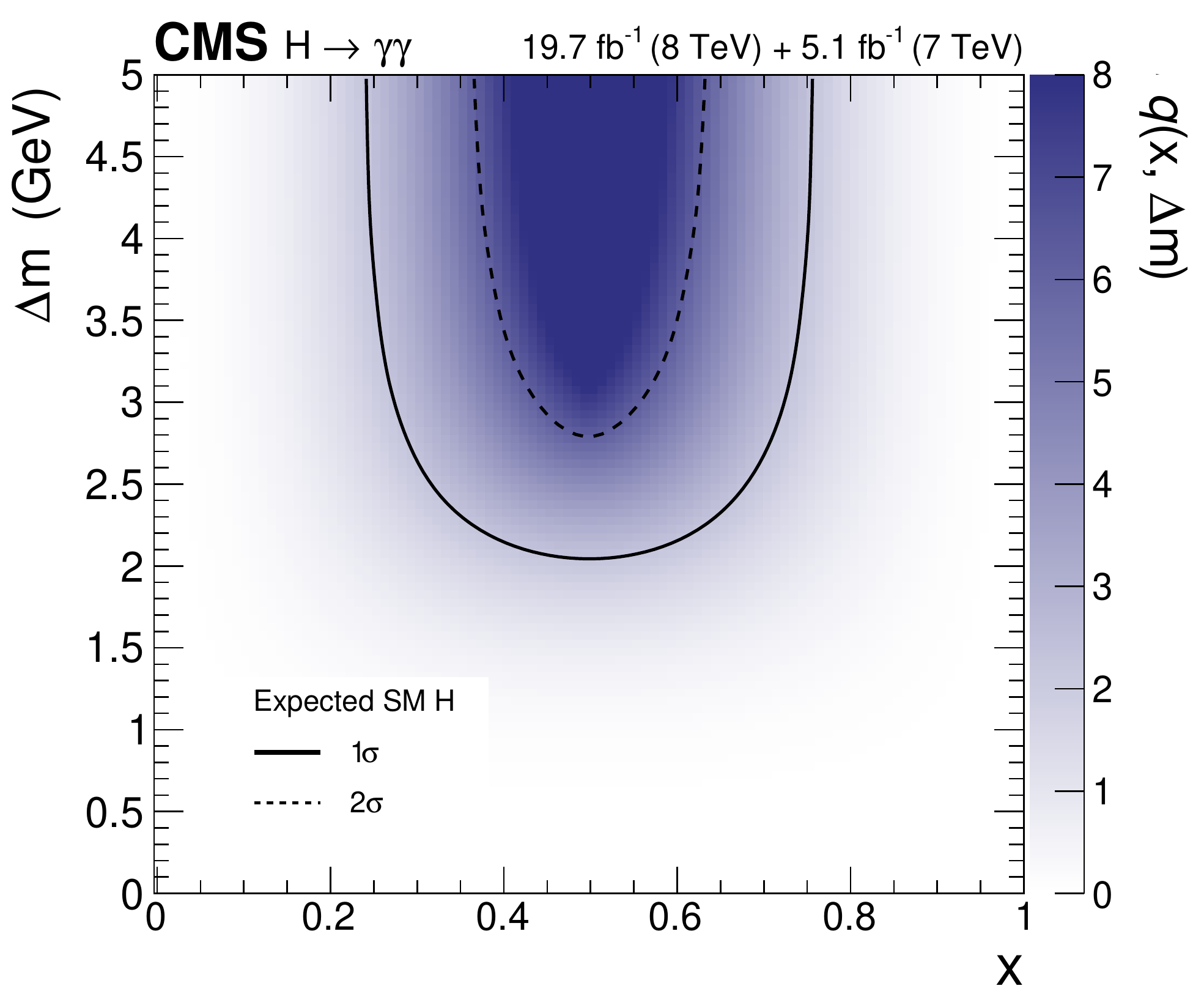} 
      \includegraphics[width=\cmsFigWidth]{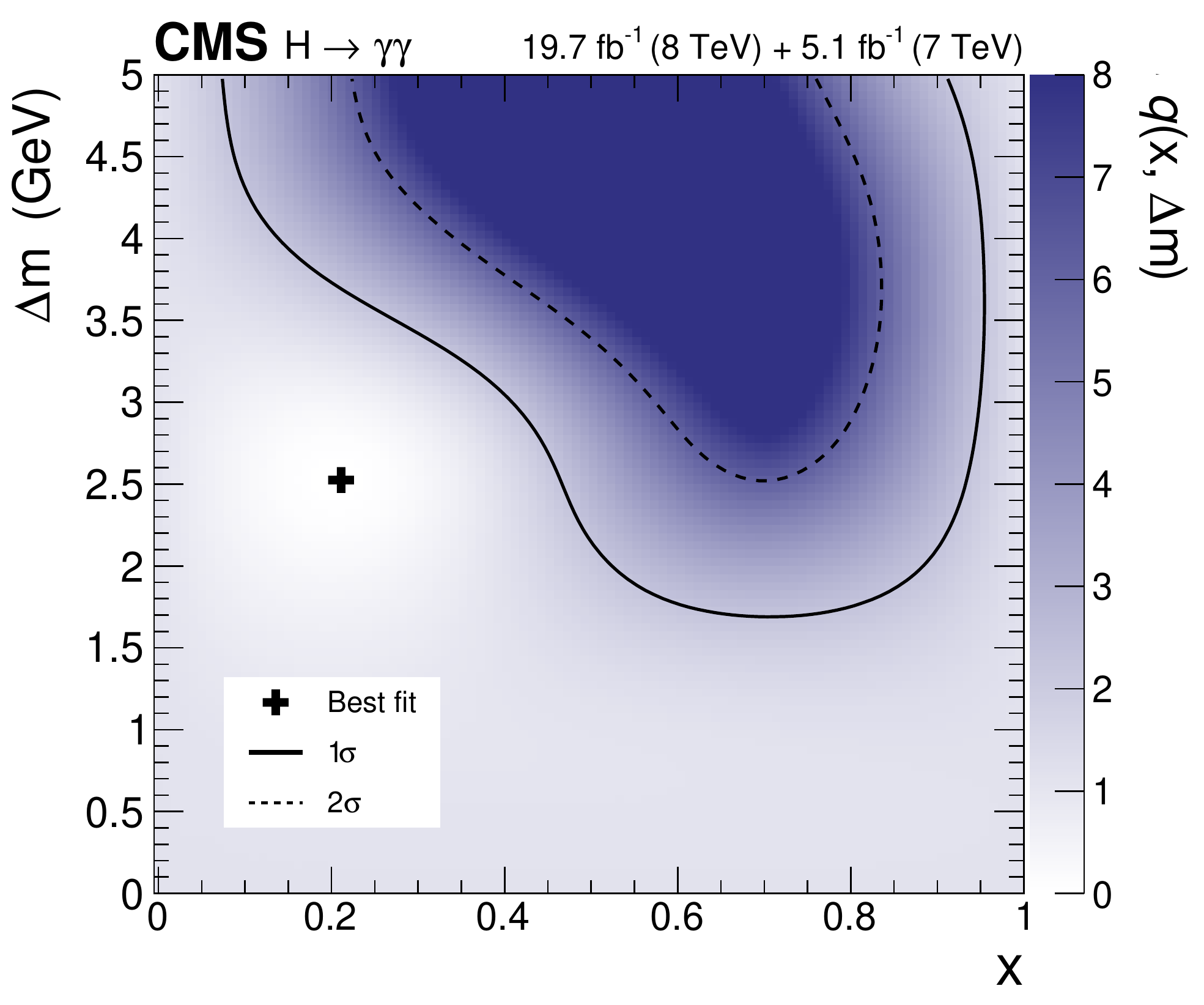} 
    \caption{Map of the values of the likelihood ratio $q(x,\Delta m)$ for two near
    mass-degenerate states parameterized by $x$ (the fraction of signal in the lower mass state)
    and $\Delta m$ (the mass difference between the states).
    The black cross shows the best-fit value, and the lines correspond to the $1\sigma$ and $2\sigma$ uncertainty contours
    for the SM (single state) expectation (upper plot) and the observation (lower plot).
}
    \label{fig:twoDegenHiggs} 
\end{figure}

The shaded regions in Figs.~\ref{fig:another-higgs-limit} and~\ref{fig:another-higgs-limit-proc}, where the expected sensitivity
to a second Higgs boson is severely degraded due to the presence of the already observed state, are probed by a dedicated search using
the high resolution of the diphoton channel to provide sensitivity to a pair of states separated by only a few \GeV.
The signal model is re-parameterized with two signals, having masses $m_{\PH'}$ and $m_{\PH'}+\Delta m$.
The relative strengths of the two signals, parameterized by the variable $x$, is allowed to vary such that the two signals are
modulated by $\mu x$ and $\mu(1-x)$ respectively, where $\mu$ is the total signal strength and $x$ is the fraction of signal contained in the state lower in mass.
A two-dimensional scan of $\Delta m$ and $x$ is obtained, while allowing both $m_{\PH'}$ and $\mu$ to vary as free parameters in the fit.
Figure~\ref{fig:twoDegenHiggs} shows the expected (upper plot) and observed (lower plot) negative-log-likelihood ratio in the $(x, \Delta m)$ plane.
Sensitivity is expected in regions where $\Delta m$ is close to or greater than the experimental mass resolution
and where the two signal strengths are similar.
The black cross shows the best-fit value, and the lines correspond to the $1\sigma$ and $2\sigma$ uncertainty contours
for the SM (\ie a single state).
It can be seen that a region of the parameter space is disfavoured at more than $2\sigma$: where the ratio of the signal strengths
is between 0.2 and 0.8 and the mass difference is greater than values ranging between 2.5 and 4\GeV depending on the ratio of the signal strengths.
The somewhat asymmetrical shape of the excluded region and the position of the best-fit value, are a
reflection of the slightly asymmetrical mass peak seen in Fig.~\ref{fig:mgg-weighted}, also reflected in the figures showing the local $p$-value,
and exclusion limit as a function of $\mH$.

\subsection{Testing spin hypotheses}
\label{sec:spin}

The Landau--Yang theorem forbids the direct decay of a spin-1 particle
into a pair of photons~\cite{Landau1948,Yang1950}.
However, it is of interest to compare the hypothesis of a spin-2 ``graviton-like'' model with minimal couplings,
$\twomp$,~\cite{Gao2010}, to that of a spin-0 SM-Higgs-boson-like, $\zerop$, model.
As the $\twomp$ is just one of many possible realizations of the spin-2 tensor structure, an attempt has been
made to make the analysis as model independent as possible.
Tests have been performed for hypotheses
in which the \twomp resonance is produced entirely by gluon-fusion ($\cPg\cPg$),
in which it is produced entirely by quark-antiquark annihilation ($\cPq\cPaq$), and for cases in which
it is produced by a mixture of the two processes.
The cosine of the scattering angle in the Collins--Soper frame, \costhetastar
~\cite{CollinsSoper1977}, is used to discriminate between the two hypotheses.
The angle is defined, in the diphoton rest frame, as that between the collinear photons
and the line that bisects the acute angle between the colliding protons:

\begin{equation}
  \costhetastar=2\times\frac{E^{\gamma2}p^{\gamma1}_{z}-E^{\gamma1}p^{\gamma2}_{z}}{\mgg\sqrt{\mgg^{2}+(\ptgg)^{2}}},
\end{equation}

where $E^{\gamma1}$ and $E^{\gamma2}$ are the energies of the leading and subleading photons,
$p^{\gamma1}_{z}$ and $p^{\gamma2}_{z}$ are the $z$ components of their momenta,
and $\mgg$ and $\ptgg$ are the invariant mass and transverse momentum of the diphoton system.
In the rest frame of a spin-0 boson the decay photons are isotropic,
and so, before the acceptance requirements, the distribution of \costhetastar is uniformly flat under the \zerop hypothesis.
In general this is not the case for the decay of a spin-2 particle.

To increase the sensitivity, the events are categorized using
the same four diphoton event classes used in the cut-based analysis, described in Section~\ref{sec:cut-based},
but without the addition classification based on $\ptgg$ used there.
Within each diphoton class, the events are binned in \abscostheta to discriminate between the different spin hypotheses.
The events are thus split into 20 event classes, four $(\eta, \RNINE)$
diphoton classes with five \abscostheta bins each, for both the 7 and 8\TeV datasets, giving a total of 40 event classes.

\begin{figure}[hbtp]
	\centering
	\includegraphics[width=\cmsFigWidth]{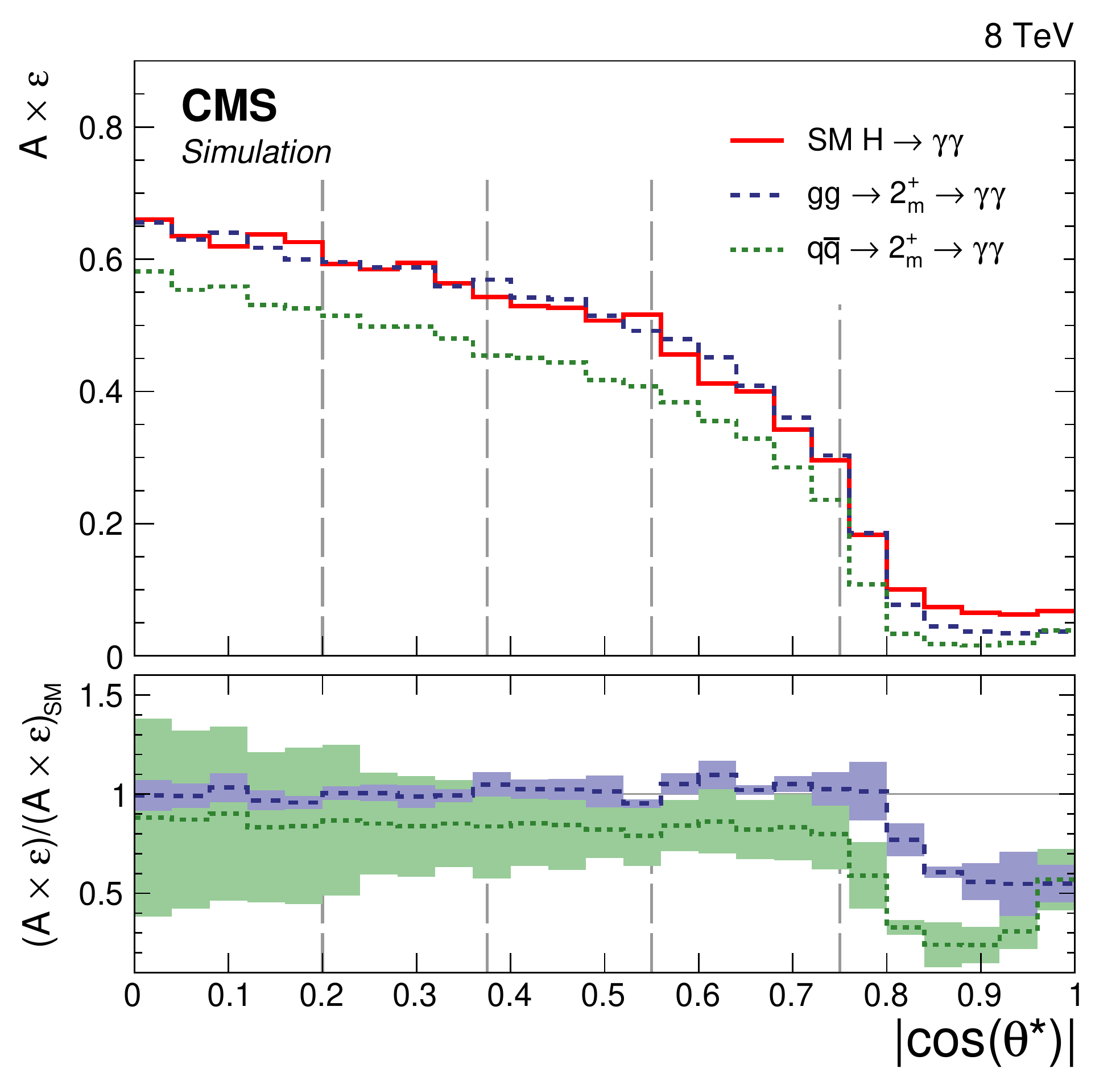} 
  \caption{Product of acceptance and efficiency $A\times\epsilon$ for \zerop (all SM production modes),
\twomp (gluon-fusion) and  \twomp ($\cPq\cPaq$ production) as a function of $\abscostheta$, as calculated for the 8\TeV dataset.
The value of $A\times\epsilon$ for the \twomp models divided by $A\times\epsilon$ for SM is shown below, where
the bands indicate the spread of values among the four diphoton classes.
The \abscostheta bin boundaries are shown by vertical dashed lines.
}
	\label{fig:eff-acc-spin}
\end{figure}

Although the acceptance times efficiency, $A\times\epsilon$, varies considerably as a function of \abscostheta, this
variation is, for gluon-fusion production, independent of the spin-parity models tested.
This is also true in the restricted ranges of $\eta$ and $\RNINE$ defined by
the diphoton classes, which
allows the extraction of the signal yield in bins of \abscostheta in a reasonably model independent way.
Figure~\ref{fig:eff-acc-spin} shows $A\times\epsilon$ for \zerop (all SM production modes), \twomp (gluon-fusion) and  \twomp ($\cPq\cPaq$ production) as a function of $\abscostheta$, as calculated for the 8\TeV dataset. The \abscostheta bin boundaries are shown by vertical dashed lines.
The value of $A\times\epsilon$ for the \twomp models divided by $A\times\epsilon$ for SM is shown below, where
the bands indicate the spread of values among the four diphoton classes.
It can be seen that the ratio is flat, independent of \abscostheta, except at the highest
values of \abscostheta where the relative contribution from SM VBF production is significant.
The events in the region where the ratio falls from its flat level, $0.75<\abscostheta<1.0$, are collected in a separate bin,
and the \abscostheta bin boundaries for the remaining events are chosen to maintain approximately
the same event yield in each bin.

\begin{figure}[hbtp]
  \centering
    \includegraphics[width=\cmsFigWidth]{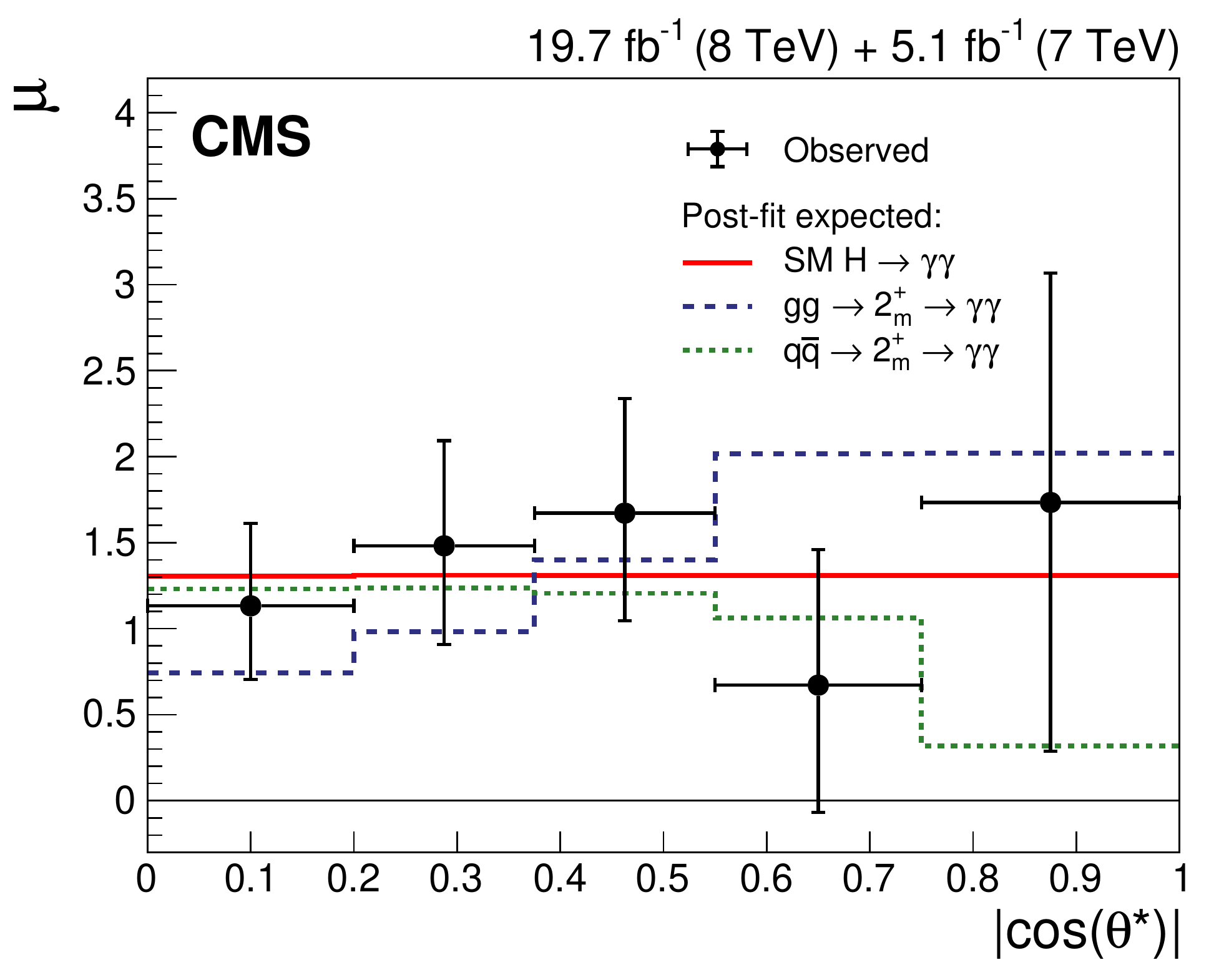} 
    \caption{Histograms showing signal strength in five bins of $\abscostheta$ expected for SM, for $\twomp$ produced
by $\cPg\cPg$, and for \twomp produced by $\cPq\cPaq$.
The signal strength observed in the data is shown by the black points.
}
    \label{fig:compat-costh}
\end{figure}

{\tolerance=800
Figure~\ref{fig:compat-costh} shows histograms of the expected signal strength, $\mu$, relative to the SM expectation
in the five bins of $\abscostheta$ for the SM, and for two $\twomp$ models:
where the $\twomp$ resonance is produced entirely by gluon-fusion ($\cPg\cPg$),
and where it is produced entirely by quark-antiquark annihilation ($\cPq\cPaq$).
The expected values in the five bins are obtained by constructing a representative pseudo-data model
in which the overall signal strength has been set to be that obtained from fitting the model in question, plus background, to the data.
When generating pseudo-experiments for a particular model, the value
of all the free parameters, including the signal nuisance
parameters, the background shape parameters, and the overall signal strength,
are set to their best-fit values obtained by fitting the model in question to the data with a single overall value of the signal strength.
The post-fit expected value of the signal strength for the SM signal model
is thus that which is observed when simultaneously fitting the 40 event classes
with a single signal strength, \ie $1.31^{+0.33}_{-0.31}$.
The observed $\mu$ values in the five bins shown in the figure are obtained from a simultaneous fit
of the SM-signal-plus-background model to the 40 event classes, with five signal strength variables
(one for each $\abscostheta$ bin) and a common $\mH$ allowed to vary.
\par}

\begin{figure}[hbtp]
  \centering
    \includegraphics[width=\cmsFigWidth]{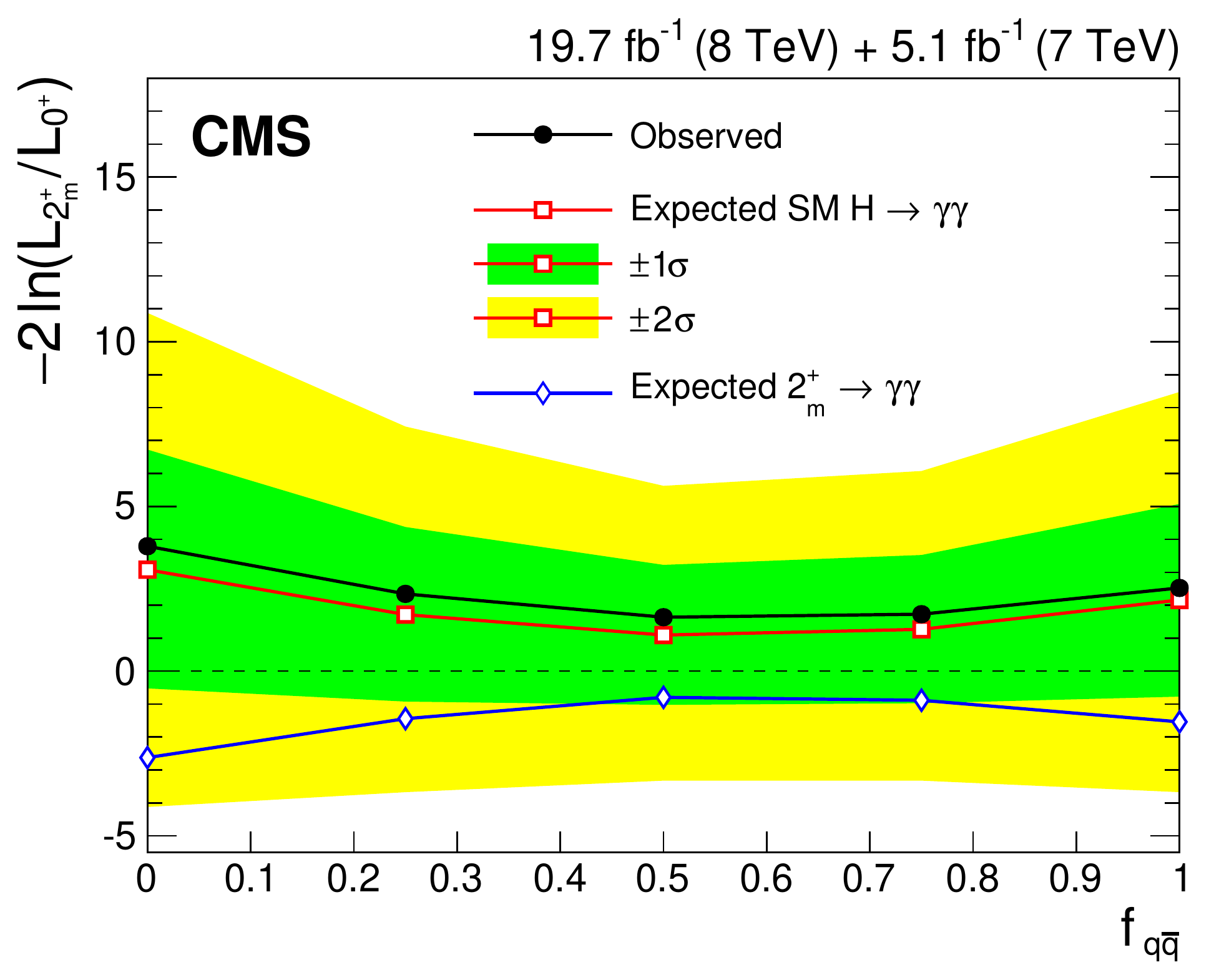} 
    \caption{Test statistic for pseudo-experiments generated under the SM, \zerop, hypothesis (open squares) and the graviton-like,
\twomp, hypothesis (open diamonds), as a function of the fraction, $\fqq$, of $\cPq\cPaq$ production.
The observed distribution in the data is shown by the black points.
}
    \label{fig:vary-fqq}
\end{figure}

The separation between the two models is extracted using a test statistic defined as twice the negative logarithm of the ratio
of the likelihoods for the $\zerop$ signal plus background hypothesis and the $\twomp$ signal plus background hypothesis when
performing a simultaneous fit of all forty event classes together,
$q=-2\,{\ln(\mathcal{L}_{2^{+}_{m} + \text{bkg}}/\mathcal{L}_{0^+ + \text{bkg}})}$.
The test is made under the assumption that the \twomp state is produced entirely by either gluon-fusion, or entirely by quark-antiquark annihilation,
or by three intermediate mixtures of $\cPg\cPg$ and $\cPq\cPaq$ spin-2 production.
The fraction of the spin-2 state produced by $\cPq\cPaq$ annihilation is parameterized by the variable $\fqq$, so that the
total signal plus background, $f(\mH)$, is given by

\begin{equation}
f(\mH)=\mu[(1-\fqq)\times S_{\cPg\cPg}^{\twomp}(\mH)+\fqq\times S_{\cPq\cPaq}^{\twomp}(\mH)]+B(\mH),
\end{equation}

where $S_{\cPg\cPg}^{\twomp}(\mH)$ is the $\cPg\cPg$-produced $\twomp$ signal, $S_{\cPq\cPaq}^{\twomp}(\mH)$ the $\cPq\cPaq$-produced $\twomp$ signal,
$\mu$ is a signal strength modifier, and $B(\mH)$ is the background.
Figure~\ref{fig:vary-fqq} shows the values of the test statistic as a function of $\fqq$.
Table~\ref{tab:spinCLs} gives the values of $1-\CLs$, expected and observed, which measures the extent to which the spin-2 model is
disfavoured, for different values of $\fqq$.
The hypothesis of the signal being $\twomp$ is disfavoured for all values of $\fqq$ tested.
When produced entirely by gluon fusion, it is disfavoured with a $1-\CLs$ value of 94\% (92\% expected).
When produced entirely by $\cPq\cPaq$ annihilation it is disfavoured with a $1-\CLs$ value of 85\% (83\% expected).
Intermediate mixtures, where there is less sensitivity to distinguish between the models, are somewhat less disfavoured.

\begin{table}[hbtp]
  \centering
    \caption{Expected and observed values of $1-\CLs$ for the \twomp signal hypothesis with
                  respect to the \zerop hypothesis, for different mixtures of $\cPg\cPg$ and $\cPq\cPaq$ production.}
    \label{tab:spinCLs}
    \begin{tabular}{l | c  c }
      \multirow{2}{*}{$\fqq$} & \multicolumn{2}{c}{$1-\CLs$} \\
                                           & expected & observed \\
      \hline
      0 & 0.92 & 0.94 \\
      0.25 & 0.78 & 0.83 \\
     0.50 & 0.64 & 0.71 \\
     0.75 & 0.69 & 0.75 \\
      1 & 0.83 & 0.85 \\
    \end{tabular}
\end{table}

\section{Summary}
\label{sec:summary}

We report the observation of the diphoton decay mode of the recently discovered Higgs boson
and measurement of some of its properties.
The analysis uses the entire dataset collected by the CMS experiment in
proton-proton collisions during the 2011 and 2012 LHC running periods.
The data samples correspond to integrated
luminosities of 5.1\fbinv at $\sqrt{s}=7\TeV$ and 19.7\fbinv at 8\TeV.
The selected events are subdivided into classes, designed to enhance the overall sensitivity and to increase the sensitivity
to individual Higgs production mechanisms, and the results of the search in all classes are
reported.

{\tolerance=800
A clear signal is observed in the diphoton channel at a mass of 124.7\GeV
with a local significance of $5.7\,\sigma$, where a significance of $5.2\,\sigma$ is expected for the standard model Higgs boson.
The mass is measured to be $124.70\pm0.34\GeV=124.70\pm0.31\stat\pm0.15\syst\GeV$, and the best-fit signal strength
relative to the standard model prediction is $1.14^{+0.26}_{-0.23}=1.14\pm0.21\stat^{+0.09}_{-0.05}\syst^{+0.13}_{-0.09}\thy$.
The best-fit values for the signal strength modifiers associated with the ggH and \ttH production mechanisms, and with the
VBF and VH mechanisms are found to be $\muhat_{\Pg\Pg\PH, \ttH}=1.13^{+0.37}_{-0.31}$
and $\muhat_\text{VBF, VH}= 1.16^{+0.63}_{-0.58}$.
\par}

A direct upper limit on the natural width of the state is set at 2.4\GeV (3.1\GeV expected) at a 95\% confidence level, and
additional SM-like Higgs bosons are excluded at a 95\% confidence level in a large fraction
of the mass range between 110 and 150\GeV.
The SM spin-0 hypothesis for the observed state is compared to a graviton-like spin-2 hypothesis with minimal couplings.
The hypothesis of the signal being $\twomp$ is disfavoured.
When produced entirely by gluon fusion, it is disfavoured with a $1-\CLs$ value of 94\% (92\% expected).

All the results are compatible with the expectations from a standard model Higgs boson.

\section*{Acknowledgements}
\hyphenation{Bundes-ministerium Forschungs-gemeinschaft Forschungs-zentren} We congratulate our colleagues in the CERN accelerator departments for the excellent performance of the LHC and thank the technical and administrative staffs at CERN and at other CMS institutes for their contributions to the success of the CMS effort. In addition, we gratefully acknowledge the computing centres and personnel of the Worldwide LHC Computing Grid for delivering so effectively the computing infrastructure essential to our analyses. Finally, we acknowledge the enduring support for the construction and operation of the LHC and the CMS detector provided by the following funding agencies: the Austrian Federal Ministry of Science, Research and Economy and the Austrian Science Fund; the Belgian Fonds de la Recherche Scientifique, and Fonds voor Wetenschappelijk Onderzoek; the Brazilian Funding Agencies (CNPq, CAPES, FAPERJ, and FAPESP); the Bulgarian Ministry of Education and Science; CERN; the Chinese Academy of Sciences, Ministry of Science and Technology, and National Natural Science Foundation of China; the Colombian Funding Agency (COLCIENCIAS); the Croatian Ministry of Science, Education and Sport, and the Croatian Science Foundation; the Research Promotion Foundation, Cyprus; the Ministry of Education and Research, Estonian Research Council via IUT23-4 and IUT23-6 and European Regional Development Fund, Estonia; the Academy of Finland, Finnish Ministry of Education and Culture, and Helsinki Institute of Physics; the Institut National de Physique Nucl\'eaire et de Physique des Particules~/~CNRS, and Commissariat \`a l'\'Energie Atomique et aux \'Energies Alternatives~/~CEA, France; the Bundesministerium f\"ur Bildung und Forschung, Deutsche Forschungsgemeinschaft, and Helmholtz-Gemeinschaft Deutscher Forschungszentren, Germany; the General Secretariat for Research and Technology, Greece; the National Scientific Research Foundation, and National Innovation Office, Hungary; the Department of Atomic Energy and the Department of Science and Technology, India; the Institute for Studies in Theoretical Physics and Mathematics, Iran; the Science Foundation, Ireland; the Istituto Nazionale di Fisica Nucleare, Italy; the Korean Ministry of Education, Science and Technology and the World Class University program of NRF, Republic of Korea; the Lithuanian Academy of Sciences; the Ministry of Education, and University of Malaya (Malaysia); the Mexican Funding Agencies (CINVESTAV, CONACYT, SEP, and UASLP-FAI); the Ministry of Business, Innovation and Employment, New Zealand; the Pakistan Atomic Energy Commission; the Ministry of Science and Higher Education and the National Science Centre, Poland; the Funda\c{c}\~ao para a Ci\^encia e a Tecnologia, Portugal; JINR, Dubna; the Ministry of Education and Science of the Russian Federation, the Federal Agency of Atomic Energy of the Russian Federation, Russian Academy of Sciences, and the Russian Foundation for Basic Research; the Ministry of Education, Science and Technological Development of Serbia; the Secretar\'{\i}a de Estado de Investigaci\'on, Desarrollo e Innovaci\'on and Programa Consolider-Ingenio 2010, Spain; the Swiss Funding Agencies (ETH Board, ETH Zurich, PSI, SNF, UniZH, Canton Zurich, and SER); the Ministry of Science and Technology, Taipei; the Thailand Center of Excellence in Physics, the Institute for the Promotion of Teaching Science and Technology of Thailand, Special Task Force for Activating Research and the National Science and Technology Development Agency of Thailand; the Scientific and Technical Research Council of Turkey, and Turkish Atomic Energy Authority; the National Academy of Sciences of Ukraine, and State Fund for Fundamental Researches, Ukraine; the Science and Technology Facilities Council, UK; the US Department of Energy, and the US National Science Foundation.

Individuals have received support from the Marie-Curie programme and the European Research Council and EPLANET (European Union); the Leventis Foundation; the A. P. Sloan Foundation; the Alexander von Humboldt Foundation; the Belgian Federal Science Policy Office; the Fonds pour la Formation \`a la Recherche dans l'Industrie et dans l'Agriculture (FRIA-Belgium); the Agentschap voor Innovatie door Wetenschap en Technologie (IWT-Belgium); the Ministry of Education, Youth and Sports (MEYS) of the Czech Republic; the Council of Science and Industrial Research, India; the HOMING PLUS programme of Foundation for Polish Science, cofinanced from European Union, Regional Development Fund; the Compagnia di San Paolo (Torino); and the Thalis and Aristeia programmes cofinanced by EU-ESF and the Greek NSRF.
\bibliography{auto_generated}   

\cleardoublepage \appendix\section{The CMS Collaboration \label{app:collab}}\begin{sloppypar}\hyphenpenalty=5000\widowpenalty=500\clubpenalty=5000\textbf{Yerevan Physics Institute,  Yerevan,  Armenia}\\*[0pt]
V.~Khachatryan, A.M.~Sirunyan, A.~Tumasyan
\vskip\cmsinstskip
\textbf{Institut f\"{u}r Hochenergiephysik der OeAW,  Wien,  Austria}\\*[0pt]
W.~Adam, T.~Bergauer, M.~Dragicevic, J.~Er\"{o}, C.~Fabjan\cmsAuthorMark{1}, M.~Friedl, R.~Fr\"{u}hwirth\cmsAuthorMark{1}, V.M.~Ghete, C.~Hartl, N.~H\"{o}rmann, J.~Hrubec, M.~Jeitler\cmsAuthorMark{1}, W.~Kiesenhofer, V.~Kn\"{u}nz, M.~Krammer\cmsAuthorMark{1}, I.~Kr\"{a}tschmer, D.~Liko, I.~Mikulec, D.~Rabady\cmsAuthorMark{2}, B.~Rahbaran, H.~Rohringer, R.~Sch\"{o}fbeck, J.~Strauss, A.~Taurok, W.~Treberer-Treberspurg, W.~Waltenberger, C.-E.~Wulz\cmsAuthorMark{1}
% ...
\vskip\cmsinstskip
% ...
1:~~Also at Vienna University of Technology, Vienna, Austria\\
2:~~Also at CERN, European Organization for Nuclear Research, Geneva, Switzerland\\
% ...

\end{sloppypar}
\end{document}